\documentclass{article}
\usepackage[utf8]{inputenc}
\usepackage{amssymb}
\usepackage[usenames,dvipsnames]{xcolor}
\usepackage{amsmath}
\usepackage{amsthm}
\usepackage{comment}
\usepackage[colorlinks=true,linkcolor=MidnightBlue,citecolor=MidnightBlue]{hyperref}
\usepackage{color}
\usepackage[normalem]{ulem}
\usepackage{tabularray}
\usepackage{multirow}
\usepackage{xcolor}
\usepackage{graphicx}
\usepackage{tabularx}
\usepackage{cleveref}
\usepackage{lscape}
\usepackage{booktabs}
\usepackage[normalem]{ulem}
\useunder{\uline}{\ul}{}
\usepackage{longtable}
\usepackage{float}
\usepackage{array}    
\usepackage{authblk}

\newcommand{\sgn}{\textrm{sgn}}

\usepackage[natbib=true,url=false,isbn=false,sorting=none,maxnames=5,giveninits=true]{biblatex}
\usepackage{pdflscape} 
\usepackage{tabularx}  
\PassOptionsToPackage{hyphens}{url}\usepackage{hyperref}

\makeatletter
\renewcommand{\maketitle}{
    \begin{center}
        {\Huge\bfseries \@title \par}      
        \vskip 1em
        {\Large \@author \par}             
        \vskip 1em
    \end{center}
}
\makeatother

\title{Signed Networks: theory, methods, and applications}
\author[1,2]{Fernando Diaz-Diaz}
\author[3,4]{Elena Candellone}
\author[1]{Miguel A. González-Casado}
\author[5,6]{Emma Fraxanet}
\author[7,8]{Antoine Vendeville}
\author[9,10,11]{Irene Ferri}
\author[12,13,14]{Andreia Sofia Teixeira}

\affil[1]{Universidad Carlos III de Madrid, Departamento de Matemáticas, Grupo Interdisciplinar de Sistemas Complejos, 28911 Leganés, Spain}
\affil[2]{Institute of Cross-Disciplinary Physics and Complex Systems, IFISC (UIB-CSIC), 07122 Palma de Mallorca, Spain}
\affil[3]{Department of Methodology and Statistics, Utrecht University, Utrecht, Netherlands}
\affil[4]{Centre for Complex Systems Studies, Utrecht University, Utrecht, Netherlands}
\affil[5]{Department of Engineering, Universitat Pompeu Fabra, Barcelona, Spain }
\affil[6]{Computational Social Sciences and Humanities Lab, Barcelona Supercomputing Center, Barcelona, Spain}
\affil[7]{Sciences Po médialab, 75007 Paris, France}
\affil[8]{Complex Systems Institute of Paris Ile-de-France CNRS, 75013 Paris, France}
\affil[9]{Departament de F\'isica de la Mat\`eria Condensada, Universitat de Barcelona (UB), c. Mart\'i i Franqu\`es, 1, 08028 Barcelona, Spain}
\affil[10]{Institut de Recerca en Sistemes Complexos (UBICS), Universitat de Barcelona (UB), Barcelona, Spain}
\affil[11]{The Roux Institute, Network Science Institute, Northeastern University, 04101 Portland, ME USA}
\affil[12]{BRAN Lab, Network Science Institute, Northeastern University London, London E1W 1LP, United Kingdom}
\affil[13]{Khoury College of Computer Sciences, Northeastern University, Boston, MA 02115, United States}
\affil[14]{LASIGE, Departamento de Inform\'{a}tica, Universidade de Lisboa, Campo Grande 1749-016, Lisboa, Portugal}

\date{\vspace{-5ex}}

\begin{document}

\maketitle
\tableofcontents

\newpage
\section{Introduction}

\begin{figure}
    \centering
    \includegraphics[width=0.8\linewidth]{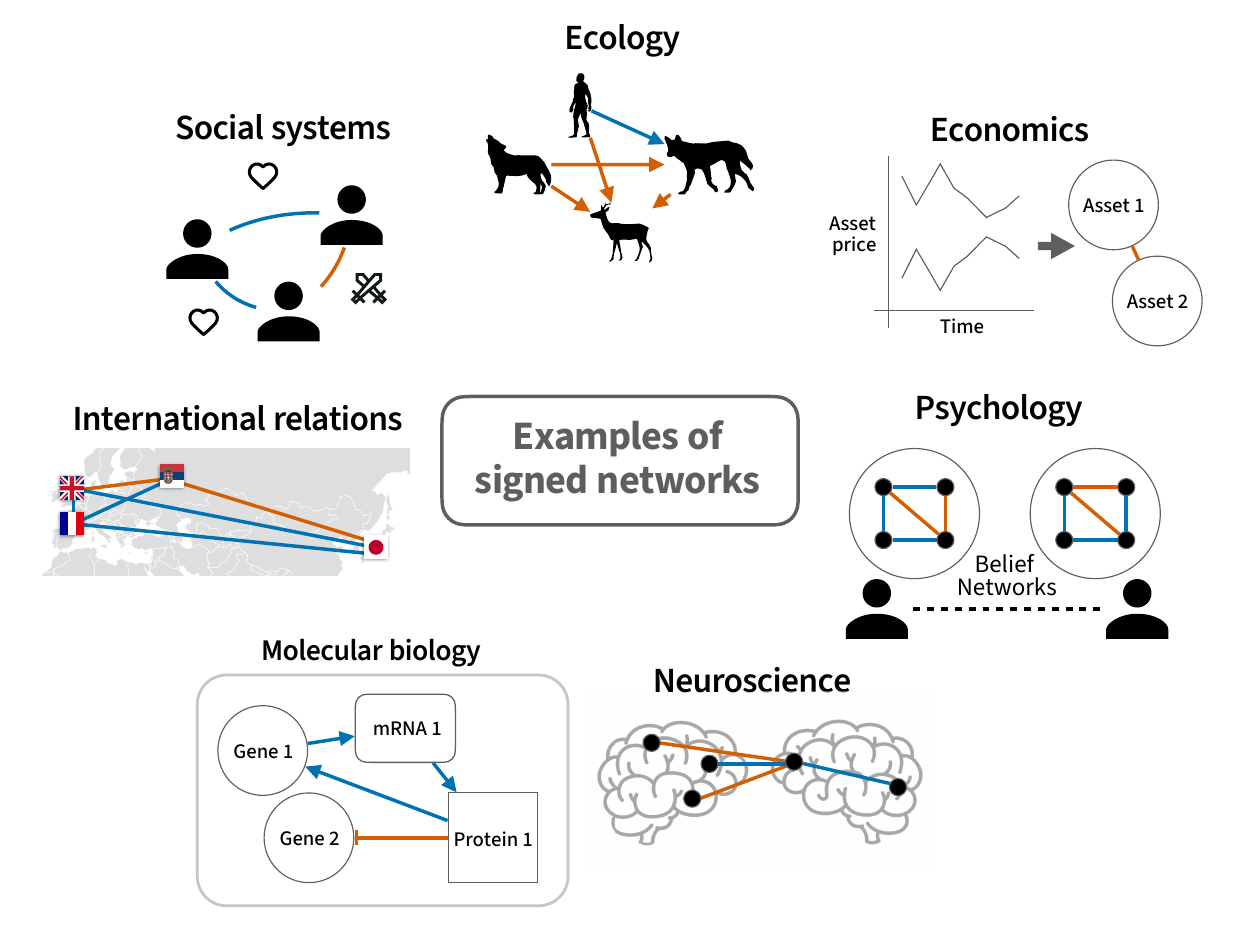}
    \caption{\textbf{Examples of signed networks across different domains.}
In social systems, nodes represent individuals and edges denote friendship/enmity or trust/distrust relationships.
In ecological systems, edges capture competitive, mutualistic, or predator–prey dynamics.
In economic systems, assets are connected based on correlations between their price time series.
In psychological systems, nodes represent beliefs or attitudes, and signed edges encode whether they reinforce or contradict each other. In biological systems, interactions between genes and proteins can be represented through activation/inhibition links.
In neuroscience and molecular systems, signed networks can represent either excitatory/inhibitory processes, such as synapses between individual neurons, or functional relationships between brain regions.
Finally, in international relations, countries may be connected by alliances or conflicts.
Figures adapted from~\citep{dorresteijnIncorporatingAnthropogenicEffects2015, aiyappaEmergenceSimpleComplex2024, markuTimeseriesTranscriptomicsGene2023, whitfield-gabrieliDefaultModeNetwork2012, diaz-diazMathematicalModelingLocal2024}.}
    \label{fig:examples-signet}
\end{figure}

Signed networks represent systems of interconnected entities, in which connections can be positive or negative. They offer a powerful framework for studying systems in which interactions are not merely present or absent, but qualitatively different: cooperative or antagonistic, supportive or conflicting, excitatory or inhibitory. Traditional network approaches that ignore negative ties or treat them as absent connections miss essential dynamics: how antagonism drives faction formation, how the interplay of cooperation and conflict creates polarisation, and how systems organise around the tension between positive and negative relationships.
Individual disciplines have long recognised the importance of antagonistic relationships (see Figure \ref{fig:examples-signet}). Ecologists study competition and predation alongside mutualism \citep{allesinaStabilityCriteriaComplex2012}. Political scientists examine rivalries and conflicts as central features of international systems \citep{hafner-burtonNetworkAnalysisInternational2009}. Sociologists have studied social systems in which positive edges can represent friendship, trust, or alliance, whereas negative edges may encode hostility, distrust, or rivalry \citep{ruiz-garciaTriadicInfluenceProxy2023}. In economics, researchers have studied financial volatility and how positive correlations between assets, indicating shared exposure to a certain risk factor, and negative correlations signal diversification or hedging opportunities \citep{souto2018capturing,bartesaghi2025global}. In Psychology and Cognitive Science, sentiments, beliefs, and attitudes are studied through networks, where positive edges represent mutually reinforcing or consistent beliefs, and negative edges may represent conflicting or opposing beliefs \citep{teixeira2021revealing,aiyappaEmergenceSimpleComplex2024}. What has been missing is a unified vocabulary and framework that allows a shared formalism to address these distinct fields.
Signed networks provide this framework. By representing systems as graphs where edges carry explicit positive or negative signs, we can identify structural patterns and dynamical processes that go beyond a single application. A signed network of international alliances and rivalries, as well as a social network of friendship and enmity, may represent different systems, but they can exhibit the same mathematical properties, be analysed with the same computational methods, and follow similar organising principles. 

\subsection*{The evolution of signed network science}

The intellectual foundations of signed networks emerged over 75 years ago from the concept of balance. Psychologist Fritz Heider \citep{heiderAttitudesCognitiveOrganization1946} proposed that individuals experience psychological tension in certain configurations of relationships. For instance, when a person has two friends who are enemies with each other, or likes someone who dislikes their friend. Social psychologist Dorwin Cartwright and mathematician Frank Harary \citep{cartwrightStructuralBalanceGeneralization1956} formalised this insight: a network is ``balanced'' when every cycle contains an even number of negative relationships. This simple principle has profound implications. Balanced networks naturally organise into mutually antagonistic factions, with positive edges within groups and negative edges between them. This fundamental insight remains central to signed network research today.
For several decades, however, empirical work remained limited to small-scale sociometric studies \citep{sampsonNovitiatePeriodChange1968,newcombAcquaintanceProcess1961}, constrained by the difficulty of collecting signed relational data and limited computational tools. The field transformed in the early 2000s with the rise of modern network science as a discipline and with large-scale digital platforms that explicitly recorded signed interactions, such as trust/distrust on Epinions, friend/foe tags on Slashdot, or support/oppose votes on Wikipedia \citep{leskovecPredictingPositiveNegative2010}. These datasets enabled researchers to test classical theories at scale and develop new computational methods. The 2010s brought rapid methodological innovation: graph neural networks adapted for signed edges\citep{derrSignedGraphConvolutional2018}, spectral methods exploiting signed Laplacian properties \citep{kunegisSpectralAnalysisSigned2010}, stochastic block models for statistical inference \citep{jiangStochasticBlockModel2015}, maximum entropy frameworks for null models \citep{galloTestingStructuralBalance2024}, and walk-based balance measures \citep{estradaWalkbasedMeasureBalance2014, talagaPolarizationMultiscaleStructural2023}. Interest spread across disciplines, such as physics, neuroscience, ecology, computer science, and political science, each recognising that antagonistic relationships fundamentally shape system organisation.
Today, signed network science research is more active than ever, with hundreds of papers published annually across diverse fields, powerful computational methods for analysis and prediction, and growing availability of large-scale signed datasets. Applications range from understanding online polarisation \citep{vendevilleModelingEchoChamber2025, fraxanetUnpackingPolarizationAntagonism2024}, international conflicts \citep{diaz-diazMathematicalModelingLocal2024}, to the analysis of brain connectivity \citep{tanner2024multi,di2025assessing}. This growth reflects both the field's maturity and its continued relevance to pressing questions about how cooperation and conflict shape complex systems. Yet, this rapid, distributed growth has created its own challenges.

\subsection*{Purpose and scope of this book}

Existing surveys and reviews have approached some of these challenges, though each has focused on a specific part of the field. Zheng et al.~\citep{zheng2015social} briefly survey social balance in signed networks. Tang et al.~\citep{tangSurveySignedNetwork2016} survey signed network mining from social media. Zhang et al.~\citep{zhang2024signed} review representation learning on signed graphs. Aref and Wilson~\citep{aref2018measuring} compare measures of partial balance. Traag et al.~\citep{traagPartitioningSignedNetworks2018} treat the partitioning of signed networks. We draw on all of them throughout this book. However, this book differs from these earlier contributions in both scope and approach. To our knowledge, there is no single reference that brings together the mathematical foundations, the methods of inference, the dynamics both on and of signed networks, the landscape of available data, and the practical judgement required to work with signed data, while using a consistent notation and explicitly comparing competing formalisms.

This book provides a comprehensive, systematic treatment of signed networks that consolidates foundational theories, clarifies methodological choices, identifies persistent misconceptions, and discusses paths forward. We treat signed networks as a unified analytical framework while respecting that different domains may require different approaches. We emphasise methodological rigour and transparency, discussing assumptions, limitations, and appropriate contexts rather than simply cataloguing methods. We integrate across levels of analysis: static structure and temporal dynamics, processes on and of networks, microscopic mechanisms, and macroscopic patterns. We develop the whole subject in a single, consistent notation, so that a result established in one chapter can be used in the next rather than restated. Where several formalisms coexist, as they do for distances, centralities, Laplacians, balance indices, and community-detection objectives, we go beyond recording their coexistence, clarifying their meaning and context dependence.

We are selective rather than exhaustive, focusing on foundational concepts, broadly useful methods, and findings that reveal general principles.
The book builds systematically. Chapter~\ref{chapter:math-foundations} establishes mathematical foundations: formalising signed networks and presenting the adaptation of classical metrics. Chapter~\ref{chapter:balance} addresses balance theory, distinguishing cognitive from structural balance and presenting measures for quantifying balance. Chapter~\ref{chapter:structure} explores structural inference: null models, embeddings, and algorithms for sign prediction and community detection. Chapter~\ref{chapter:dynamics} examines dynamics: opinion formation, network evolution, and co-evolutionary frameworks. Chapter~\ref{chapter:collection-inference-data} tackles data collection and surveys available datasets. Chapter~\ref{chapter:pitfalls} addresses common misconceptions and errors. In Chapter~\ref{chapter:conclusions}, we discuss open challenges and promising directions. Our aim with this book is to make signed network science more accessible without sacrificing rigour, more unified without erasing important distinctions, and more self-aware about its limitations, helping researchers effectively use this powerful lens for understanding how positive and negative relationships jointly shape complex systems.

\clearpage

\section{Mathematical foundations of signed networks}\label{chapter:math-foundations}

Signed networks offer a convenient way to represent systems where cooperation and conflict coexist. However, these systems are often too large and interconnected to be analysed by hand. Fortunately, mathematical tools like matrices and vectors provide an ideal framework for capturing and analysing these interactions. 
In this chapter, we collect the essentials that allow us to construct such a framework from first principles. We first introduce the core mathematical objects that define signed networks and show how to represent them in algebraic form (Section \ref{subsec:definition_signed_networks}). We continue by presenting the fundamental network measures (degrees, clustering coefficients, distances, centralities, and motifs) now adapted to the signed case, discussing both the conceptual difficulties and the available solutions (Section \ref{subsec:basic_metrics}). Finally, in Section \ref{subsec:signed-laplacian} we present the signed Laplacian, the central operator that connects structure, spectra, and dynamics, on which much of the analysis developed in later chapters is based.

\subsection{Definition and representation of signed networks}  \label{subsec:definition_signed_networks}

\begin{figure}
    \centering
    \includegraphics[width=\linewidth]{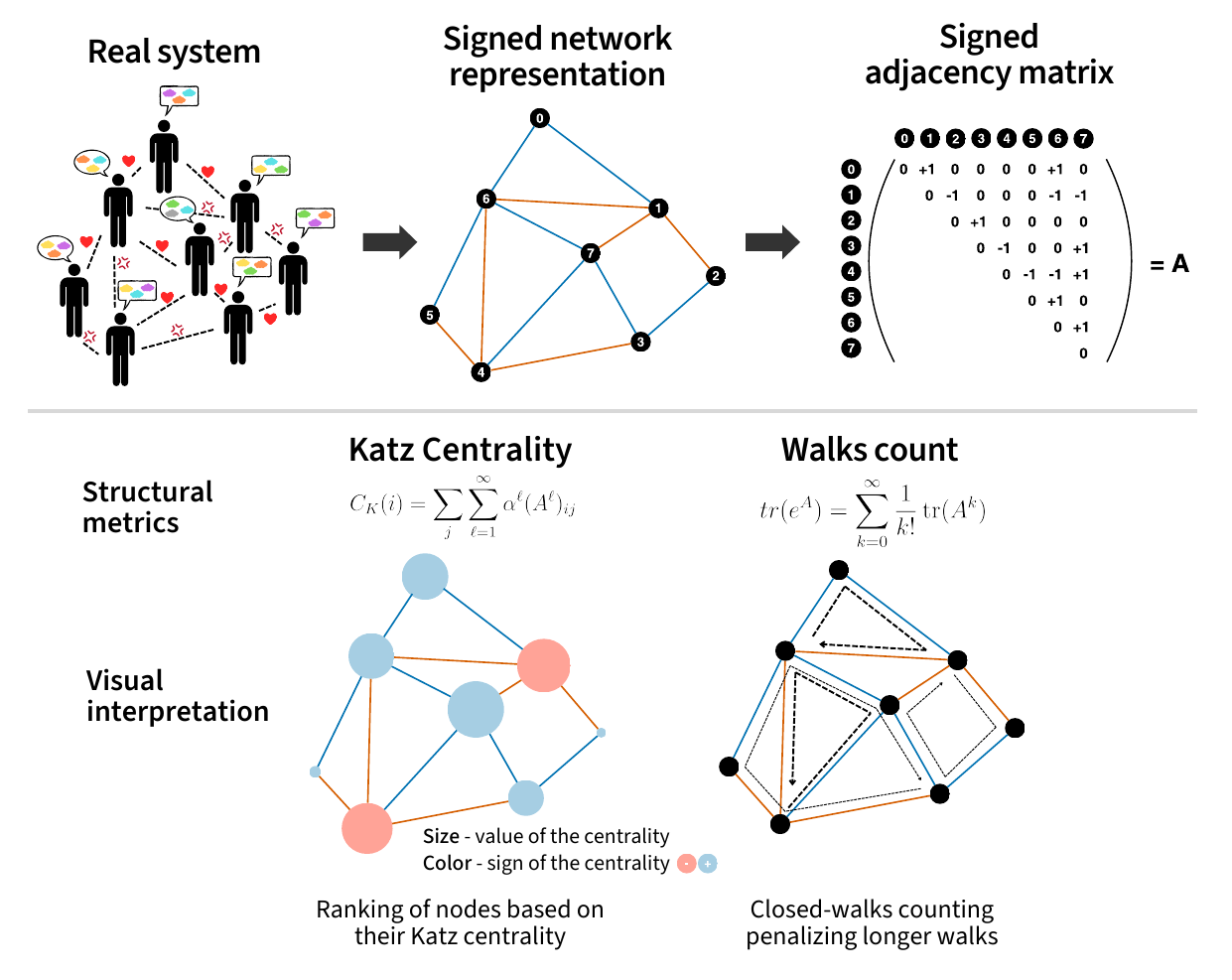}
    \caption{\textbf{Real systems as signed networks.} \textit{Top panel}: A step-by-step abstraction of a real-world system. We begin with a group of individuals, each with specific interests and opinions about others, forming a complex web of interpersonal relations involving liking and disliking. Its complexity is then simplified into a signed network, where nodes represent individuals and links encode positive or negative relationships. Finally, the network is translated into an adjacency matrix, where rows and columns correspond to nodes, and the entries take values of $+1$, $0$, or $-1$ depending on whether the relationship is positive, neutral, or negative. \textit{Bottom panel}: Illustrative examples of how the adjacency matrix is used to compute structural properties of the network. Specifically, we show the Katz centrality calculation and the count of closed walks, alongside visual representations of their meaning.}
    \label{fig:signed-math-representation}
\end{figure}

Let us introduce the fundamental mathematical concepts and terms, and some of the key concepts used throughout this book. From a mathematical point of view, networks (whether signed or unsigned) are modelled as \textit{graphs}~\citep{newmanNetworks2018}. A graph $G =(V, E)$ consists of a set $V=\{v_i\}$ of nodes (also called vertices) and a set $E=\{e_{ij}\}$ of edges (also called links) that connect pairs of nodes. An edge connecting the nodes $v_i$ and $v_j$ is denoted by $e_{ij}=(v_i,v_j)$. The number of nodes is denoted by $n$ and the number of edges by $m$. Nodes can represent entities such as people, genes, countries, or more abstract constructs like stocks or time series. Edges capture pairwise interactions between these nodes: social connections between individuals, regulatory relationships between genes, or correlations between time series, among others.

On a \textit{signed graph}, edges carry a sign attribute~\citep{zaslavskySignedGraphs1982} expressing a negative or positive type of relation between the nodes. Beyond signs, edges may also have directionality to allow for the modelling of asymmetrical relations between nodes~\citep{mackay2020directed}, and weights to quantify varying levels of importance across relations~\citep{barratArchitectureComplexWeighted2004,newmanAnalysisWeightedNetworks2004}. The interplay between these attributes can lead to rich emergent phenomena; for instance, biological circuits leverage both sign and directionality to generate oscillatory dynamics~\citep{alon2019introduction}. In this book, we focus on the specific role of signed edges, and we avoid details regarding the role of directionality and weights.

A \textit{walk} is a sequence of edges, where consecutive edges share a common node. The \textit{length} of the walk is the number of edges it traverses. Walks may traverse both nodes and edges multiple times. If no edge is repeated, the walk is called a \textit{path}. A \textit{closed walk} starts and ends at the same node, and a closed path is called a \textit{cycle}.  In a signed graph, walks, paths, and cycles possess a new attribute, called the sign. The \textit{sign} of a walk is the product of the signs of its edges. In other words, a walk is \textit{positive} if it contains an even number of negative edges and \textit{negative} if this number is odd.

A graph is most commonly represented by its \textit{adjacency matrix} \( A \), which encodes node connections. In the signed case, \( A_{ij} = +1 \) for positive edges, \( A_{ij} = -1 \) for negative edges, and \( A_{ij} = 0 \) when no edge exists. 
Additionally, the adjacency matrix admits a decomposition into positive and negative components:
\(
A = A^+ - A^-, \ \text{where} \ A^+_{ij} = \max(A_{ij},0), \ A^-_{ij} = \max(-A_{ij},0).
\)
This decomposition into non-negative matrices is useful to mitigate potential issues arising from negative entries. The matrices \( A^+ \) and \( A^- \) also define the adjacency matrix of the underlying unsigned graph, which is the graph obtained by removing all edge signs: $\vert A\vert = A^+ + A^-$. 

The adjacency matrix enables the application of techniques from linear algebra and spectral graph theory to analyse graphs~\citep{biggsAlgebraicGraphTheory1993,zaslavskyMatricesTheorySigned2013}. When the graph is unsigned, the $(i,j)$\textsuperscript{th} element of $A^k$ counts the number of walks of length $k$ between nodes $v_i$ and $v_j$. When the graph is signed, it counts the difference between the number of positive and negative walks. These are given respectively by
\begin{align*}
    (W^+)^{(k)}_{ij} &= \frac{1}{2} \big( (|A|^k)_{ij} + (A^k)_{ij} \big), \\
    (W^-)^{(k)}_{ij} &= \frac{1}{2} \big( (|A|^k)_{ij} - (A^k)_{ij} \big).
\end{align*}
Note that these expressions differ from the naïve guess  
\(
(W^+)^{(k)}_{ij} = (A^+)^k_{ij}, \ (W^-)^{(k)}_{ij} = (A^-)^k_{ij}.
\)
That is, \( A^+ \) does not count positive walks, nor does \( A^- \) count negative walks. This simple result shows that the complex coupling between the positive and negative subnetworks cannot be treated independently in most cases.

\subsection{Basic metrics}  \label{subsec:basic_metrics}

Centrality measures allow us to characterise local and global properties of networks. However, typical network metrics, such as degree, centrality measures, shortest-path distances, clustering coefficients, and network motifs~\citep{newmanNetworks2018}, are not straightforward to generalise for signed networks. Some of these cease to be well-defined (e.g., the shortest-path distance), while others admit different generalisations suitable to different contexts. In this section, we explore how to approach each of these metrics and centrality measures in signed networks in a principled manner.

\subsubsection{Signed degree and its distribution}

In an unsigned network, the degree of a node $v_i$, denoted by $k_i$, counts the number of edges adjacent to the node. In a social network, the degree usually quantifies the popularity or influence of a person. Studying the distribution of degrees $P(k)$ can then help answer questions such as whether social influence is concentrated in the hands of a few or more equally spread in the population. When the network is signed, the degree is not characterised by a single number, but by a pair \( (k^+, k^-) \), denoting the number of positive and negative edges incident to the node, respectively. The two-dimensional nature of the degree causes the degree distribution of a signed network to become bivariate \( P(k^+, k^-) \). Analysing or visualising bivariate distributions can be challenging, so it is common to consider the marginal distributions, \( P(k^+) \) and \( P(k^-) \). However, one should keep in mind that marginalising erases any correlation between \( k^+ \) and \( k^- \), which may contain valuable structural information. An alternative is to analyse the \textit{absolute} or \textit{underlying} degree \(
k = k^+ + k^-\), and the \textit{net} or \textit{signed} degree \(k_{\text{net}} = k^+ - k^-\), through their joint probability distribution \( P(k, k_{\text{net}}) \) and its marginals. Asymmetries have been found between the distributions \( P(k^+) \) and \( P(k^-) \)~\citep{szellMeasuringSocialDynamics2010}, suggesting fundamental differences in the formation mechanisms of positive and negative links, and challenging the use of $P(k)$ as an appropriate coarse-graining.

\subsubsection{Signed clustering coefficient}

The (unsigned) clustering coefficient measures local cohesiveness by quantifying the abundance of triangles. This is a key structural property of networks that plays a central role in the famous small-world phenomenon~\citep{Watts1998}. For the signed case, the (global) clustering coefficient can be computed as~\citep{kunegisSlashdotZooMining2009}:
\begin{align*}
    C =  \frac{\text{tr}(A^3)}{\sum_i k_i(k_i-1)} = \frac{6(t^+-t^-)}{\sum_i k_i(k_i-1)} ,
\end{align*}
where $k_i$ is the absolute degree of node $v_i$. This index captures the difference between the number of positive triangles $t^+$ and negative triangles $t^-$, with the remaining terms\footnote{In the presence of self-loops, one must use the normalisation constant $k_i^2/2$ instead~\citep{kunegisSlashdotZooMining2009}.} ensuring that it remains bounded between $+1$ and $-1$. Clustering can also be studied at the node level, quantifying the difference between the number of positive triangles ($t^+$) and negative triangles ($t^-$) \textit{containing node $v_i$}, relative to all two-paths passing through $v_i$~\citep{costantiniGeneralizationClusteringCoefficients2014}:
\begin{align*}
    C_{i} =  \frac{(A^3)_{ii}}{k_i(k_i-1)}= \frac{2(t_i^+-t_i^-)}{k_i(k_i-1)}.
\end{align*}
Conveniently, the global clustering $C$ is a weighted average of the local clusterings $C_i$ of every node in the network:
\begin{align*}
    C = \frac{\sum_i C_i k_i(k_i-1)}{\sum_j k_j(k_j-1)}.
\end{align*}
The two indices ($C$ and $C_i$) are equivalent when the graph is regular (all nodes have the same degree). However, in networks with degree heterogeneity or imbalances between positive and negative links, these indices can differ significantly.

Both $C$ and $C_i$ penalise negative triangles ($t^-$) and unclosed triangles. To isolate the effect of negative triangles while controlling for network topology, Kunegis introduced the \textit{relative clustering coefficient}~\citep{kunegisSlashdotZooMining2009}, defined as the ratio between the signed and unsigned clustering coefficients:
\begin{align*} C^{rel} = \frac{C(A)}{C(|A|)}= \frac{t^+-t^-}{t^++t^-}. 
\end{align*} 
This approach allows for a direct comparison between clustering patterns in signed networks and their unsigned counterparts.
Similar reasoning allows for the definition of a relative local clustering coefficient.

\subsubsection{Signed distances}  \label{subsec:distances}

Another fundamental concept in network analysis is the distance between pairs of nodes. This measure plays a central role in various applications, including the definition of various centrality metrics~\citep{newmanNetworks2018,borgattiCentralityNetworkFlow2005}, community detection~\citep{newmanFindingEvaluatingCommunity2004}, and network geometry~\citep{bogunaNetworkGeometry2021}. Formally, a distance is a function that takes two nodes as input and returns a real number that satisfies four key properties. 
\begin{enumerate}
    \item Symmetry: the distance between $v_i$ and $v_j$ must be equal to the distance between $v_j$ and $v_i$.
    \item Non-negativity: the distance function must always return a non-negative number
    \item The ``identity of indiscernibles'': the distance is zero if and only if the two nodes are identical, meaning the function assigns zero distance exclusively to an object and itself.
    \item The triangle inequality\footnote{Intuitively, the triangle inequality means that the shortest way to travel between two nodes is always the direct route, so if we take a detour through $v_j$, the total distance will be at least as long as going straight from $v_i$ to $v_k$, and usually longer.}: $d(v_i,v_j)+d(v_j,v_k)\geq d(v_i,v_k)$.
\end{enumerate}

The most common distance measure for networks is the \textit{shortest-path} distance, defined as the minimum number of edges one must traverse to go from one node to another. In weighted networks, this generalises to the \textit{minimum-weight} distance, computed as the smallest sum of edge weights along the paths joining two nodes.

In the case of signed networks, however, these measures fail to satisfy the properties of a distance~\citep{diazdiazthesis}. To understand the reasons, consider a graph made of three nodes $v_1,v_2, v_3$, the first two linked by a negative edge and the last two by a positive one. Using the minimum-weight distance we obtain $d(v_1,v_2)<0$, violating the non-negativity condition, and $d(v_1,v_3)=0$, contradicting the identity of indiscernibles. These issues are not just theoretical; they can result in algorithms selecting infinitely long paths as optimal routes.

A natural way to overcome these limitations is to use \textit{generalised distances}~\citep{bogunaNetworkGeometry2021}. These distances extend beyond simple shortest-path measures by incorporating the collective influence of all paths between two nodes, often through dynamic processes such as diffusion, random walks, or epidemic spreading.
One prominent example is the \textit{resistance distance}, which originated from electrical circuit theory~\citep{chandraElectricalResistanceGraph1989}. Here, the distance between two nodes corresponds to the effective resistance in a hypothetical circuit where a unit current (1 amp) is injected into $v_i$ and extracted from $v_j$, with every edge acting as a 1-ohm resistor. Alternatively, this measure has a probabilistic interpretation: it is equal to the {expected commuting time} for a random walker travelling from $v_i$ to $v_j$ and back. Mathematically, the resistance distance is given by

\begin{equation*}
    d_{\text{res}}(i,j) = \sqrt{(L^\dagger)_{ii} +(L^\dagger)_{jj} - 2(L^\dagger)_{ij} },
\end{equation*}
where $L$ is the \textit{graph Laplacian}~\citep{devriendtEffectiveResistanceMore2022}, discussed in detail in Section \ref{subsec:signed-laplacian}, and $L^\dagger$ is its Moore-Penrose pseudoinverse, defined as: $L^\dagger = (L+\gamma J)^{-1}-\gamma^{-1}J$, with $\gamma$ a regularisation parameter and $J$ an all-ones matrix\footnote{The Moore–Penrose pseudoinverse generalises the notion of the matrix inverse and remains well defined even when some eigenvalues are equal to zero, as in the Laplacian. For nonsingular matrices, it coincides with the ordinary inverse.}. 
The function $d_{res}$ satisfies all properties of a distance, even when including negative edges~\citep{kunegisApplicationsStructuralBalance2014}. Moreover, it is Euclidean, meaning it can be interpreted as the usual ``straight-line length'' in some abstract Euclidean space.

The effective resistance is not the only generalised distance that retains good properties in signed networks. Other relevant ones include the \textit{diffusion distance}~\citep{dedomenicoDiffusionGeometryUnravels2017}, which captures how similar two nodes are based on the spread of a diffusive process over time; and the \textit{communicability distance}~\citep{estradaComplexNetworksEuclidean2012,diazdiaz2025}, which measures the ease of information transfer between nodes using the exponential of the adjacency matrix. The best distance metric to use depends on the application. Resistance distance is the default when no particular dynamic process is specified. Diffusion distance is more appropriate for problems where one needs to tune the scale at which nodes are compared, because it depends on a parameter that controls how far a signal spreads before the comparison is made. The communicability distance instead weighs all walks between two nodes, with shorter ones counting more, and is a natural choice for analysing flows that propagate through walks.

Despite the mathematical challenges discussed, algorithms exist to compute the minimum-weight distance in signed directed graphs. Two important examples are the Bellman-Ford and Floyd-Warshall algorithms~\citep{cormenIntroductionAlgorithmsFourth2022}. Both algorithms are based on dynamic programming and are preferred over Dijkstra's algorithm when negative edges are present. Apart from computing minimum-weight paths, these algorithms also detect $\Sigma$-negative cycles (cycles where the sum of edge weights is negative)\footnote{Note that this notion is different from negative cycles, where the \textit{product} of edge weights is negative. $\Sigma$-negative cycles appear in diverse applications. For instance, in exchange-rate networks, currencies can be represented as nodes and exchange rates as edge weights (typically using the negative logarithm of the rate). In this setting, a $\Sigma$-negative cycle indicates an arbitrage opportunity.}.

\subsubsection{Signed centralities}  \label{subsubsect:centralities}

Centrality is a fundamental concept in network analysis, widely used to identify the most structurally important nodes in a network. Classical measures such as degree, closeness, betweenness, and eigenvector centrality have been extensively studied in unsigned networks. Central nodes can be defined as those with many connections (degree centrality), those who are at a short distance from others (closeness centrality), those who lie on many shortest paths (betweenness centrality), or, in a circular way, as those with connections with many other central nodes (eigenvector centrality). These measures can often be interpreted in terms of network flow processes~\citep{borgattiCentralityNetworkFlow2005}. In signed networks, the notion of centrality becomes more nuanced, as flow-based interpretations become less applicable. Depending on the context, centrality in signed networks can represent many concepts, such as influence, power, trustworthiness, control, cohesiveness, or balance.

The simplest approach is to extend classical centrality measures by treating positive and negative links equivalently. This perspective views centrality as a measure of structural importance without distinguishing between different types of relationships. A similar approach, used for instance in P2P networks~\citep{kamvarEigenTrustAlgorithmReputation}, is to ignore negative links entirely and compute classical centrality measures on the positive subgraph. This assumption fails to distinguish between a negative link and the absence of a link, discarding structural information. Another approach computes centralities separately for positive and negative subgraphs and subtracts them~\citep{shahriariRankingNodesSigned2014}. While this accounts for negative links, it ignores their interactions with positive ones, leading to outputs that are often difficult to interpret.

Thus, these straightforward adaptations of unsigned centrality measures often fail to provide interpretable results. A simpler way forward is to start by asking what kind of importance we want to capture, rather than adapting existing formulas. Once this notion is clear, suitable mathematical definitions can follow. For instance, in politically charged networks, centrality can represent power or control, and the mathematical formulation changes depending on which notion one aims to capture~\citep{smithPowerPoliticallyCharged2014}. If we interpret centrality as power, positive edges increase it by reinforcing influence and access to allies, while negative edges reduce it by introducing threats and opposition. If, instead, we view centrality as control, the effect reverses: a node may gain centrality when its allies face many negative edges, as they rely more on it for stability and mediation~\citep{smithPowerPoliticallyCharged2014}. 

Other perspectives of centrality in signed networks include measures quantifying trust or cohesiveness. Trust is particularly relevant in social networks, recommendation systems, and online marketplaces, where an individual’s credibility is key~\citep{guhaPropagationTrustDistrust2004,tangSurveySignedNetwork2016}. Trust-based centrality measures emphasise the asymmetry between trust and distrust, often weighing negative endorsements more heavily~\citep{guhaPropagationTrustDistrust2004}. In some other applications, being central means being embedded in a cohesive cluster of mutual support~\citep{bonacichCalculatingStatusNegative2004,gromovSocialBalancebasedCentrality2025}. In this framework, negative links can increase centrality by reinforcing cohesion within a group through shared adversaries. Centrality then reflects how well an individual contributes to the stability of a coalition rather than mere connectivity. Finally, centrality in signed networks can also be linked to balance theory (see Chapter~\ref{chapter:balance}). Some classical measures, like subgraph centrality, naturally function as local balance indicators when applied to signed networks~\citep{diaz-diazMathematicalModelingLocal2024}.

In what follows, we discuss the mathematical formulations behind some important centrality measures adapted from unsigned networks.

\paragraph{Degree centrality} Degree centrality is straightforward to adapt to signed networks using the net degree. Alternatively, one can use the ratio of net degree to underlying degree to obtain a relative centrality measure.

\paragraph{Closeness centrality} Since shortest paths are not well-defined in signed networks, closeness centrality relies on alternative distance measures. One approach replaces the shortest-path distance with a generalised distance $d$. The generalised closeness centrality is then~\citep{estradaCommunicabilityCosineDistance2024}
\begin{align*}
    C_C(i)= \left( \sum_{j\neq i}d(i,j) \right)^{-1}.
\end{align*}

\paragraph{Katz centrality} Katz centrality extends degree centrality by considering both direct and indirect connections, penalising walks of length $\ell$ by an attenuation factor $\alpha^\ell$ ($\alpha<1$). In unsigned networks, it is defined as~\citep{newmanNetworks2018}\footnote{The Katz centrality is only well-defined when $\alpha$ is smaller than the inverse of the spectral radius of $A$, $\rho(A)$.}

\begin{equation*} \label{eq:katz}
C_K(i) = \sum_{j} \sum_{\ell=1}^{\infty} \alpha^\ell (A^\ell)_{ij},
\end{equation*}

where $C_K(i)$ counts all walks of all lengths $\ell$ starting from node $i$, with more importance given to the shortest ones. In signed networks, Katz centrality was first introduced in~\citep{chiangExploitingLongerCycles2011} and later formalised in~\citep{gomez-ambrosiMeasuringNetEffects2025}. The same formula applies, but its interpretation differs: in unsigned networks, all walks contribute positively to a node’s influence, whereas in signed ones, walks with an odd number of negative edges contribute negatively. The resulting value thus reflects the net level of power of a node. Unlike in the unsigned case, Katz centrality can take negative values, indicating that the node has, on average, stronger antagonistic relations. Variants with different weightings on positive and negative edges have also been explored~\citep{everettNetworksContainingNegative2014}.

\paragraph{Eigenvector centrality} The core idea of eigenvector centrality is that a node is considered important if it is connected to other important nodes~\citep{bonacichUniquePropertiesEigenvector2007}. In the signed case, the most natural generalisation assumes that a node's importance increases when it is connected to important nodes through positive edges, and decreases when connected via negative edges to important nodes. Mathematically, this is captured by the following self-referential equation:
\begin{align} \label{eq:eigenvector_centrality}
    x_i = \lambda^{-1} \sum_{j}  \left(A_{ij}^+ x_j - A_{ij}^- x_j\right) = \lambda^{-1} (A x)_i
\end{align}
where $x_i$ is the eigenvector centrality of node $i$, and $\lambda$ is a normalisation constant. The term $A^+_{ij}x_j$ represents the positive contributions to node $i$'s centrality from important neighbours, while $A^-_{ij}x_j$ subtracts contributions from negatively connected important nodes. Eq.~\eqref{eq:eigenvector_centrality} can be rewritten as the eigenvector equation $Ax=\lambda x$, justifying the denomination of eigenvector centrality. In general, there are several linearly independent eigenvectors fulfilling the eigenvalue equation, but we select the one corresponding to the largest eigenvalue $\lambda$. 

In signed networks, the interpretation of eigenvector centrality depends critically on whether the network is structurally balanced (see Section~\ref{subsec:math-propr-balanced}). In fact, in a balanced network, the adjacency matrix is guaranteed to have a dominant eigenvector associated with the largest eigenvalue, which will be non-degenerate~\citep{bonacichCalculatingStatusNegative2004}. Unlike in the unsigned case, the components of this eigenvector can take both positive and negative values, and their signs align with the two balanced factions of the network. Eigenvector centrality can thus be interpreted as a measure of a node’s adherence to its faction, with the magnitude reflecting its structural importance and the sign indicating its faction.

When the network is unbalanced, the largest eigenvalue can become degenerate, and the dominant eigenvector is no longer uniquely defined. Everett et al.~\citep{everettNetworksContainingNegative2014} illustrate this issue with two signed networks whose eigenvector centralities are ill-defined. Even so, the measure often gives useful results when the network is close to being balanced~\citep{diazdiazthesis} (see Section~\ref{sec:balance_indices}).

\paragraph{PageRank} Originally designed by Google for ranking webpages~\citep{pagePagerankCitationRanking1999}, PageRank centrality is akin to a directed version of eigenvector centrality that incorporates a damping factor \(\alpha\) to ensure that every node receives a minimal level of importance, even if it has no incoming edges. While eigenvector centrality can be interpreted as the stationary distribution of a random walk on the graph, PageRank centrality is the stationary distribution of a random walk with teleportations happening at rate $\alpha$. In unsigned networks, PageRank satisfies the recursive equation:

\begin{equation*}
C_P(i) = \frac{(1 - \alpha)}{n} + \alpha \sum_{j} \frac{A^T_{ij} C_P(j)}{k_j^{out}},
\end{equation*}

where \(k_j^{out}\) is the out-degree of node \(j\). This formulation captures the intuition that a node’s importance is determined by the importance of the nodes linking to it, weighted by their out-degree. In signed networks, the most natural generalisation replaces \(A\) with the signed adjacency matrix while still using the underlying out-degree~\citep{kunegisSlashdotZooMining2009}. This minimal modification keeps the core properties of PageRank while taking into account that negative links penalise centrality. However, note that several properties get lost when including negative edges. For example, some PageRank values may become negative, so they can no longer be interpreted as a probability distribution.

\paragraph{Other approaches} Beyond adaptations of classical measures, several alternative approaches to centrality in signed networks have been proposed. Approaches considering signed links as indicators of trust or distrust include \textit{PageTrust}~\citep{kerchovePageTrustAlgorithmHow2008}, inspired by PageRank, and \textit{exponential ranking}~\citep{traagExponentialRankingTaking2010}. \textit{Bias and prestige centrality} assume that the trustworthiness and reputation of a node should be assessed separately~\citep{mishraFindingBiasPrestige2011}, much like the HITS algorithm~\citep{kleinbergWebGraphMeasurements1999}. Finally, Liu et al. proposed a framework reminiscent of signed PageRank, but distinguishing between a node’s total centrality (i.e., its overall influence on the network), and its net centrality (which reveals whether that influence is mainly positive or negative)~\citep{liuSimpleApproachQuantifying2020}.

\begin{table}[htbp]
\centering
\caption{Comparison of degree, closeness, Katz, eigenvector, and PageRank centrality in signed networks. }
\label{tab:signed-centralities}
\small
\begin{tabular}{p{2cm}p{3cm}p{4cm}p{3cm}p{2.2cm}}
\toprule
\textbf{Measure} & \textbf{Definition} & \textbf{Meaning of the centrality} & \textbf{Meaning of a negative value} & \textbf{Well-defined when...} \\
\midrule
Degree (net) &
$k_i^+ - k_i^-$ &
Net support in the immediate neighbourhood &
More antagonistic than supportive ties &
Always \\
\addlinespace
Closeness &
$\left(\sum_{j \neq i} d(i,j)\right)^{-1}$ &
Proximity to all nodes, depending on the chosen distance &
Never negative &
A signed distance $d$ is available \\
\addlinespace
Katz &
$\displaystyle \sum_j \sum_{\ell=1}^{\infty} \alpha^\ell (A^\ell)_{ij}$ &
Net aggregation of walks of all lengths starting at this node (shorter walks count more) &
Negative walks dominate; net hostility or distrust &
$\alpha < 1/\rho(A)$ \\
\addlinespace
Eigenvector &
Leading eigenvector of $A$ &
Alignment with one of two factions, under near balance &
Membership in the opposite faction &
$G$ is close to structurally balanced \\
\addlinespace
PageRank &
Stationary scores of a teleporting walk on signed $A$ &
Net aggregation of walks of all lengths arriving at this node, damped and reweighted by out-degree
 &
Negative walks dominate; net hostility or distrust  &
$\alpha < 1$ \\
\bottomrule
\end{tabular}
\end{table}

\begin{figure}
    \centering
\includegraphics[width=0.8\linewidth]{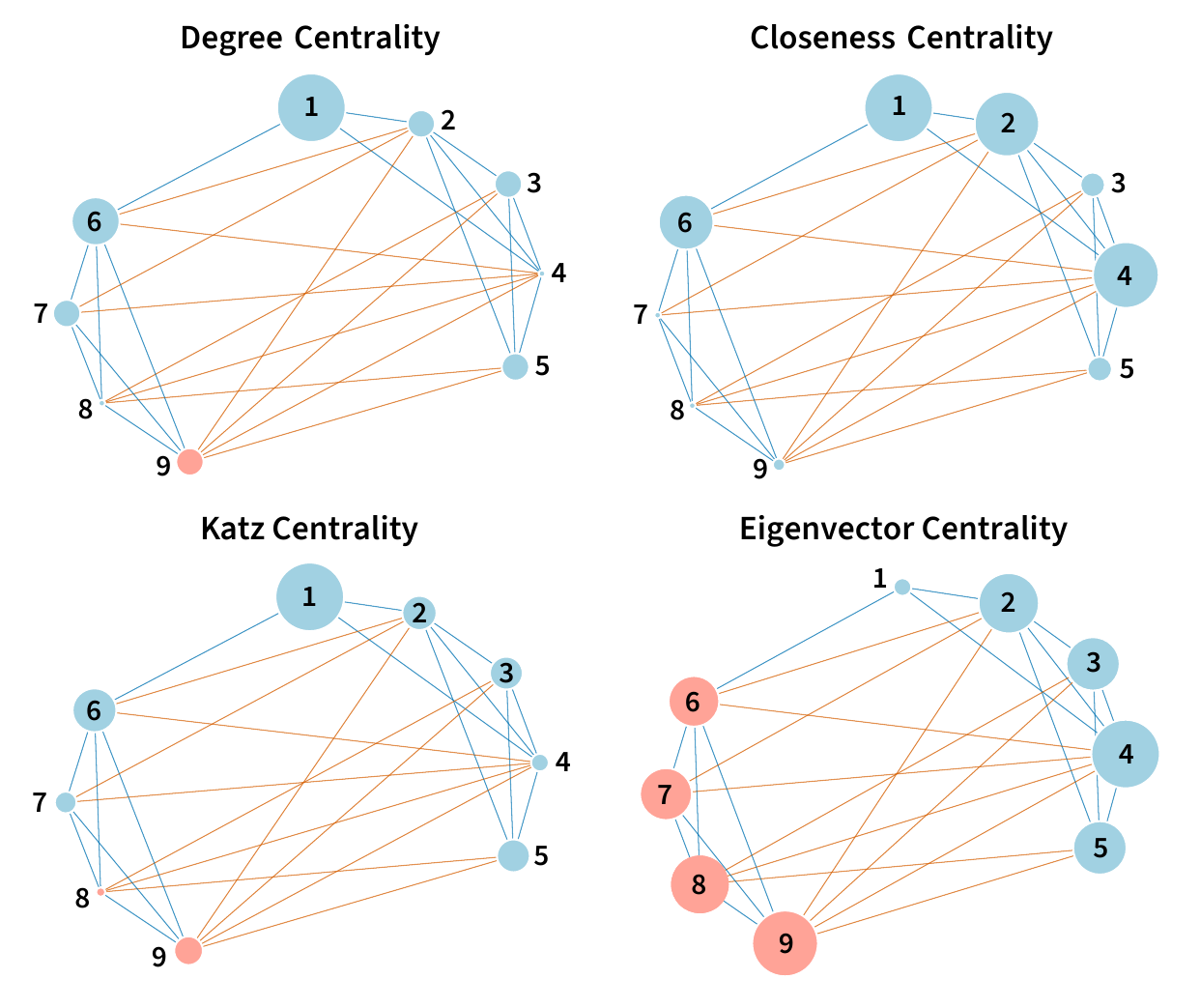}
    \caption{\textbf{Degree, closeness, Katz, and eigenvector centrality in signed networks.} Node size is proportional to the absolute value of the corresponding centrality, while colour indicates its sign (blue for positive values, red for negative ones). We omit PageRank because the graph is undirected.}
    \label{fig:centr-measures}
\end{figure}

Table \ref{tab:signed-centralities} compares five centrality measures, making explicit what each one means by centrality, how a negative value should be read, and the conditions under which the measure remains well-defined. Let us now explore visually, to illustrate the differences in interpretation of the different centrality measures, the network depicted in Figure \ref{fig:centr-measures}. It consists of two opposing factions, nodes 1–5 and 6–9, with only positive edges within each group and mostly negative edges between them. Only one edge, between nodes 1 and 6, breaks this pattern. Intuitively, the centrality measures should reflect this structural feature in their rankings.

Degree centrality ranks node 1 as the most central because it has no negative edges incident to it. Node 9, in contrast, is incident to more negative than positive edges and therefore obtains a negative centrality value, making it the least central. Closeness centrality favours nodes that are at short distances to all others through positive paths; thus, nodes near the positive bridge $e_{1,6}$ and hubs of the largest factions have the highest closeness values. Katz centrality increases when a node connects through positive edges to already central nodes, and decreases when it connects through negative edges to central nodes. As a consequence, the smaller faction is assigned lower Katz values, with some of its nodes being assigned negative values (indicating a net lack of power). 
Finally, eigenvector centrality assigns positive values to all nodes within one of the factions and negative ones to the other. Node 1, for example, shows the smallest magnitude of centrality, as it participates in the only link breaking the factional structure. Thus, the centrality quantifies the node’s alignment with a given faction.

\subsubsection{Signed motifs}

A network motif is a recurring subgraph pattern that appears significantly more frequently in a given network than in randomised counterparts~\citep{miloNetworkMotifsSimple2002}. These motifs are fundamental building blocks of complex networks and provide insights into their structural and functional properties. The detection of motifs typically involves searching for all possible small subgraphs, such as triads (three-node subgraphs) or tetrads, and comparing their frequency to a null model to ensure statistical significance. One of the simplest ways to quantify this significance is through the Z-score:

\begin{equation*}
Z(m) = \frac{O(m) - \bar{O}_{\text{null}}(m)}{\sigma_{\text{null}}(m)},
\end{equation*}

where $O(m)$ is the observed count of a motif, and $\bar{O}_{\text{null}}(m)$ and $\sigma_{\text{null}}(m)$ are the mean and standard deviation obtained from randomized versions of the network that preserve its basic connectivity patterns (see Section~\ref{subsec:null-models} for details on randomized null models for signed networks). Positive values of $Z$ indicate that the motif is overrepresented, while negative values indicate that the motif is underrepresented. A typical threshold for significance is $|Z| > 1.96$, corresponding to a 0.05 significance level.

Motif analysis has been extensively conducted in unsigned networks, but many questions remain open in signed networks, due to three challenges. First, the presence of both positive and negative edges increases the number of possible motifs substantially. Second, the null model employed must be adapted to correctly address both positive and negative edges. For example, for some datasets, it may need to preserve not only the degree sequence but also the proportion of positive and negative connections, to avoid distortions of the signed structure. Third, statistically evaluating the abundance of a motif often requires large and varied network data, which is unsigned in the vast majority of cases (see Section~\ref{subsec:common-data} for freely available signed network datasets). 

Despite these challenges, studies have identified and classified significant motif patterns in signed networks. Two complementary frameworks are commonly used to interpret these motifs: structural balance and status theory. In the balance framework, signed motifs are divided into balanced and unbalanced types (see Chapter \ref{chapter:balance}). The former contain only cycles with an even number of negative edges, whereas the latter include cycles with an odd number of negative edges. Analyses of empirical signed networks often report that nearly all balanced motifs involving three or four nodes are overrepresented, while nearly all unbalanced motifs are underrepresented~\citep{haoProperNetworkRandomization2023}. However, one must keep in mind that these statistical patterns depend strongly on the type of data and the choice of null model~\citep{galloTestingStructuralBalance2024}, and the evidence across the literature remains mixed. 

Moreover, status theory~\citep{leskovecSignedNetworksSocial2010} provides a hierarchical interpretation of signed motifs. It assumes that each node occupies a latent status level, and that the sign and direction of edges reflect relative positions within this hierarchy. A positive directed edge indicates that the source node perceives the target node as having a higher status, whereas a negative edge implies the opposite. Status theory explains asymmetries in social networks, such as the tendency of lower-status individuals to form positive links with higher-status individuals~\citep{leskovecSignedNetworksSocial2010}. 

Motifs can help to understand the mechanisms that affect network evolution (see Section~\ref{subsec:dynamics-of-networks}). If a mechanism favours certain local configurations, these configurations tend to appear more often than expected in a suitable null model; that is, as motifs. Therefore, they act as ``clues'' of the processes that operate underneath. For example, balance and status are two common mechanisms that drive the dynamics of the network, so balanced and status-consistent motifs tend to be abundant in empirical data. However, it is important to keep in mind that they are by no means the only mechanisms. Others, like homophily or differential popularity, can increase the abundance of different motifs. An important open avenue is expanding motif analysis beyond the two paradigmatic views of balance and status.

\subsection{Signed Laplacian}\label{subsec:signed-laplacian}

A central tool in graph theory and network analysis is the Laplacian matrix. Its applications range from community detection to the modelling of dynamical processes on networks. To extend these applications to signed networks, we must modify the Laplacian to properly account for negative edges. The Laplacian of an unsigned network (\textit{unsigned Laplacian}) is defined as the difference between the degree matrix and the adjacency matrix, with entries given by
\[
(L_u)_{ij} = \begin{cases} 
\sum_{k} A_{ik}, \quad \text{if} \quad i=j, \\ 
-A_{ij}, \quad \text{if} \quad i\neq j .
\end{cases}
\]
A key question in extending the Laplacian to signed networks is how to define the degree of each node. One option is to employ the underlying degree, while another is to use the net degree. The first choice results in a matrix commonly known as the \textit{opposing Laplacian}~\citep{houLaplacianEigenvaluesSigned2003, houBoundsLeastLaplacian2005}, while the latter leads to the \textit{repelling Laplacian}~\citep{shiDynamicsSignedNetworks2019}.

The opposing Laplacian is given by:
\[
(L_o)_{ij} = \begin{cases} 
\sum_{k} |A_{ik}|, \quad \text{if} \quad i=j, \\ 
-A_{ij}, \quad \text{if} \quad i\neq j. 
\end{cases}
\]
This matrix has the advantage of being always positive semidefinite, meaning all its eigenvalues are non-negative~\citep{kunegisSpectralAnalysisSigned2010}. This ensures that any associated dynamics are stable (see Section \ref{subsec:dynamics-on-networks}). Moreover, it has a zero eigenvalue if and only if the graph is balanced. In such cases, the signs of the components of the eigenvector corresponding to the zero eigenvalue can be used to determine the balanced faction structure of the network\footnote{A faction is a set of nodes whose mutual connections are mostly positive and whose connections to nodes outside the set are mostly negative. A balanced faction is an ideal case where \textit{every} link within the set is positive and \textit{every} link to other sets is negative. See Section \ref{subsec:math-propr-balanced} for more details.}~\citep{kunegisSpectralAnalysisSigned2010}. This property makes the opposing Laplacian particularly useful for uncovering faction structures and community partitions in signed networks ~\citep{cucuringuSPONGEGeneralizedEigenproblem2019} (see Section \ref{subsec:comm-det})\footnote{It is instructive to compare the spectrum of \(L_o\) with that of \(L_u\). The smallest eigenvalue of \(L_o\) plays the same role as the second smallest eigenvalue of \(L_u\): both vanish when the network is balanced or disconnected, respectively, and their associated eigenvectors uncover balanced factions or connected components. When these eigenvalues are small but nonzero, the network is nearly balanced or nearly disconnected, and the corresponding eigenvector reveals faction or community structures, respectively.}.

In contrast, the repelling Laplacian is defined as:
\begin{equation*}\label{eq:repelling_laplacian}
(L_r)_{ij} = \begin{cases} 
\sum_{k} A_{ik}, \quad \text{if} \quad i=j, \\ 
-A_{ij}, \quad \text{if} \quad i\neq j .
\end{cases}
\end{equation*}
This slight alteration shifts the focus from structural balance to the dynamical behaviour of the network~\citep{panLaplacianDynamicsSigned2016}. Like the unsigned Laplacian, the repelling Laplacian always has a zero eigenvalue, with its associated eigenvector being the constant vector. Unlike the opposing Laplacian, $L_r$ is not necessarily positive semidefinite. This means that the repelling Laplacian is better suited for capturing instabilities and repulsive interactions that may lead to divergent behaviour in the network.

Both of these variants find a wide range of applications. The opposing Laplacian is instrumental in tasks such as spectral embedding for node clustering, random walks that explore the network structure, and the analysis of generalised distances like resistance distance~\citep{tianSpreadingStructuralBalance2024, kunegisApplicationsStructuralBalance2014, altafiniConsensusProblemsNetworks2013}. On the other hand, the repelling Laplacian has been employed in applications that require an understanding of dynamical stability, such as the analysis of electrical and mechanical networks where interactions can be repulsive, as well as consensus and synchronisation models where the stability of the evolving state is a central concern~\citep{shiDynamicsSignedNetworks2019, fontanPseudoinversesSignedLaplacian2023}.

Table \ref{tab:laplacian} provides a summary of the differences and applications of the opposing and repelling Laplacians. For completeness, a summary of the usual (unsigned) Laplacian is also included.

\clearpage
\begin{longtblr}[
  caption = {Comparison between the unsigned, opposing, and repelling Laplacians. $G$ is assumed to be undirected and connected.},
  label = {tab:laplacian},
]{
  colspec = {X X X X},
  width = \linewidth,
  cell{1}{1} = {cmd=\textbf},
  cell{1}{2} = {cmd=\textbf},
  cell{1}{3} = {cmd=\textbf},
  cell{1}{4} = {cmd=\textbf},
}
\hline
 & Unsigned Laplacian, \(L_u\) & Opposing Laplacian, \(L_o\) & Repelling Laplacian, \(L_r\) \\
\hline
\textbf{Formula} &
\((L_u)_{ij} =
\begin{cases}
\sum_k |A_{ik}|, & \text{if } i=j;\\
-|A_{ij}|, & \text{if } i\neq j.
\end{cases}\) &
\((L_o)_{ij} =
\begin{cases}
\sum_k |A_{ik}|, & \text{if } i=j;\\
-A_{ij}, & \text{if } i\neq j.
\end{cases}\) &
\((L_r)_{ij} =
\begin{cases}
\sum_k A_{ik}, & \text{if } i=j;\\
-A_{ij}, & \text{if } i\neq j.
\end{cases}\) \\
\textbf{Associated property} & Unsigned community structure & Signed community structure & Signed community structure \\
\textbf{Positive semidefinite?} & Yes & Yes & Not in general \\
\textbf{Off-diagonal elements} & All negative & Mixed & Mixed \\
\textbf{Associated incidence matrix?} & Yes & Yes & No \\
\textbf{Zero eigenvalue} & Always present & Only if \(G\) balanced & Always present \\
\textbf{Eigenvector associated with zero eigenvalue} & Constant vector, \(\vec{1}\) & Indicator vector of factions (only if \(G\) balanced) & Constant vector, \(\vec{1}\) \\
\textbf{Smallest non-constant eigenvector} & Reveals communities (Fiedler vector; associated with 2\textsuperscript{nd} smallest eigenvalue) & Reveals approximately balanced factions (associated with smallest eigenvalue) & Reveals approximately balanced factions (associated with 2\textsuperscript{nd} smallest eigenvalue) \\
\textbf{Applications} &
Node embeddings~\citep{vonLuxburg2007a}, random walks~\citep{masuda2017random}, linear consensus~\citep{degrootReachingConsensus1974}, synchronisation~\citep{arenasSynchronizationComplexNetworks2008}, generalised distances (resistance distance)~\citep{devriendtEffectiveResistanceMore2022,chandraElectricalResistanceGraph1989}, analysis of electrical and spring networks~\citep{chandraElectricalResistanceGraph1989,lovaszRandomWalksGraphs1993}, kernel functions~\citep{foussRandomWalkComputationSimilarities2007}. &
Node embeddings~\citep{kunegisSpectralAnalysisSigned2010,cucuringuSPONGEGeneralizedEigenproblem2019}, random walks~\citep{tianSpreadingStructuralBalance2024}, linear consensus~\citep{altafiniConsensusProblemsNetworks2013}, generalised distances (resistance distance)~\citep{kunegisApplicationsStructuralBalance2014}, kernel functions~\citep{kunegisSpectralAnalysisSigned2010}. &
Node embedding (SHEEP)~\citep{babulSHEEPSignedHamiltonian2024}, analysis of electrical and spring networks~\citep{fontanPseudoinversesSignedLaplacian2023}, linear consensus, synchronisation~\citep{panLaplacianDynamicsSigned2016,shiDynamicsSignedNetworks2019}. \\
\hline
\end{longtblr}

\newpage

\section{Cognitive and structural balance} \label{chapter:balance}

In the previous chapter, we treated signed networks as a mathematical abstraction independent of the specific system they represent. This approach has the advantage of yielding general results applicable to a wide variety of systems, but it also forgets about the unique characteristics of those individual systems. Furthermore, it overlooks the historical origins of the field in the study of social systems, which influence the terminology, concepts, and methods we use nowadays. This observation becomes particularly relevant when discussing the concept of balance. Balance is a central concept in signed networks that can be traced back to the origins of the field \citep{heiderAttitudesCognitiveOrganization1946,cartwrightAmbivalenceIndifferenceGeneralizations1970}. 
In a nutshell, it refers to the way connections in a system are arranged to produce a sense of consistency or stability for the individuals that form that system. As an example, a configuration where the friend of my friend is my friend and the enemy of my enemy is my friend would be considered balanced.

The signed network literature contains two closely related yet distinct notions often referred to as ``balance'': cognitive balance, introduced by Fritz Heider~\citep{heiderAttitudesCognitiveOrganization1946}, and structural balance, introduced by Cartwright and Harary~\citep{hararyNotionBalanceSigned1953,cartwrightStructuralBalanceGeneralization1956}. Both frameworks aim to capture the idea that certain configurations of positive and negative relations are more stable or consistent than others; nevertheless, they diverge significantly in scope and interpretation. \textit{Heiderian balance}, rooted in cognitive psychology, focuses on how individuals perceive and respond to patterns of relations in their immediate social environment. It emphasises subjective (local) experience and psychological tension as drivers of network change. In contrast, \textit{Hararian balance} abstracts away individual perception to define balance as a global structural property of a signed network. This formalisation enables the application of graph-theoretical tools and extends the notion of balance beyond social contexts to any domain involving antagonistic interactions. This chapter compares these two perspectives, clarifying common misconceptions and identifying contexts where each framework is most applicable. It then surveys several mathematical properties of balanced networks. Finally, it introduces a range of balance indices to quantify Hararian balance in empirical data.

\subsection{Heiderian (cognitive) balance}\label{subsec:Heiderian_balance}

\begin{figure}[ht]
\centering
\includegraphics[width=\linewidth]{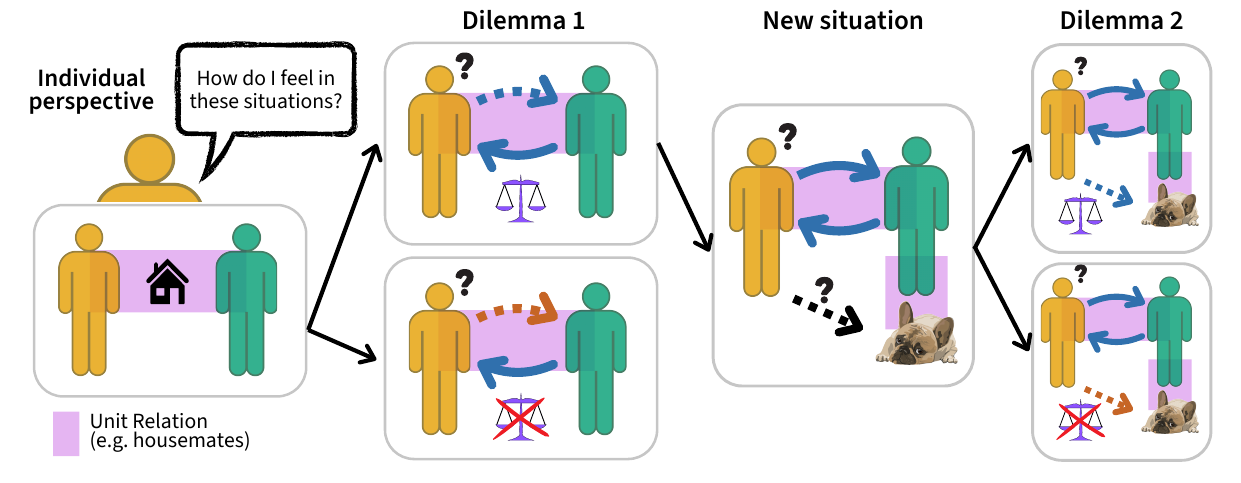}
\caption{\textbf{Heiderian balance at play.} The figure illustrates how cognitive balance operates in a real system, emphasising the local perspective and the interaction between unit relations and attitudes. We consider two individuals living in the same household, hence connected by a unit relation. From the viewpoint of the leftmost person (depicted in yellow), two situations are evaluated in terms of Heiderian balance. In the first, \textit{Dilemma 1}, the other person (depicted in green) expresses liking toward the focal individual; according to Heiderian balance, this configuration is balanced only if the feeling is reciprocated. In the second situation, \textit{Dilemma 2}, the housemate introduces a pet into the home. Given the structure of unit relations and attitudes in this system, balance is achieved if the focal person also likes the pet.}
\label{fig:heider}
\end{figure}

We first explore Heider's notion of balance, which we refer to as \textit{cognitive balance}. In his foundational work~\citep{heiderAttitudesCognitiveOrganization1946}, Heider analysed systems composed of two or three elements, which can be either people (denoted by $p$, $o$, $q$) or impersonal elements like objects or concepts (denoted by $x$). He distinguished between two types of pairwise relations between elements: \textit{unit relations} and \textit{attitudes}. A unit relation, written as $pUq$, refers to an objective connection between $p$ and $q$ that goes beyond personal perception, such as belonging, similarity, or spatial proximity. These relations are symmetric, even if not equivalent in meaning from both directions (for instance, $p$ belongs to $q$ always implies that $q$ contains $p$). In contrast, an attitude, written as $pLq$, expresses a subjective evaluation like liking or trusting. Unlike unit relations, attitudes can be asymmetric because they are rooted in the local individual perception. Importantly, any relation $U$ or $L$ can be negated, so that, if $L$ means liking, $p\neg Lq$ means that $p$ does not like $q$. In summary, the differentiation between unit relations and attitudes refers to the differentiation between the material configuration and the psychological configuration between two or more elements.

Using these notions, Heider introduced intuitive notions about when a configuration is balanced. For instance, a basic intuition is that a configuration involving three elements is balanced if a person's attitude toward the endpoints of a unit relation is consistent. For example, if $p$ likes intelligent people ($pLx$) and $o$ is intelligent ($oUx$), then the configuration is balanced if $p$ likes $o$ ($pLo$). Heider's initial proposal is that a system is balanced when all elements of a unit are connected by positive attitudes, or when negative attitudes appear only between elements that do not belong to the same unit. Both attitudes and unit relations can change to reduce imbalance and resolve psychological tension. 

This idea can be extended to two-element systems, so that configurations like ($pLx$, $pUx$) or ($p\neg Lx$, $p\neg Ux$) are balanced ($p$ likes what it is connected to, or dislikes what it is not). The same logic applies when both elements are people: for instance, ($pLo$, $pUo$) is balanced ($p$ likes $o$, who belongs to its family), as is mutual liking ($pLo$, $oLp$) (see \textit{Dilemma 1} in Figure~\ref{fig:heider}) or mutual dislike ($p\neg Lo$, $o\neg Lp$). In systems with three elements, the number of possible configurations increases. A configuration like ($pLo$, $oUx$, $pLx$) is balanced: $p$ likes $o$, $o$ is related to $x$, and $p$ also likes $x$ (see \textit{Dilemma 2} in Figure~\ref{fig:heider}). Another familiar example is ($pLo$, $oLx$, $pLx$), which illustrates transitivity: if $p$ likes $o$ and $o$ likes $x$, then $p$ also likes $x$. Moreover, the triangle  ($pLo$, $o\neg Lx$, $p\neg Lx$), reminiscent of the adage ``the enemy of my enemy is my friend'', is also balanced according to Heider's conceptualisation. An illustrative example of cognitive balance at play is provided in Figure \ref{fig:heider}. 

With this relatively simple formulation, Heider provides us with a powerful framework to explain how a system can experience tensions depending on its configuration, and evolve (or not) towards balanced states. Interpretability goes well beyond signed triads. In fact, many well-known social mechanisms can be understood as instances of Heiderian balance (see Figure~\ref{fig:mechanisms} for a visual representation of such mechanisms, and section~\ref{subsubsec:mechanisms-governing} for a more detailed overview). For example, homophily (the tendency of people to like those with whom they share attributes) can be expressed in Heiderian terms as ($pUo$, $pLo$), while heterophobia (the tendency to dislike others with different attributes) is represented by ($p\neg Uo$, $p\neg Lo$). Similarly, transitivity corresponds to ($pLo$, $oLq$, $pLq$) and reciprocity to ($pLo$, $oLp$). Differential popularity also fits this logic: a person $q$ may attract more positive links because they possess a desirable attribute $x$ ($qUx$), which $p$ values ($pLx$), thus leading to $pLq$. Even mechanisms like conflict aversion can be captured: if $p$ dislikes conflict ($p\neg Lx$) but is engaged in one ($pUx$), Heider’s balance predicts that $p$ will seek to resolve or abandon the conflict, even if it forms part of a structurally balanced triangle. 

The core idea of Heiderian balance lies in its focus on the cognitive domain, that is, the psychological perceptions of individuals within a system. The key implication is that Heiderian balance represents an inherently local operationalisation of balance. Put simply, the question is not whether the triad $poq$ is balanced in some absolute sense, but whether it is perceived as balanced from $p$’s perspective. Naturally, by aggregating the perspectives of $p$, $o$, and $q$, we can evaluate balance at the level of the triad itself. However, this overlaps with the second notion of balance, structural balance, which we introduce in the next section. We will return to the contrast between local versus global, and cognitive versus structural balance, once both perspectives have been introduced.

Moreover, it is worth mentioning that some studies have focused on empirically validating the existence of this ``subjective experience'', namely, the perception that certain configurations feel more tense or less stable, and its limitations. For instance,~\citep{morrissetteExperimentalStudyTheory1958} assesses stress levels of respondents in certain hypothetical configurations of a system, and~\citep{whiteLimitationBalanceTheory1977} shows the importance of individuals feeling personally involved in the situation to test any hypothesis related to Heider's balance. Other examples that explore this cognitive perspective of balance in hypothetical situations are~\citep{millerStructuralBalanceDyads1972},~\citep{horowitzInductionForcesDiscussion1951},~\citep{chiangTriadicBalanceBrain2020}, which explores with neuroimaging differences in the brain when people are in unbalanced vs. balanced situations, or~\citep{jordanBehavioralForcesThat1953}, which explores the existence of conflict aversion beyond balance.

Heider's notion of balance has some limitations~\citep{cartwrightStructuralBalanceGeneralization1956}. First, negative relations in his framework are ambiguous, as they can represent either the absence of a positive relation or an explicitly antagonistic relation. For instance, $p\neg Lq$ could mean that $p$ simply does not consider $q$ a friend, or that $p$ and $q$ are actual enemies. The notion of balance in the system as a whole is ambiguous itself: a single directional positive link can be balanced with respect to transitivity but unbalanced with respect to reciprocity.  Second, Heider’s formulation is not trivially generalisable beyond pairs and triads, while most social or complex networks involve larger groups of interacting entities. Finally, his work was restricted to cognitive or attitudinal relations between elements, even though the underlying ideas had the potential to be generalised to arbitrary signed relations.

\subsection{Hararian (structural) balance}  \label{subsec:Hararian_balance}

Heider’s notion of balance is grounded in the individual and local perceptions of people within a system of two or three elements. In his framework, balance is described conceptually rather than through formal rules. In this setting, the key insight of Cartwright and Harary was to recognise that Heider’s theory, though expressed in psychological and local terms, pointed towards general structural properties that could be formalised within the language of graph theory~\citep{cartwrightStructuralBalanceGeneralization1956}\footnote{Harary first introduced structural balance as a purely mathematical property of graphs~\citep{hararyNotionBalanceSigned1953}. It was only three years later that he and Cartwright fully developed the connections between this mathematical notion of balance and Heider's theory~\citep{cartwrightStructuralBalanceGeneralization1956}.}. This new conceptualisation of balance, often referred to as Hararian or structural balance, quickly gained popularity and has, over time, somewhat overshadowed Heider’s original and broader notion of balance.

Cartwright and Harary’s core realization was that, within Heider’s framework, if (1) we restrict attention to systems in which relations represent only attitudes, ignoring unit relations, (2) disregard the directionality of attitudes so that liking or disliking is treated as mutual and, importantly, no longer local\footnote{In their original work~\citep{cartwrightStructuralBalanceGeneralization1956}, the authors considered directed graphs; however, most applications of structural balance theory are concerned with undirected networks.}, and (3) evaluate whether a triad is balanced from a global perspective (that is, disregarding individual perceptions and considering the system as a whole), then balanced triads are precisely those containing an even number of negative relations, whereas unbalanced triads contain an odd number. They then extended this principle beyond triads, proposing that the same rule could apply to any cycle in a graph: a cycle is balanced if it has an even number of negative edges, and unbalanced if the number is odd. Finally, they generalised this to networks by defining a network as balanced if and only if all its cycles are balanced. With these steps, we arrive at the following definition:

\textbf{Definition (balanced network).}  
An undirected signed graph is \emph{balanced} if all its cycles contain an even number of negative edges; otherwise, it is unbalanced.

By construction, this new notion of balance avoids the ambiguity of defining a negative link merely as the negation of a positive attitude (non-liking versus disliking), and it also allows the idea of balance to be generalised to larger graphs beyond dyads and triads. Moreover, its abstract nature allows the concept of balance to be extended beyond psychology to any scientific field where antagonistic relations are present. However, translating Heider's ideas into mathematical terms inevitably disregards some important aspects of Heider's original formulation. For example, no distinction is made between attitudes and unit relations, nor between material and cognitive entities. In addition, because Harary’s theory is formulated for signed graphs, it does not account for structures that Heider would have considered balanced but that do not involve explicit negations of relations (a notable example is transitivity: removing an edge from an all-positive triangle does not lead to an unbalanced structure in Harary's sense, but may do so under Heider's notion).

\subsection{Comparison between Heiderian and Hararian balance}\label{subsec:hei-har}

\begin{figure}
    \centering
    \includegraphics[width=\linewidth]{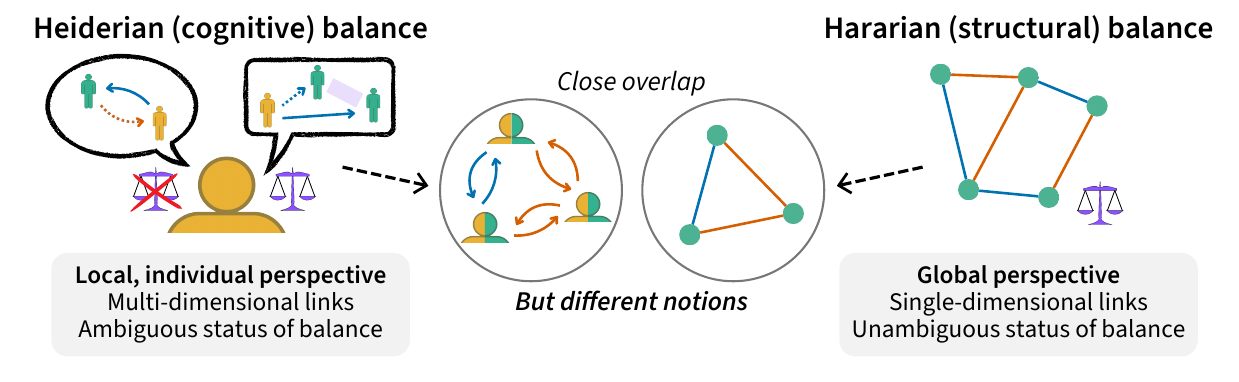}
    \caption{\textbf{Cognitive vs structural balance.} The leftmost diagram illustrates cognitive balance, which is characterised by a local perspective: whether a situation is balanced depends on the viewpoint of a specific individual rather than a global assessment. Links in this framework are multi-dimensional, representing both attitudes (positive or negative) and unit relations that can reflect various types of connections. As a result, the overall balance of the system is inherently ambiguous, as it depends on the aggregation of all local perspectives, which may or may not agree. In contrast, the rightmost diagram depicts structural balance, which adopts a global perspective: the system is either balanced or unbalanced as a whole, independently of individual viewpoints. Here, links represent only attitudes (positive or negative), and the status of balance is unambiguous due to this global, uniform interpretation. The central panel illustrates this conceptual contrast by showing a signed triad analysed under both frameworks. Although the underlying structure seems equivalent, both notions are fundamentally different.}
    \label{fig:balance-comparison}
\end{figure}
Although Heiderian and Hararian balance are often treated interchangeably in the literature, they rest on different conceptual foundations and can even lead to contradictory predictions. Heider's notion of balance is rooted in cognitive psychology: it aims to model how individuals seek internal consistency. In contrast, Harary's formulation of structural balance discards psychological interpretation and formalises balance as a global mathematical property of a network. The fundamental difference between these two notions lies in the local versus global conception of balance. At first glance, this distinction may seem minor, but it profoundly changes how a system is studied. This becomes clearer when we consider the closest point of intersection between both frameworks: a triad of individuals connected by relations of liking and disliking. From Heider’s perspective, we have six potential attitudes (directed relations from each individual toward the other two members of the triad). For each person, we can define a local perception of balance. That is, each individual holds their own view of whether the triad is balanced, and these views may differ depending on what they perceive from their local standpoint. Moreover, these local notions of balance are multi-dimensional: one individual may perceive tension because a relationship is not reciprocated, while at the same time judging the triad to be balanced due to transitivity with the third person. By contrast, in Harary’s framework, relations are rarely directed, and there are only three of them in a triad. Here, the definition of balance is unambiguous and can be uniquely determined simply by counting the number of negative relations. Thus, in Harary’s perspective, we can definitively state whether a triad is balanced or not, whereas in Heider’s notion, the focus is on how local perceptions may generate tension for the individuals involved. An illustrative comparison between both notions is depicted in Figure \ref{fig:balance-comparison}.

It is worth noting that when the study of structural balance is restricted to triads, but longer cycles are ignored, the term social balance is sometimes used. With this terminology, structural balance can be viewed as a generalisation of social balance when extended to larger cycles in broader networks. However, structural balance is not strictly a generalisation of cognitive balance. Rather, it is an operationalisation of Heider’s core idea at the collective level, allowing it to be scaled and formalised mathematically. 

The difference in scope has important practical implications. Heiderian balance provides a powerful framework for modelling and interpreting social behaviour, even outside signed networks, because it captures a wide range of psychosocial mechanisms such as reciprocity, transitivity, homophily, heterophobia, and conflict aversion.  However, it has a limited scope for understanding phenomena beyond social systems, and it is not suitable for operationalisation and mathematical implementation. Harary’s structural balance, by contrast, trades cognitive detail for generality. Its abstraction allows balance to be analysed in any signed network, social or otherwise, and scales naturally to larger structures. Both theories are thus complementary: Heider provides fine-grained insight into cognitive mechanisms, while Harary’s provides general tools applicable to any system and able to describe large-scale structures. 

Because of these differences, applying one framework in contexts suited for the other can be problematic, and it can lead to contradicting predictions. For example, conflict aversion, a form of Heiderian balance, may favour dissolving negative links, even if they take part in balanced cycles in Harary's sense. Unfortunately, empirical studies have blurred the line between the two, often interpreting results that support Heider’s cognitive mechanisms as evidence for structural balance in the Hararian sense or viceversa. This conflation is partly historical: Cartwright and Harary originally presented structural balance as a formal generalisation of Heider’s ideas, which encouraged researchers to treat them as equivalent despite their conceptual divergence. A constructive way forward is to recognise that these approaches can complement rather than compete with each other. An example is the work of Doreian and Hummon~\citep{doreianEvolutionHumanSigned2004,hummonDynamicsSocialBalance2003}, who developed a dynamic model that integrates both perspectives: Heiderian balance at the local cognitive level and structural balance at the group level. Most of the focus in this review will be on Harary’s structural balance, although Heiderian balance remains important and can be useful in certain contexts.

\subsection{Mathematical properties of balanced networks}\label{subsec:math-propr-balanced}

Harary's mathematical definition of balance has the advantage of unlocking the full machinery of graph theory. With this framework, we can derive results and properties that hold for any signed graph, regardless of its size or the details of the complex system it represents. One of the most striking examples of this generality is Harary's theorem~\citep{hararyNotionBalanceSigned1953}, which states that any balanced graph, regardless of its size, can be divided into two antagonistic groups.

\begin{figure}[h!]
\centering
\includegraphics[width=0.75\linewidth]{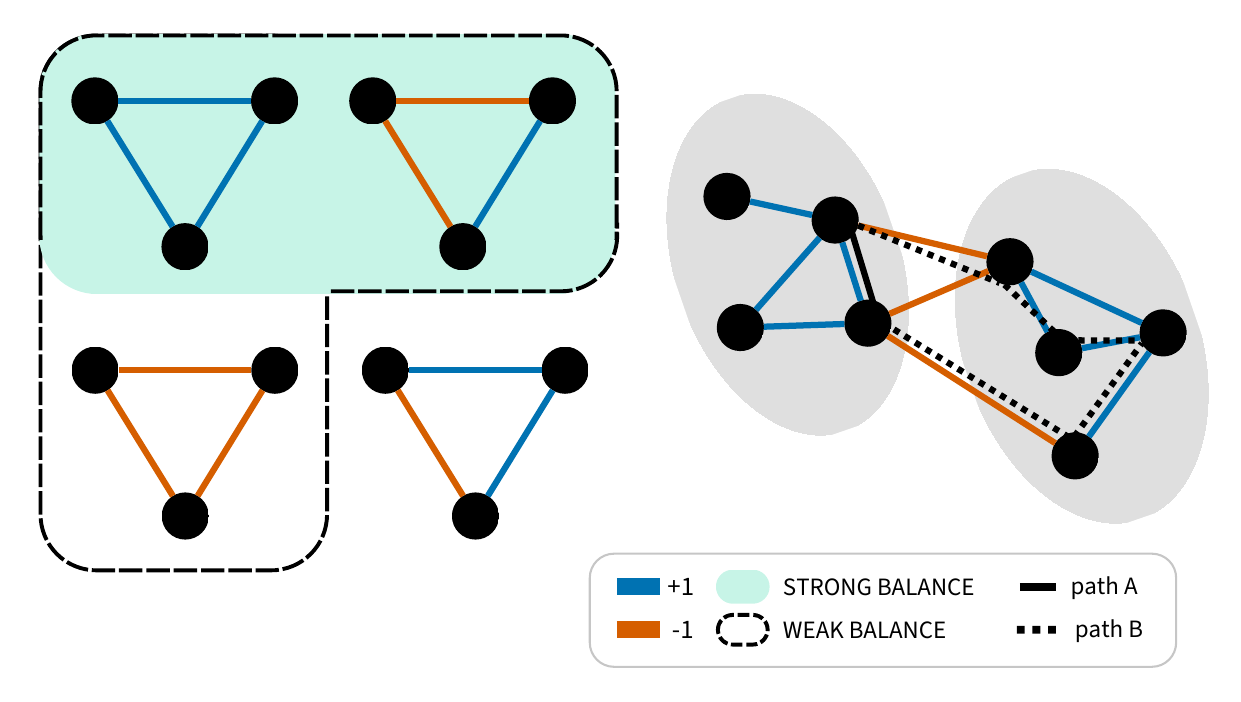}
\caption{\textbf{Overview of structural balance. }
Left: the four possible triads in an undirected signed network. 
Balanced triads are those with an even number of negative edges \((+++)\) and \((+--)\). 
Weakly balanced triads include these two plus the \((---)\) triad. 
Right: illustration of Harary’s structure theorem. 
The network shown is balanced, so it satisfies the three statements in the theorem. 
First, every cycle has an even number of negative edges. 
Second, all paths between the same two nodes have the same sign (for example, paths A and B are both positive). 
Third, the node set can be divided into two balanced factions (marked by grey ellipses) such that positive edges lie within groups and negative edges lie between groups.}
\label{fig:balance}
\end{figure}

\textbf{Theorem (Harary):}  
Let \( G \) be an undirected signed graph. Then the following statements are equivalent:  \begin{enumerate}
    \item \( G \) is balanced (i.e., every cycle in \( G \) has a positive product of its edge signs).
    \item  For any pair of nodes \( v_i \) and \( v_j \), all paths connecting them have the same sign (given by the product of the signs of their edges).
    \item The node set \( V \) can be partitioned into two disjoint subsets in such a way that all positive edges join nodes within the same subset while all negative edges join nodes in different subsets.  
\end{enumerate}

In other words, the defining property of balanced networks is that they admit a natural bipartition into two factions, often referred to as the \textit{balanced factions}. Thus, in a balanced network, every node has either a relation of effective alliance or enmity with any other node in the network, with no ambiguous relations. This immediately makes tasks like link sign prediction trivial: for any pair of disconnected nodes \(i,j\) in a balanced graph, any structural-balance-based algorithm will predict a positive connection when the nodes belong to the same balanced faction, and negative otherwise (see Section~\ref{subsubsec:ML-link-pred} for an overview on sign prediction methods). 
Furthermore, structural balance has strong algebraic consequences. First, the adjacency matrix $A$ and its unsigned counterpart $|A|$ are cospectral; that is, they have identical eigenvalue spectra~\citep{acharyaSpectralCriterionCycle1980}. Second, any balanced signed graph can be transformed into an equivalent unsigned graph through a node-based operation called \emph{switching}~\citep{zaslavskySignedGraphs1982}, which reverses the signs of all edges incident to selected nodes. This means that many structural and dynamical properties of balanced signed networks can be studied using the rich theory of unsigned graphs~\citep{zaslavskyMatricesTheorySigned2013}.

With respect to dynamics, balance guarantees a coherent propagation of information. Coherent means that the effect of a signal does not depend on the particular path it takes: because all paths between any two nodes have the same sign, any piece of information arrives at the target node with the same polarity, independently of the chosen path. This typically gives rise to polarised dynamical states, where the dynamical states of nodes belonging to opposite balanced factions have opposite sign~\citep{altafiniConsensusProblemsNetworks2013} (see Section~\ref{subsec:dynamics-on-networks}).

In terms of energy minimisation, balanced configurations correspond to ground states of the Ising Hamiltonian\footnote{In statistical physics, the Ising model describes interacting spins that can take two values, $+1$ or $-1$. The Hamiltonian encodes the energy of a configuration, and ground states are those with minimal energy. Interpreting factions as spin values, a balanced partition minimises the Hamiltonian, so balance corresponds to an energy minimum.}. This provides a solid connection between structural balance and statistical physics, and more concretely, the theory of spin glasses~\citep{facchettiComputingGlobalStructural2011,facchettiExploringLowenergyLandscape2012}. For instance, one can use simulated annealing techniques to find the balanced factions, or even extend these techniques to unbalanced networks that are close to a balanced state~\citep{traagCommunityDetectionNetworks2009}. The energy-minimisation interpretation of balance also provides a suggestive explanation as to why many social and biological systems gravitate toward balanced structures~\citep{facchettiComputingGlobalStructural2011}.

Harary’s characterisation of balance has a counterpart, known as \emph{antibalance}~\citep{hararyStructuralDuality1957}. A graph is antibalanced if it becomes balanced after flipping the sign of every edge. Equivalently, an antibalanced graph satisfies that every cycle has an even number of \textit{positive} edges. In matrix terms, antibalance means that the spectrum of \(A\) is the negative of the spectrum of \(|A|\). Antibalanced networks can also be partitioned into two antibalanced factions, so that \textit{negative} edges lie within each group and \textit{positive} edges across the two groups.

\begin{figure}
    \centering
    \includegraphics[width=\linewidth]{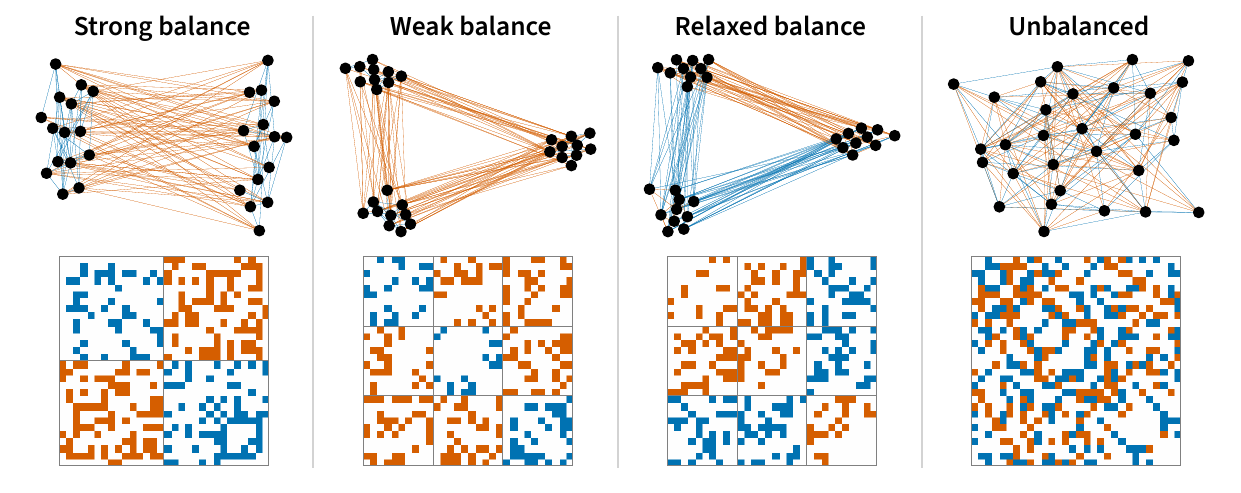}
    \caption{\textbf{Examples of strongly balanced, weakly balanced, relaxed-balanced, and unbalanced signed networks, together with their adjacency matrices.} In the strongly balanced case, nodes can be grouped in two factions, with all edges within a faction being positive and all edges between factions being negative. This results in an adjacency matrix with two positive blocks in the main diagonal and two off-diagonal negative blocks. In the weakly balanced case, more than two factions are allowed, resulting in several positive diagonal blocks in the adjacency matrix, with negative off-diagonal blocks. For relaxed-balanced networks, the adjacency matrix can be permuted into blocks that are each entirely positive or entirely negative, without requiring positive blocks to lie on the diagonal or negative blocks off the diagonal. Finally, unbalanced networks lack any consistent block structure, so signs are scattered across the matrix.}
    \label{fig:balance-types}
\end{figure}

Arguably, the most important criticism of Harary’s theory is that it is too restrictive for many practical applications. In reality, most signed networks do not fulfil the conditions for strict balance. In some cases, however, they do satisfy a weaker form of balance, whereby nodes can be divided into more than two mutually antagonistic factions. This observation motivated Davis to introduce the notion of \emph{weak balance}~\citep{davisClusteringStructuralBalance1967}\footnote{Davis originally termed this concept ``clustering''; today ``clustering'' typically refers to community‑detection algorithms or clustering coefficients.}:
A signed graph is \emph{weakly balanced} if it contains no cycle with exactly one negative edge\footnote{To distinguish the two notions of structural balance, Harary’s original definition is often called \emph{strong} balance.}.
Weakly balanced graphs fulfil an analogue of Harary's theorem, called Davis’s structure theorem: a signed graph is weakly balanced if and only if its nodes can be partitioned into \(k\ge2\) disjoint factions such that every positive edge lies within a faction and every negative edge lies between factions (see Figure \ref{fig:balance-types}). In this sense, weak balance generalises Harary’s bipartition by allowing more than two balanced factions, thus capturing multipolar configurations (e.g., multiple rival coalitions) that strong balance cannot.

More recently, Doreian and Mrvar~\citep{doreianPartitioningSignedSocial2009} argued that even weak balance remains too rigid for many empirical networks. Accordingly, they proposed the \emph{relaxed structural balance} block model. 
A signed graph is relaxed-balanced if its adjacency matrix can be permuted into a block structure containing only three block types: positive (all links are positive), negative (all links are negative), or null (no links), without restriction on block location (see Figure \ref{fig:balance-types}).
Relaxed balance accounts for phenomena invisible under strict or weak balance, such as mediation (actors bridging hostile groups), differential popularity (actors receiving uniformly positive links), and internal hostility (groups where all members dislike each other).

\subsection{Balance indices}  \label{sec:balance_indices}

Balanced signed networks have many appealing theoretical properties, like the existence of a partition of the nodes into balanced factions. In practice, however, perfect balance is too restrictive. It requires the complete absence of negative cycles, a condition that becomes increasingly unlikely as network size and complexity grow. As a result, empirical networks rarely satisfy the exact criteria of balance. This motivates the question of whether one can measure to what extent they lean toward balance; in other words, whether their level of imbalance is significantly lower than what a suitable null model would predict. To answer this question, several quantitative indices have been proposed over the years. The key idea is to identify a network property that correlates with the number of negative cycles while being computationally easier to evaluate. In this review, we consider four popular approaches: those based on \textit{motif counts}, \textit{frustrated edges}, \textit{algebraic properties}, and \textit{closed walks}.

The simplest way to measure balance uses \textit{motifs}. It relies on the idea that small subgraphs can serve as proxies for global structure. Harary himself pioneered this approach, proposing cycles of size order $N$ as a reasonable indicator of the level of balance~\citep{harary1955local}. Today, triangles are the most common choice, since they are the smallest cycles that can reveal balance. The idea is to compare how many of them are balanced and how many are not~\citep{galloTestingStructuralBalance2024, haoProperNetworkRandomization2023, arefMultilevelStructuralEvaluation2020}. A convenient way to do so is computing the trace of the cube of the adjacency matrix, $\mathrm{tr}(A^3)$. To control for the total number of triangles, one can use the ratio
\[
T = \frac{\mathrm{tr}(A^3)}{\mathrm{tr}(|A|^3)}.
\]
This quantity reflects the difference between positive and negative triangles relative to the total number of triangles. One can also easily adapt this metric to measure balance locally: 
\[
T_i = \frac{(A^3)_{ii}}{(|A|^3)_{ii}}.
\]
This approach has clear advantages: it is very simple and fast to compute. Yet this simplicity is also its main limitation. Structural balance is a global property, so consistency at the level of triangles does not guarantee balance at larger scales. A network can have only balanced triangles but still be globally unbalanced when longer cycles exist. For this reason, motif-based measures are better interpreted as indicators of local consistency rather than full measures of balance.

The \textit{frustration-based approach}~\citep{facchettiComputingGlobalStructural2011, facchettiExploringLowenergyLandscape2012, aref2018measuring, arefBalanceFrustrationSigned2019, arefMultilevelStructuralEvaluation2020}, is rooted in spin glass theory and statistical mechanics~\citep{mezardSpinGlassTheory1987}, and treats the network as a system of interacting spins. Such systems can exhibit frustration, a phenomenon that occurs when no configuration of spins can satisfy all interactions at once. In other words, no matter how the spins are aligned, at least one positive edge connects spins that point in opposite directions, or at least one negative edge connects spins that point in the same direction. Consider, for example, a triangle where all edges are negative: each pair of nodes prefers to point in opposite directions, but three spins cannot all be opposite, so at least one interaction must remain unsatisfied. Even in the most favourable configuration, understood as the one that satisfies as many interactions as possible, some edges remain in conflict (frustrated). The minimum number of such edges is called the \textit{frustration index} of the network. Equivalently, the frustration index is the minimum number of edges that must be removed or have their sign flipped to obtain a system with no frustration. 

When no edges are frustrated in the optimal configuration, then all positive edges link aligned spins and all negative edges link opposite spins. This condition is exactly the definition of a balanced network, with the spin direction indicating the balanced factions. Hence, the presence of frustration is equivalent to imbalance, and the frustration index measures how far a network is from perfect balance. 

Frustration-based approaches have several advantages. They establish a direct link between imbalance and a physically meaningful quantity, namely, the number of unsatisfied interactions. They also allow us to use theoretical concepts from spin glass physics, such as energy landscapes and optimisation principles, as well as powerful computational tools originally developed in statistical mechanics and combinatorial optimisation. Their main limitation lies in computation. Determining the frustration index is an NP-hard problem, so exact solutions are not feasible for large networks, and one must rely on heuristic or approximate algorithms~\citep{facchettiComputingGlobalStructural2011}.

The \textit{algebraic approach} assesses balance through spectral properties of signed matrices. The most common choice is the opposing Laplacian (see Section \ref{subsec:signed-laplacian}). Just as the second smallest eigenvalue of the unsigned Laplacian measures connectivity in an unsigned network, the smallest eigenvalue \(\mu_1\) of the opposing Laplacian serves as an indicator of imbalance~\citep{kunegisSpectralAnalysisSigned2010}. It is equal to zero if and only if the network is perfectly balanced, while larger values indicate an increasingly unbalanced structure. An advantage of this spectral viewpoint is that it connects structural balance to network dynamics such as synchronisation, diffusion, or consensus processes. In particular, \(\mu_1\) controls the rate at which diffusive dynamics would converge to the trivial state where all node variables vanish~\citep{altafiniConsensusProblemsNetworks2013}. Highly unbalanced networks have large \(\mu_1\) and therefore decay rapidly, whereas perfectly balanced networks never decay to that state at all.

The last approach we will discuss uses \textit{closed walks} as proxies for cycles\footnote{This assumption has proved valid for a wide range of empirical systems~\citep{talagaPolarizationMultiscaleStructural2023}.}~\citep{kirkleyBalanceSignedNetworks2019, estradaWalkbasedMeasureBalance2014, estradaRethinkingStructuralBalance2019}. It compares the number of positive and negative closed walks through powers of the adjacency matrix (Section \ref{subsec:definition_signed_networks}), weighting their contributions by length. This is done by assigning a penalty \(c_k\) to walks of length \(k\):

\begin{align} \label{aux:walk-based-balance}
    \kappa = \sum_{k=0}^\infty c_k\, \mathrm{tr}(A^k),
\end{align}

where the trace ensures that only closed walks are counted. As with the triangle index, one typically normalises by the total number of closed walks in the unsigned network so that the index lies in \([0,1]\):

\begin{align*}
    \kappa = \frac{\sum_{k=0}^\infty c_k\, \mathrm{tr}(A^k)}{\sum_{k=0}^\infty c_k\, \mathrm{tr}(|A|^k)}.
\end{align*}

A value $\kappa=1$ indicates perfect balance\footnote{This follows from the fact that \(A\) and \(|A|\) are cospectral if and only if the network is balanced.} Several choices of \(c_k\) appear in the literature, such as \(c_k = \alpha^k\)~\citep{singhMeasuringBalanceSigned2017} or \(c_k = \alpha^k / k\)~\citep{kirkleyBalanceSignedNetworks2019}. The most common choice is \(c_k = \beta^k / k!\), where \(\beta > 0\) controls the penalty for longer walks~\citep{estradaWalkbasedMeasureBalance2014}. With this choice, Eq.~\eqref{aux:walk-based-balance} becomes the matrix exponential \(\mathrm{tr}(e^{\beta A})\).

This framework admits a node-level extension that quantifies the contribution of each node to global balance~\citep{diaz-diazMathematicalModelingLocal2024,diazdiazthesis}:

\begin{align*}
    \kappa_i = \frac{\sum_{k=0}^\infty c_k\, (A^k)_{ii}}{\sum_{k=0}^\infty c_k\, (|A|^k)_{ii}}.
\end{align*}

The walk-based approach has notable advantages. Closed walks can be counted efficiently using matrix powers, and the matrix exponential formulation allows us to leverage well-established theoretical properties and efficient numerical algorithms~\citep{molerNineteenDubiousWays2003}. It also creates links with statistical mechanics, where \(\beta\) plays the role of an inverse temperature, and with diffusion processes, where \(\beta\) represents time~\citep{estrada2012Physics}. However, closed walks have an important limitation: they allow backtracking, that is, repeatedly moving back and forth along the same edge instead of exploring genuine cycles. As a result, long backtracking walks may dominate shorter ones, even when penalisation factors are applied. Recent advances have proposed principled methods to mitigate this effect with remarkable success~\citep{talagaPolarizationMultiscaleStructural2023}.

Table \ref{tab:balance_indices_detailed} summarises the main characteristics of the four balance measures discussed above.

\begin{longtblr}[
  caption = {Comparison of four approaches to measure balance in signed networks. $n$ is the number of nodes and $m$ is the number of edges. },
  label = {tab:balance_indices_detailed},
]{
  colspec = {X X X X X},
  width = \linewidth,
  rowhead = 1,
  cell{1}{1} = {cmd=\textbf},
  cell{1}{2} = {cmd=\textbf},
  cell{1}{3} = {cmd=\textbf},
  cell{1}{4} = {cmd=\textbf},
  cell{1}{5} = {cmd=\textbf},
}
\hline
Approach & Motif-based & Frustration-based & Algebraic & Walk-based \\
\hline
\textbf{Specific example} & Triangle index,  \(T=\frac{\mathrm{tr}(A^3)}{\mathrm{tr}(|A|^3)}\) & Frustration index & Smallest eigenvalue of the opposing Laplacian, $\mu_1$ & Estrada–Benzi index, \(\kappa=\frac{\mathrm{tr}(e^{\beta A})}{\mathrm{tr}(e^{\beta |A|})}\) \\

\textbf{Origin} & Social balance, motif enumeration & Spin-glass theory, statistical mechanics & Spectral theory, dynamical processes & Combinatorics, diffusion processes \\

\textbf{Bounds} & \([-1,1]\) & \([0,m]\) & \([0,2m/n]\) & \([0,1]\) \\

\textbf{Perfect balance distinguisher} & No (false positives possible) & Yes (zero iff balanced) & Yes (zero iff balanced) & Yes (one iff balanced) \\

\textbf{Tunable parameters} & No & No & No & Yes ($\beta$) \\
\hline
\end{longtblr}

\clearpage

\section{Network structure and inference}\label{chapter:structure}
\begin{figure}
    \centering
    \includegraphics[width=0.75\linewidth]{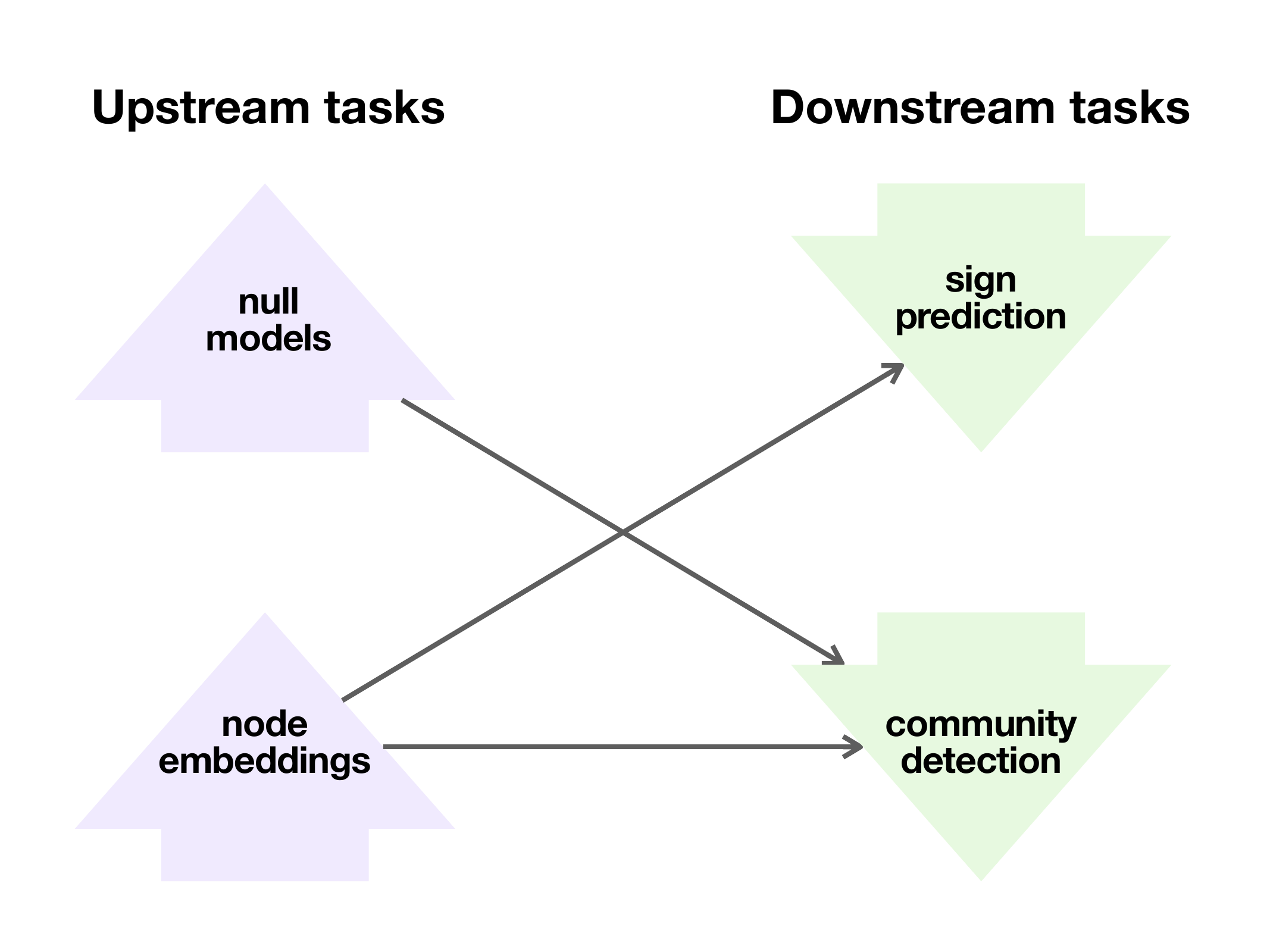}
    \caption{\textbf{Overview of upstream and downstream tasks in signed structural analysis}. Upstream methods such as null models and embeddings inform downstream applications, including sign prediction and community detection.}
    \label{fig:structure-inference-scheme}
\end{figure}

Historically, the approach to the study of networks has been two-fold: on the one hand, we are interested in investigating their complex structures, while on the other hand, we might want to study their evolution over time and the dynamical processes occurring on them~\citep{BOCCALETTI2006}. These two aspects are strongly interconnected, as the structure of a network can influence its dynamics (e.g., opinions will spread differently in a highly clustered network than in a non-clustered one) and vice versa (e.g., a change in opinion might break a link or alter its sign).

In this chapter, we focus on the structural properties of signed networks, while Chapter~\ref{chapter:dynamics} provides an overview of their dynamics. The network structure contains a wealth of information: by analysing it, we can identify antagonistic factions, predict missing links, find recurring motif patterns, and more. To obtain such information, we often need to perform intermediate, more fundamental, upstream tasks~\footnote{Upstream tasks refer to those tasks that are required to be completed first, to approach subsequent tasks, i.e., the downstream tasks.}.

Here, we will describe two such upstream tasks: comparisons with a null model and node embeddings. Null models, described in Section~\ref{subsec:null-models}, help us understand whether the patterns we find are significant, rather than features that even randomised versions of the same network would have. This is a fundamental step for distinguishing noise from statistical evidence, for example, when assessing structural balance in a signed network. On the other hand, node embeddings (Section~\ref{subsec:node-embeddings}) assign a vector to each node to reduce the system’s description to an $n\times f$ matrix, where $f\leq n$ is the dimension of each embedding. 
These upstream tasks serve as the basis for downstream, higher-level ones, such as sign prediction and community detection, as shown in Figure~\ref{fig:structure-inference-scheme}. For instance, node embedding tasks often precede sign prediction, and most community detection methods compare against null models or cluster node embeddings. 

Ultimately, these upstream steps serve a single purpose: to enable downstream tasks that test our hypotheses or describe the system. Two central downstream tasks of this kind are sign prediction and community detection~\footnote{We distinguish community detection from clustering, as the former is typically applied to networks, whereas the latter can be performed on non-networked data.}. Sign prediction (Section \ref{subsubsec:ML-link-pred}) is the signed-network extension of link prediction. The goal is to predict the sign of an edge based on certain features, thereby also allowing us to assess how far a given real-world network is from perfect balance. Sign prediction methods draw on physics, spectral graph theory, and machine learning. Traditional approaches use topological features (such as triads, degree patterns, or balance measures) and apply models like logistic regression or random forests, which remain interpretable and theoretically grounded. Deep learning methods, mainly Graph Neural Networks, learn node representations directly from structure and attributes through iterative aggregation of neighbour information, which captures complex dependencies but reduces interpretability. On the other hand, community detection (Section \ref{subsec:comm-det}) aims to uncover the mesoscale organisation of nodes into groups that follow certain assumptions (in the case of signed networks, large in-group cohesion and out-group antagonism) or that arise from underlying generative mechanisms.

\subsection{Null models}\label{subsec:null-models}

In network analysis, as in many areas of science, a model is only meaningful if we can assess whether the patterns it captures are statistically significant rather than the consequence of noise or inherent constraints. To do so, we often compare observed data to a null model, a randomised version of the network that preserves some of its basic features but lacks the specific structure for which we are testing. In this sense, null models act as baselines and help us answer the question: \textit{Is this pattern special, or is it something we’d expect to see anyway?}

In signed networks, the most common use of null models is to detect the presence of strong or weak structural balance in empirical networks. 
Real-world networks often deviate from a state of perfect balance; therefore, a statistical analysis is required to determine the significance of specific patterns compared to a null model. Part of the debate over whether social systems exhibit strong or weak structural balance comes from conflicting results caused by differences in the null models employed~\citep{maozWhatEnemyMy2007,kirkleyBalanceSignedNetworks2019}.
In what follows, we present null models that have mostly been developed for assessing balance. However, null models also play an important role in other contexts, for instance, to predict the creation of enmities, detect communities, or assess the importance of homophily and heterophily in signed networks, and can be used in such cases, beyond structural balance. For additional details on how to choose appropriate null models, see \citep{hobsonGuideChoosingImplementing2021}.

\begin{figure}[ht]
    \centering
    \includegraphics[width=0.75\linewidth]{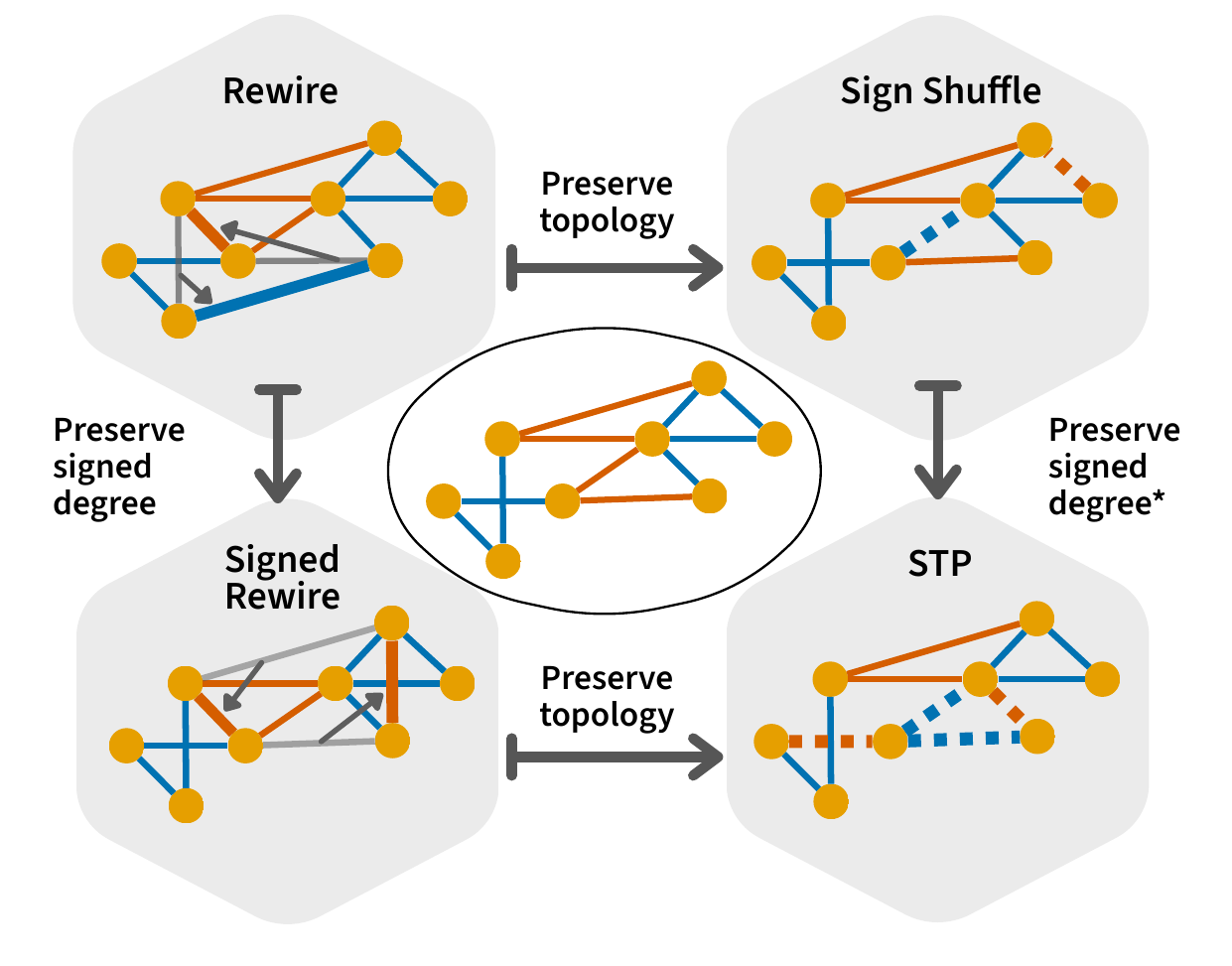}
    \caption{\textbf{Null models for signed networks.} The figure shows the different null models that can be used in signed networks, depending on the constraints that are preserved. The top row shows simple re-wiring or sign shuffle null models, while the bottom row depicts two models that preserve the signed degree. The asterisk (*) indicates that the signed degree in the STP model is preserved on average. Models in the right column preserve the underlying topology. All models preserve the ratio of negative edges in the network. This figure is an adaptation of Figure 2 from~\citep{haoProperNetworkRandomization2023}.}
    \label{fig:null-models}
\end{figure}

Following the classification presented in \citep{haoProperNetworkRandomization2023}, null models can be organised by the constraints they preserve, such as topology, node degree, or signed degree. Figure~\ref{fig:null-models} illustrates examples of the different null models presented below. In addition to the classification by Hao et al.~\citep{haoProperNetworkRandomization2023}, we also consider null models based on conditional probability methods, stochastic block models, and maximum-entropy formulations.

\paragraph{Re-wire model}
This model disrupts topology while maintaining the underlying degree (i.e., the number of neighbours of each node, regardless of sign). One way to implement this null model is to randomly select two edges and interchange their source or target nodes with equal probability. To avoid multi-edges, this change is only accepted if the newly defined edge does not exist in the network. While this model is widely used in unsigned networks because it preserves node degree, it is less suitable for signed networks, because it disrupts the heterogeneity of nodes’ signed degrees and may conceal patterns related to the distribution of positive and negative links.

\paragraph{Sign shuffle model~\citep{szellMultirelationalOrganizationLargescale2010, leskovecSignedNetworksSocial2010}}
Here, the topology remains fixed while the signs of the edges are randomly redistributed. An analogous approach, shown to provide similar results in terms of balance~\citep{facchettiComputingGlobalStructural2011}, is to draw the edge signs as independent and identically distributed variables from a Bernoulli distribution.
While this model preserves the total ratio of positive versus negative edges, it assumes a homogeneity in the individual ratios of each node (i.e., it assumes all nodes have the same proportion of positive edges). This assumption is dangerous because in most real-life networks there is a clear heterogeneity between individuals, some of them being more ``friendly'' than others~\citep{haoProperNetworkRandomization2023} (see Section~\ref{subsubsec:mechanisms-governing}). Moreover, in~\citep{facchettiComputingGlobalStructural2011} it is discussed how signed degree homogeneity generates more unbalanced networks and thus results in conclusions in favour of balance theory when comparing real data with such models. The explanation for this is that nodes with a lot of enmities are easier to ``cast out'' than uniformly distributed conflicts in the system. Another issue with this null model, brought forward by \citep{fengTestingBalanceSocial2022}, is that it assumes that positive and negative edges are interchangeable, therefore testing a stronger statement than just balance. To solve this issue, the authors propose a modified permutation-based null model that stratifies the uniform permutation to edges with the same level of embeddedness.

\paragraph{Signed re-wire model~\citep{liConstructingRefinedNull2021}}
This model follows the same logic as its unsigned counterpart, explained above. However, to maintain the signed degree, it rewires the positive and negative edges separately. An equivalent implementation is to ``cut'' all edges in half and reattach them randomly, as long as the two halves that are reattached have the same sign\footnote{In the case of a directed network, the reattachment must also follow the direction of the edge.}. This approach was used, for example, in the evaluation of balance in ecological networks \citep{saizEvidenceStructuralBalance2017}. An alternative model, proposed in \citep{derrSignedNetworkModeling2018}, is the signed Chung-Lu model, which preserves the degrees of the nodes on average rather than exactly. The authors employ this model to generate synthetic signed networks that reproduce both the global sign distribution and the proportion of balanced triangles observed in real data.

\paragraph{Signed Degree and Topology Preserving (STP) Model~\citep{haoProperNetworkRandomization2023}}

Hao et al.~\citep{haoProperNetworkRandomization2023} introduced a model that preserves both the network topology and the signed node degree. It is based on the maximum entropy framework \citep{squartiniAnalyticalMaximumlikelihoodMethod2011} and provides the exact probabilities of occurrence of random graphs with both constraints. Note that, while it does preserve the signed node degree on average, it does not preserve the exact, individual signed degrees for each node. 

\paragraph{Conditional-probability approach~\citep{lernerStructuralBalanceSigned2016}}
Lerner et al.~\citep{lernerStructuralBalanceSigned2016} proposed a different approach in terms of conditional probabilities, specifically the probability of a link having a particular sign given the presence of the link itself, instead of marginal probabilities. This approach, while it has a different implementation, has a similar conceptualisation to the topology-preserving models. The decomposition of the marginal probability of a negative link, then, would be given by:

\begin{equation}
    P(A_{ij} = -1) = P\left(|A_{ij}| = 1\right) \cdot P\left(A_{ij} = -1 \mid |Y_{ij}| = 1\right)
\end{equation}
\label{eq:conditional-probability}

where $A_{ij}$ and $|A_{ij}|$ are the signed and unsigned adjacency matrices, respectively. This decomposition allows for distinguishing between the likelihood of forming a link and its tendency to be positive or negative. The authors argue that this framework provides clearer interpretations of structural balance by avoiding the confounding effects of link formation mechanisms such as common neighbours.

\paragraph{Stochastic block models~\citep{jiangStochasticBlockModel2015,Yang_Zhao_Liu_2015,galloAssessingFrustrationRealworld2024}}

A common null model for community detection is the stochastic block model (SBM). It assumes that nodes can be grouped into latent blocks and that the probability of observing a positive or negative edge depends only on the blocks of the endpoints. The model parameters  $p^+_{rs}$ and $p^-_{rs}$ form a block matrix that specifies the probabilities of positive and negative links between nodes belonging to blocks $r$ and $s$, respectively. Jiang et al.~\citep{jiangStochasticBlockModel2015} proposed a generalised signed SBM where they differentiate positive and negative block matrices. Yang et al.~\citep{Yang_Zhao_Liu_2015, yangCommunityMiningSigned2007} then developed a Bayesian signed SBM which explicitly accounted for edge density and frustration in the generative model. Finally, Gallo et al.~\citep{galloAssessingFrustrationRealworld2024} derived the signed SBM from an exponential random graph framework and connected it to the theory of relaxed balance (see Section \ref{subsec:math-propr-balanced}).

\paragraph{Maximum-entropy approaches~\citep{galloTestingStructuralBalance2024,becattiEntropybasedRandomizationRating2019,hao2025social,galloPatternsLinkReciprocity2024}}

Gallo et al.~\citep{galloTestingStructuralBalance2024} have extended exponential random graph models to signed networks with both global and local constraints. Their global Signed Configuration Model with Fixed Topology (SCM-FT) imposes the topology as a hard constraint, while the positive and negative degrees of each node are matched only on average, as in the STP model explained above. The authors also define three additional variants, aligned with the types of constraints explained above: a homogeneous model with varying topology (Signed Random Graph Model or SRGM), with the same constraints as the Re-wire model, a homogeneous model with fixed topology (SRGM -FT), similar to the Sign shuffle model, and a heterogeneous model with varying topology (SCM), similar to the Signed re-wire model. A bipartite version of the SCM can also be recovered from the Bipartite Score Configuration Model in \citep{becattiEntropybasedRandomizationRating2019},
and extensions for directed signed networks can be found in \citep{hao2025social} and \citep{galloPatternsLinkReciprocity2024}.

\begin{table}[htbp]
\centering
\caption{Comparison of models based on preservation properties. All models preserve the total number of positive and negative edges.}
\label{tab:model_comparison}
\begin{tabularx}{\linewidth}{l X X X}
\toprule
\textbf{Model} & \textbf{Preserves Topology} & \textbf{Preserves Degree} & \textbf{Preserves Signed Degree} \\
\midrule
Re-wire model & No & Yes & No \\
Sign shuffle model & Yes & Yes & No \\
Signed re-wire model & No & Yes & Yes \\
Signed degree and topology preserving model & Yes & Yes & Yes (on average) \\
\bottomrule
\end{tabularx}
\end{table}

The choice of a null model strongly shapes the interpretation of structural balance and other patterns in signed networks. In
\citep{haoProperNetworkRandomization2023}, the authors argue that matching the signed degrees is a relevant and often forgotten feature, while \citep{lernerStructuralBalanceSigned2016} shows that results may support or contradict a hypothesis of balance depending on whether conditional or marginal probabilities of edge signs are considered. Similarly, \citep{galloTestingStructuralBalance2024} finds that conclusions about strong or weak balance depend on whether heterogeneous sign propensities are included and on whether the network topology is preserved.

Beyond testing balance, null models serve broader purposes such as identifying mesoscale organisation, evaluating link or sign prediction, or constructing bipartite projections \citep{gallo2025statistically}. Their design should always reflect the constraints of the system under study. Re-wire models are useful in scenarios where individuals are free to decide \textit{who} and \textit{how} to interact with other individuals (e.g., an unconstrained social system like social media). If users' friendliness and hostility are homogeneous, it is safe to use models that do not preserve signed degrees. However, if there are different tendencies in the way users interact with each other, the signed re-wire model would be more appropriate. Conversely, if the individuals cannot choose \textit{who} they interact with but they choose the \textit{how} (e.g., a high school), we would prefer topology-preserving models like a sign shuffle model, or models such as STP or SCM-FT. Ideally, one can go a few steps further and design a tailored null model when information on the allowed connections is available, since generally the topology of a network is not completely arbitrary nor strongly determined. In that case, these constraints could be incorporated in maximum-entropy-based models similar to the STP or SCM-FT \citep{galloTestingStructuralBalance2024}.

\subsection{Node embeddings}\label{subsec:node-embeddings}

Node embedding methods are a powerful branch of the \textit{representation learning} field, used to learn compact, low-dimensional representations of nodes, such that proximity in the embedding space reflects the underlying network structure. In this section, we review the methods specifically developed for signed networks, with a summary provided in Table~\ref{tab:node-embeddings}. For a more complete overview of the field, we refer to a recent survey~\citep{zhang2024signed}.

The core principle underlying many of these methods is synthesised in the title of one of the first papers in this domain: \textit{When Everybody Can Sit Closer to Friends than Enemies}~\citep{kermarrecSignedGraphEmbedding2014}. The goal of these methods is to embed the nodes in a space so that the distance between positively connected nodes (friends) is smaller than between negatively connected nodes (enemies). However, in general graphs, this ideal positioning cannot be achieved, as Kermarrec and Thraves~\citep{kermarrecSignedGraphEmbedding2014} proved even for simple graphs. Consequently, most algorithms relax this objective and instead aim to minimise the number of violated constraints by optimising equivalent spectral, probabilistic, or geometric criteria. Early work in this domain focused on Euclidean and spectral approaches, while more recent contributions have leveraged advances in natural language processing, probabilistic modelling, hyperbolic geometry, and graph neural networks (GNNs). 

One of the first contributions to node embeddings for signed networks was proposed by Kunegis et al.~\citep{kunegisSpectralAnalysisSigned2010}, who introduced a spectral approach to represent nodes in a $k$-dimensional Euclidean space.  In a nutshell, the eigenvectors corresponding to the $k$ smallest eigenvalues of the opposing Laplacian are used to embed nodes in $\mathbb{R}^k$, so that Euclidean proximity reflects many positive links and few negative ones. This spectral embedding is particularly useful for community detection in signed graphs, as we discuss in Section~\ref{subsec:comm-det}. A related Euclidean embedding proposed in~\citep{diazdiaz2025} leverages the fact that the communicability distance is Euclidean to embed the nodes within a hypersphere.

Another spectral embedding technique, inspired by physical arguments, is the SHEEP algorithm by
Babul and Lambiotte~\citep{babulSHEEPSignedHamiltonian2024}. The method is based on a
Hamiltonian where positive edges act as springs pulling nodes together and negative
edges act as repulsive forces. 

\begin{equation}\label{eq:SHEEP}
\pi^\top L_r \pi = \sum_{i,j} A^+_{ij} \lVert \pi_i - \pi_j \rVert^2 - \sum_{i,j} A^-_{ij} \lVert \pi_i - \pi_j \rVert^2 ,
\end{equation}

Minimising this Hamiltonian under a spherical constraint
reduces to an eigenvalue problem involving the repelling Laplacian (Eq. \ref{eq:repelling_laplacian}). 
The method has been successfully applied to empirical data, including detecting opinion polarisation in online social networks~\citep{candellone2025negative}.

A different family of methods takes inspiration from word embedding techniques like  Mikolov's \textit{word2vec} model~\citep{mikolovEfficientEstimationWord2013} and adapts them to networks. In word2vec, words are mapped to vectors by scanning sentences with sliding windows and adjusting the vectors so that words appearing in similar contexts lie close together. Grover and Leskovec adapted this idea to graphs with node2vec~\citep{groverNode2vecScalableFeature2016}, replacing sentences by random walks on the network and words by nodes. Nodes that frequently co-occur along walks thus become close in the embedding space. The first methods to extend this principle to signed graphs were \textit{SNE}~\citep{yuanSNESignedNetwork2017} and \textit{Sign2vec} ~\citep{islamDistributedRepresentationsSigned2018}. In these works, walks are adapted to incorporate edge signs, so that the learned vectors reflect both proximity and polarity. More recently, the stance embeddings model (\textit{SEM})~\citep{pougue-biyongLearningStanceEmbeddings2022} added edge attributes, enabling embeddings that account for topical differences. Alternatively, one can map the signed network into an unsigned one without losing the information carried by negative edges. This approach, used in pipelines like ROSE~\citep{javariROSERolebasedSigned2020}, allows the direct use of methods such as node2vec while retaining the information contained in the negative edges in the embedding. Other approaches to signed network embedding are based on correlations between signed random walks~\citep{huangPOLEPolarizedEmbedding2022}, Bayesian approaches~\citep{maraCSNEConditionalSigned2020}, hyperbolic geometry~\citep{songHyperbolicNodeEmbedding2021,du2025beyond}, or latent distance models that structure embeddings within polytopes~\citep{nakisCharacterizingPolarizationSocial2023,nakisHybridMembershipLatent2023}.

The introduction of Graph Neural Networks (GNNs) opened a new phase for representation learning on signed graphs and extended node-embedding ideas into deep models. One of the earliest attempts in this direction is \textit{SiNE}~\citep{wangSignedNetworkEmbedding2017}, a feedforward neural network that learns embeddings by enforcing friends to be closer than foes. This line of work has evolved toward graph convolutional neural networks~\citep{derrSignedGraphConvolutional2018,kimTrustworthinessDrivenGraphConvolutional2023}, which extend the embedding principle to layer-wise message passing on graphs, and attention mechanisms~\citep{huangSignedGraphAttention2019,yanMUSEMultifacetedAttention2021}, which adaptively weight neighbours when aggregating information. Beyond these, other neural architectures have also been explored: autoencoders~\citep{shenDeepNetworkEmbedding2020}, generative adversarial networks~\citep{chakrabortySigGANAdversarialModel2022}, and multilayer perceptron-based GNNs~\citep{heSSSNETSemiSupervisedSigned2022}. A parallel and now prominent line adapts self-supervised and contrastive learning to signed graphs, generating positive and negative views through balance-aware augmentations \cite{shuSGCLContrastiveRepresentation2021,koUniversalGraphContrastive2023,zhouAdversarialRobustnessLink2026}. A related direction builds spectral GNNs that define learnable filters directly on signed Laplacian operators \cite{liSignedLaplacianGraph2023,wangGegenNetSpectralConvolutional2025}. Other works have also targeted more complex settings, including directed signed networks~\citep{huangSDGNNLearningNode2023}, bipartite structures~\citep{huangSignedBipartiteGraph2021}, and temporal networks~\citep{sharmaRepresentationLearningContinuousTime2023}. For practical implementation, the PyTorch Geometric Signed Directed library \cite{hePyTorchGeometricSigned2023} provides a unified software framework covering many of the GNN-based methods discussed above.

In summary, node embeddings provide a distilled representation of the complex structure of signed networks, where positive relations are typically interpreted as attractive forces and negative relations as repulsive ones. This allows the embedding to capture the essential patterns of cooperation and antagonism in a low-dimensional space.

\subsection{Sign prediction}\label{subsubsec:ML-link-pred}
Link prediction~\citep{kumarLinkPredictionTechniques2020} refers to the task of inferring missing edges in a network from its observed structure. In unsigned networks, this problem has received extensive attention due to its wide range of applications, like forecasting future interactions between agents and reconstructing incomplete datasets \citep{cimini2021reconstructing}. When extending link prediction to signed networks, the problem splits into two related but distinct tasks: determining whether an edge exists between two nodes, and, assuming it exists, predicting its sign. In this book, we focus exclusively on the latter, as link prediction is often not signed-network-specific. Existing approaches to sign prediction can be grouped into four main categories. The first relies on structural balance, using measures of imbalance to infer the most aligned sign assignment. The second treats sign prediction as a classification problem, where edges are represented through structural features and their signs are predicted with statistical learning models. A third family uses matrix factorisation or completion, exploiting low-rank structure in the adjacency matrix to recover missing signs. Finally, deep learning methods learn edge representations directly from the graph, often through neural architectures that capture non-linear patterns. In the following, we describe a few methods from each category, and a comprehensive list can be found in Table~\ref{tab:sign-prediction}.

Methods based on balance theory attribute a sign to an edge $(i,j)$ according to its consistency with the structural balance in the rest of the network~\citep{guhaPropagationTrustDistrust2004}. In this framework, the product of edge signs along a path encodes indirect influence: an even number of negative edges favours a positive relation, while an odd number favours a negative one. To aggregate these influences, one considers all $i \to j$ walks, typically under-weighting longer walks. A common formalisation uses powers of the signed adjacency matrix with a penalisation factor $c_k$, akin to the walk-based balance indices discussed in Section~\ref{sec:balance_indices}:
\[
\Gamma \;=\; \sum_{k = 1}^\infty c_k A^k.
\]
Here, $\Gamma_{ij}$ measures the (weighted) difference between the number of positive and negative walks between $i$ and $j$; therefore, $\operatorname{sgn}(\Gamma_{ij})$ can be used as the edge sign prediction, with $|\Gamma_{ij}|$ providing a confidence score. Several penalisation schemes have been explored, such as exponential decay $c_k=\beta^k$~\citep{guhaPropagationTrustDistrust2004, chiangExploitingLongerCycles2011, singhMeasuringBalanceSigned2017, kirkleyBalanceSignedNetworks2019} and factorial decay $c_k=\tfrac{1}{k!}$~\citep{kunegisSlashdotZooMining2009,diazdiaz2025}. Beyond adjacency-based formulations, alternatives like the effective resistance and diffusion kernel have also been proposed~\citep{kunegisSlashdotZooMining2009, kunegisSpectralAnalysisSigned2010}. Another line of work treats positive and negative edges separately; for example, Guha et al.~\citep{guhaPropagationTrustDistrust2004} reported that the best results arise when the negative influence propagates at most once, leading to expressions such as $\Gamma=(A^+)^kA^-$. Despite these variations, the underlying logic is always that the local patterns of balanced and unbalanced relations act as ``votes'' on the most consistent sign between two nodes.

The second category of sign prediction methods frames it as a classification problem, where each edge is described by a vector of structural features that are then passed to a supervised learning algorithm. The pioneering work of Leskovec et al.~\citep{leskovecPredictingPositiveNegative2010} introduced this methodology using logistic regression. They proposed two main families of edge features. The first are \textit{degree-based} features, which include the positive and negative in- and out-degrees of each endpoint, together with the edge embeddedness (defined as the number of common neighbours shared by the two endpoints). The second are \textit{triad-based} features, obtained by counting how many triads of each type include the edge under consideration. Logistic regression then uses these features to predict the edge sign. This approach yields interpretable models, since the learned features' weights can be directly related to balance and status theories. Its predictive power, however, depends strongly on embeddedness: triad features become informative only when the endpoints share many neighbours, so performance deteriorates for edges with low embeddedness. Moreover, because it relies exclusively on degree- and triad-based features, the method cannot capture balance patterns that emerge through longer cycles. 

To address these limitations, Chiang et al.~\citep{chiangExploitingLongerCycles2011} extended the feature construction so that the representation of an edge $(i,j)$ includes counts of walks forming cycles of length up to $k$. For example, to capture the influence of 4-cycles, they considered terms such as $(A^+)^4_{ij}$, $(A^+)^3_{ij}(A^-)_{ij}$, $(A^+)^2_{ij}(A^-)^2_{ij}$, $(A^+)_{ij}(A^-)^3_{ij}$, and $(A^-)^4_{ij}$. Incorporating these higher-order cycle features into logistic regression significantly improved predictive performance, particularly for edges with low embeddedness where triadic features provide no information. Building on the idea of going beyond triads, Papaoikonomou et al.~\citep{papaoikonomouPredictingEdgeSigns2014} proposed to represent edges through frequent subgraph patterns (i.e., motifs) in their local ego-graphs, including them as a feature in the support vector machine model used for classification. Their results showed that these motifs carry strong discriminative power. More recently, Derr et al.~\citep{derrBalanceSignedBipartite2019} addressed the case of bipartite signed networks, which by definition cannot have triangles or any loop of odd length. They replaced triad-based features with bipartite motifs such as ``caterpillars'' and ``butterflies''~\footnote{Signed butterflies are cycles of length 4, and caterpillars are paths of length 3 that can become a butterfly by adding one edge.}, and used them in a classifier to predict edge signs.

The third category for sign prediction in signed networks treats the adjacency matrix $A$ as a partially observed matrix and approximates it by a low-rank factorisation $A \approx U V^\top$, where $U \in \mathbb{R}^{n \times d}$ and $V \in \mathbb{R}^{n \times d}$ ($d\ll n$) \citep{forsati2015pushtrust, tang2016recommendations, derrBalanceSignedBipartite2019}. The factorisation is attained by minimising a loss $\ell$ over the observed entries, for example~\citep{forsati2015pushtrust}
\[
\min_{U,V} \sum_{(i,j)\in \Omega} \ell\big(A_{ij}, u_i^\top v_j\big) + \lambda(\|U\|_F^2+\|V\|_F^2),
\]
where $\Omega$ is the set of known signs and $\ell$ encourages $u_i^\top v_j$ to be positive for friendly edges and negative for antagonistic ones. $\|U\|_F^2+\|V\|_F^2$ is a regularisation term that prevents overfitting by penalising excessively large embeddings, and whose importance is determined by the parameter $\lambda$. After optimisation, the sign of an unobserved edge is predicted by
\(
\hat{A}_{ij} = \operatorname{sign}(u_i^\top v_j),
\)
where $u_i$ (the $i$-th row of $U$) and $v_j$ (the $j$-th row of $V$) represent the latent vectors of nodes $i$ and $j$. Alternatively, instead of assigning a fixed $\pm1$ value to each interaction, probabilistic variants weight edges through a logistic or hinge-based function.
A related method, introduced by Hsieh et al.~\citep{hsiehLowRankModeling2012}, builds on the observation that the adjacency matrix of weakly balanced complete networks has rank equal to the number of factions. This motivates formulating sign prediction as a matrix completion problem: given a partially observed $A$, the task is to recover the missing entries under a low-rank constraint. In practice, because the rank function is non-convex, it is typically replaced by the convex trace norm. The resulting matrix is used to predict the missing signs. Similarly, Ye et al.~\citep{ye_predicting_2013} introduced a matrix-based approach with transfer learning. In their framework, two networks (the source and the target) are embedded into a shared latent space. Structural patterns are first learned from the well-labelled source and then used to guide sign prediction in a sparsely labelled target.

Finally, the last category of techniques for sign prediction is based on deep learning approaches. Ruiz-Garcia et al~\citep{ruiz-garciaTriadicInfluenceProxy2023} used neural networks to predict the sign of social ties among teenagers from a measure of imbalance, the triadic influence. Interestingly, adding node-level features such as gender or prosociality did not improve prediction, meaning that the structure of connections between individuals provides more information about their specific characteristics. Alternatively, GNN-based models have been adapted to message passing to handle both positive and negative edges. For instance, SELO encodes subgraphs around each link to learn edge embeddings~\citep{fangSignedSubgraphEncoding2023}, SIHG places embeddings in hyperbolic space with attention mechanisms that separate friends and foes~\citep{luoInterpretableSignedLink2021}, and Chen et al. introduce graph augmentation strategies to filter unreliable negative edges and improve robustness~\citep{chenBalancingAugmentationEdgeUtility2023}.

\subsection{Community detection}\label{subsec:comm-det}

Community detection aims to detect mesoscopic modular structures in large, complex systems; this is, to find groups of nodes in a network that share common characteristics~\citep{fortunatoCommunityDetectionGraphs2010}. This task has attracted considerable attention in the scientific literature, mainly for two reasons: first, because communities reduce the complexity of a network and let us focus on interpretable modules that reflect the underlying organisation, and second, because of the algorithmic challenges of designing a robust and efficient community detection method. In social networks, these modules might correspond to interest groups, political factions, or geographic clusters. In scientific collaboration networks, they are typically identified with scientific fields and research topics~\citep{girvanCommunityStructureSocial2002}. In brain networks, they often align with functionally specialised regions involved in specific cognitive tasks~\citep{spornsModularBrainNetworks2016}. In ecological and metabolic networks, communities can reveal trophic levels or biochemical pathways~\citep{gauzensTrophicGroupsModules2015}.

Most methods for community detection in unsigned networks build on the idea that nodes within the same group tend to be more densely connected than nodes in different groups~\citep{fortunatoCommunityDetectionGraphs2010}. For instance, when considering face-to-face contacts, a person is more likely to interact with coworkers than with people working in another location. However, this structural view alone may not always capture meaningful community structure, especially when additional information, such as node or edge attributes, is available. These attributes can uncover groupings that topology alone might miss. 

\begin{figure}
    \centering
    \includegraphics[width=\linewidth]{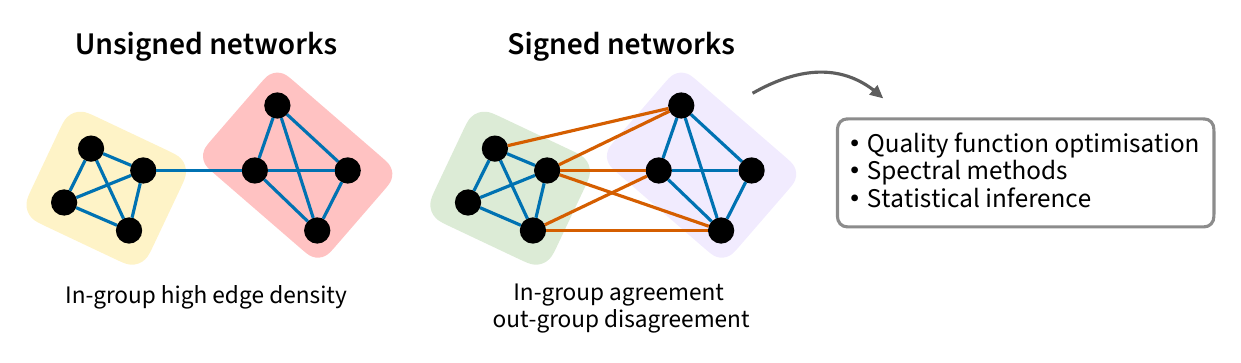}
    \caption{\textbf{Community detection differs in unsigned and signed networks.} In unsigned networks, one typically assumes groupings that maximise internal edge density and minimise connections across groups. In signed networks, the goal is instead to maximise positive edges within each group and negative edges between them.}
    \label{fig:comm-det-unsigned-signed}
\end{figure}

This limitation becomes evident in signed networks, where the information carried in the edge sign is relevant to detecting and interpreting hostile communities. This leads to a new conceptualisation of a signed network community as a group of nodes with internal agreement and external disagreement\footnote{To distinguish between these two conceptualisations of community, the term \textit{faction} is often used to refer to communities in signed networks.}. In this perspective, a good partition minimises the number of negative edges within communities and the number of positive edges between them\footnote{Note that this objective is equivalent to minimising the frustration of a given partition, a topic discussed in Section \ref{sec:balance_indices}}. This definition of community is closely connected to structural balance theory: a network with a clear signed community structure typically exhibits a high level of structural balance. In the limit of perfectly balanced networks, the community structure becomes trivial: the network can be partitioned into mutually antagonistic balanced factions, characterised by only positive links within each faction and only negative links between them. Figure~\ref{fig:comm-det-unsigned-signed} shows the difference between community detection in unsigned networks, where the goal is typically to maximise within-group edge density, and signed networks, where the goal is typically frustration minimisation. It is important to recognise that community detection always relies on assumptions about the type of structure we expect, and different assumptions can reveal different communities with different interpretations~\citep{peelGroundTruth}.

Finding communities by minimising frustration is typically an NP-hard problem, which makes exact solutions computationally infeasible for large networks. As a result, most approaches rely on approximations. Several algorithms originally developed for unsigned networks have been generalised to signed ones, such as Infomap~\citep{esmailianCommunityDetectionSigned2015}, walk-based strategies~\citep{suAlgorithmBasedPositive2017}, and matrix completion methods~\citep{hsiehLowRankModeling2012}. Broadly, most methods proposed in the literature can be grouped into three main families: heuristic algorithms that optimise a \textit{quality function}, \textit{spectral methods} that exploit the spectrum of the Laplacian or adjacency matrix, and approaches based on \textit{statistical inference}. 
\Cref{tab:community-detection-quality,tab:community-detection-spectral,tab:community-detection-stat,tab:community-detection-other} show a comprehensive view of the methods analysed in this book. In the remainder of this section, we outline the most popular strategies and representative methods for community detection in signed networks. For comprehensive comparisons and further methodological details, we refer the reader to \citep{tangSurveySignedNetwork2016,chiangPredictionClusteringSigned2013a,tomassoAdvancesScalingCommunity2022,traagPartitioningSignedNetworks2018,schaubStructuredNetworksCoarsegrained2019,candellone2025community}.

\subsubsection{Quality function optimisation}\label{subsubsec:qualityfunction}

\begin{figure}
    \centering
    \includegraphics[width=\linewidth]{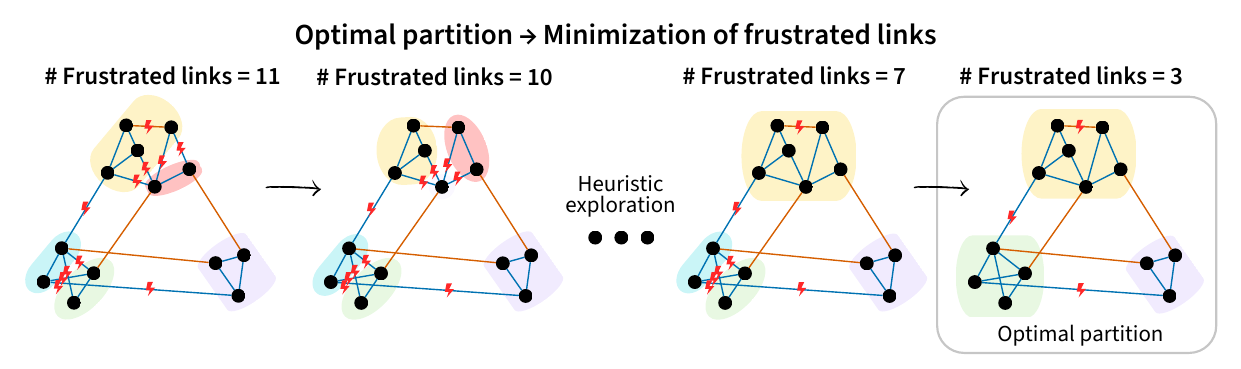}
    \caption{\textbf{Optimal partition via frustration minimisation.} Illustration of the algorithmic procedure for identifying the partition of a signed network that minimises the number of frustrated links (negative links within communities and positive links between communities). The process begins with an initial partition, which contains several frustrated links (highlighted with red lightning symbols). Through heuristic exploration, the algorithm iteratively adjusts the partition to reduce frustration, ultimately converging on an optimal configuration with only three frustrated links. }
    \label{fig:comm-det-frustration}
\end{figure}

Quality-function optimisation methods for community detection find network partitions that optimise a scalar objective, referred to as a quality function. This function quantifies the extent to which a given partition aligns with the expected structure of communities. The underlying assumption is that meaningful partitions correspond to extremal (minimal or maximal) values of this function. These methods usually proceed in two stages: first, the design of a quality function that encodes the desired structural features (such as density of internal links, frustration, or structural balance) and second, the development of an algorithm that explores different partitions to identify the one that optimises the objective.  Table~\ref{tab:community-detection-quality} shows a summary of these methods' references. 

We first explore the most important quality functions proposed to detect communities in signed networks. The first one was introduced by Doreian and Mrvar~\citep{doreianPartitioningApproachStructural1996}. Their formulation combines two criteria: the number of positive links between different communities \( m^+(C_a, C_b) \), i.e., the out-group agreement, and the number of negative links within each community \( m^-(C_a) \), i.e., the in-group disagreement. The objective function to minimise is defined as a weighted average of these two quantities:
\begin{equation}\label{eq:doreian-mrvar-commdet}
    P(C) = \alpha \sum_{C_a} m^-(C_a) + (1 - \alpha) \sum_{C_a \neq C_b} m^+(C_a, C_b),
\end{equation}
where \( 0 \leq \alpha \leq 1 \) controls the relative importance of the two terms. Note that, for $\alpha=0.5$, this quality function reduces to the frustration index of a signed network \citep{hararyMeasurementStructuralBalance1959}. An illustration of the algorithmic procedure to find the partition of the network via frustration minimisation is depicted in Figure \ref{fig:comm-det-frustration}.
The authors later extended this approach to bipartite (two-mode) networks, adapting the quality function to account for the structural properties specific to such systems~\citep{mrvarPartitioningSignedTwoMode2009}, and also to detect structures consistent with relaxed structural balance~\citep{doreianPartitioningSignedSocial2009} (see Section~\ref{subsec:math-propr-balanced}).

A conceptually different approach to quality-function optimisation was later introduced through extensions of modularity for signed networks~\citep{newmanFindingEvaluatingCommunity2004}. The modularity function compares the observed edge density within communities to that expected under a null model in which connections are randomised but node degrees are preserved. Formally, given a partition \( C \) of the node set, the modularity is defined as
\begin{equation*}
    Q(C) = \frac{1}{2m} \sum_{i,j} \left[ A_{ij} - \frac{k_i k_j}{2m} \right] \delta(C_i, C_j),
\end{equation*}
where \( A_{ij} \) is the adjacency matrix, \( k_i \) is the degree of node \( i \), \( m \) is the total number of edges, and \( \delta(C_i, C_j) \) equals 1 if nodes \( i \) and \( j \) are in the same community and 0 otherwise. The first term rewards internal connections, while the second subtracts the expected number of such connections under the null model (in this case, the configuration model). Modularity can assume values in the range $[ -\frac{1}{2}, 1]$. A high value of \( Q \) indicates that the partition captures a dense internal structure.

To extend this idea to signed networks, Gómez, Jensen, and Arenas~\citep{gomezAnalysisCommunityStructure2009} proposed treating positive and negative links separately, each with its own null model. The signed modularity is defined as
\begin{equation*} \label{eq:signed-modularity}
    Q^\pm(C) = \frac{1}{2m^+ + 2m^-} \sum_{i,j} \left[ A_{ij} - \left( \frac{k_i^+ k_j^+}{2m^+} - \frac{k_i^- k_j^-}{2m^-} \right) \right] \delta(C_i, C_j),
\end{equation*}
where \( A_{ij} \in \{-1, 0, +1\} \), \( m^+ \) and \( m^- \) are the total number of positive and negative edges, and \( k_i^+ \), \( k_i^- \) are the positive and negative degrees of node \( i \) and \( j \), respectively. This formulation rewards positive links within communities while penalising negative ones, and adjusts the baseline expectation accordingly. The separation of the null models implicitly assumes that positive and negative interactions have distinct structural roles and must be benchmarked independently.

A more general framework was proposed by Traag and Bruggeman~\citep{traagCommunityDetectionNetworks2009}, rooted in statistical mechanics. Extending the work of Reichardt and Bornhold for unsigned networks~\citep{reichardtStatisticalMechanicsCommunity2006}, they interpret the quality function as the Hamiltonian of a Potts model that separates the contribution of positive and negative edges in the same spirit as Gómez et al.~\citep{gomezAnalysisCommunityStructure2009}.
\begin{equation}  \label{eq:traag_bruggerman}
    H(C) = -\sum_{i,j} \left[ A_{ij} - \left( \gamma^+ p^+_{ij} - \gamma^- p^-_{ij} \right) \right] \delta(C_i, C_j),
\end{equation}
where \( A_{ij} \) is the signed adjacency matrix, \( p^+_{ij} \) and \( p^-_{ij} \) are the expected positive and negative weights between nodes \( i \) and \( j \) under the null model, and \( \gamma^+, \gamma^- \) are resolution parameters controlling the relative contribution of each term. Typically, the chosen null model is the configuration model, so that \( p_{ij}^\pm = \frac{k_i^\pm k_j^\pm}{2m^\pm} \). Under this choice, and setting \( \gamma^+ = \gamma^- = 1 \), the Hamiltonian reduces to the modularity function from~\citep{gomezAnalysisCommunityStructure2009}. Alternatively, choosing \( \gamma^+ = \gamma^- \) and \( \alpha = \frac{1}{2} \), the Hamiltonian becomes the frustration index (Eq.~\ref{eq:doreian-mrvar-commdet}). This shows that the quality function in Eq.~\ref{eq:traag_bruggerman} unifies and generalises both frustration-based and modularity-based formulations into a single quality function. Moreover, the interpretation of the quality function as a Hamiltonian makes it ideally suited for use in conjunction with simulated annealing minimisation algorithms. Finally, the parameters $\gamma^\pm$ can be interpreted as resolution parameters that circumvent the infamous resolution limit of the modularity function \citep{fortunatoResolutionLimitCommunity2007}. The main disadvantage of the method is the necessity to tune those two additional parameters. Some subsequent works have proposed ways to choose these parameters in a principled way (see~\citep{esmailianCommunityDetectionSigned2015}). 

It is important to note that quality function-based algorithms, especially those based on modularity, inherit the same limitations observed in the unsigned case~\citep{peixotoDescriptiveVsInferential2023}, such as the resolution limit and the tendency to overfit the data~\citep{guimeraModularityFluctuationsRandom2004}. Despite these drawbacks, modularity remains a popular quality function for signed graphs (see e.g.~\citep{anchuriCommunitiesBalanceSigned2012} and~\citep{amelioCommunityMiningSigned2013}).

Once a quality function is fixed, optimisation algorithms are used to find the best partition. Since optimising these functions is NP-hard~\citep{brandesModularityClustering2008}, heuristic methods are typically required. Two of the most popular ones are the Louvain algorithm (and its extensions such as Leiden \citep{traag2019louvain}) and simulated annealing. The Louvain algorithm ~\citep{blondelFastUnfoldingCommunities2008} is a greedy, hierarchical method that iteratively optimises modularity by locally reassigning nodes to communities, aggregating communities into super-nodes, and then repeating the process. In Louvain, nodes cannot move into communities of non-neighbouring nodes, a restriction that in signed networks often leads to fragmented partitions, particularly when negative edges are abundant. To solve this issue, Pougué-Biyong et al. have recently proposed SignedLouvain~\citep{pougue-biyongSignedLouvainLouvainSigned2024}, a modification of the Louvain algorithm that allows nodes to explore communities two steps away via negative links. This targeted relaxation lets the algorithm recover balanced mesoscopic structures while remaining fast and consistent.

Simulated annealing, on the other hand, offers a more flexible approach that is particularly effective when optimising global energy-like functions such as frustration~\citep{kirkpatrickOptimizationSimulatedAnnealing1983}. It allows for configurations that increase the energy of the systems during early phases, to escape poor local optima that hinder greedy algorithms. This makes it suitable for frustration-based formulations, which often exhibit a rugged energy landscape. Nonetheless, its computational cost is higher, and it lacks the fast multi-scale aggregation mechanism of Louvain. In practice, Louvain-type methods work when optimising modularity-like scores, while simulated annealing provides a robust, albeit slower, alternative for global criteria or when precise partitions are needed in small to medium-sized graphs. Other approaches to minimise quality functions include the Parallel Iterative Local Search algorithm~\citep{mendoncaRelevanceNegativeLinks2015}, and integer programming approaches~\citep{arefModelingComputationalStudy2020,arefDetectingCoalitionsOptimally2020,arefBalanceFrustrationSigned2019}

\subsubsection{Spectral methods}\label{subsubsec:spectral}

Spectral methods for community detection infer clusters by analysing the eigenvectors of graph-based matrices, usually the Laplacian. The idea is to use the eigenvectors associated with the smallest (nontrivial) eigenvalues of the Laplacian to embed the network’s nodes in a low-dimensional space, and then identify communities as compact groups of points in that space~\citep{vonLuxburg2007a}.

In the case of signed networks, the Laplacian of choice is the opposing Laplacian, $L_o$, because of its convenient mathematical properties. As discussed in Chapter~\ref{subsec:signed-laplacian}, $L_o$ is positive semidefinite and has a zero eigenvalue if and only if the network is balanced. When the latter condition occurs, the corresponding eigenvector identifies the balanced factions.
In the more general unbalanced case, the eigenvector associated with the smallest eigenvalue indicates a partition of the network into two (unbalanced) factions. 

Building on this principle, Kunegis et al.~\citep{kunegisSpectralAnalysisSigned2010,kunegisApplicationsStructuralBalance2014} proposed a natural extension of spectral methods to signed networks, arguing that it preserves the mathematical and physical interpretations that make spectral clustering effective in the unsigned case.  
Formally, they conceptualised communities as partitions that minimise the signed ratio cut,  
\begin{align} \label{eq:signed-ratio-cut}
\text{SignedRatioCut}(C_1, C_2) &= \left( \frac{1}{|C_1|} + \frac{1}{|C_2|} \right) \Big( \text{Cut}^{+}(C_1, C_2) + \tfrac{1}{2} \big( \text{Cut}^{-}(C_1, C_1) + \text{Cut}^{-}(C_2, C_2) \big)\Big),
\end{align}  
where \( \text{Cut}^{\pm}(C_1, C_2) \) counts the number of positive/negative edges between \( C_1 \) and \( C_2 \), and $|C_1|$ is the number of elements in group $C_1$. This objective function penalises positive edges across groups and negative edges within groups, as the quality functions from the last section. A perfectly balanced graph achieves a signed ratio cut of zero. The term \( \left( \tfrac{1}{|C_1|} + \tfrac{1}{|C_2|} \right) \) prevents trivial solutions such as collapsing all nodes into one community or isolating single nodes.  

As in the unsigned case, minimising the signed ratio cut exactly is computationally intractable. Thus, one relaxes the problem by allowing the discrete community assignment vector (which normally takes integer values depending on group membership) to take real values. With this relaxation, minimising Eq.~\ref{eq:signed-ratio-cut} becomes equivalent to minimising the Rayleigh quotient \( \frac{x^\top L_o x}{x^\top x} \), where \( L_o \) is the opposing Laplacian. The solution of this continuous problem is given by the eigenvectors of \( L_o \). To recover a partition of the network, one interprets the entries of the eigenvectors corresponding to the smallest eigenvalues as coordinates: nodes with similar signs or magnitudes in the eigenvectors are grouped. In this way, the eigenvectors provide an approximate but efficient way to split the network into communities, bypassing the need to evaluate every possible partition.  

Building on these results, later work proposed alternative cut functions and Laplacians to address some of the limitations of Kunegis’s formulation. Chiang et al.~\citep{chiangScalableClusteringSigned2012} showed that the signed Laplacian works well for bipartitions but does not extend naturally to 
k-groups clustering. They introduced the {Balance Ratio Cut}, an objective designed to handle multiple communities more effectively, and demonstrated its advantages both analytically and numerically. More recently, Cucuringu et al.~\citep{cucuringuSPONGEGeneralizedEigenproblem2019} developed the {SPONGE} algorithm, which generalises the spectral approach by combining the Laplacians of the positive and negative subgraphs alongside extra tunable parameters. Their results show that SPONGE performs especially well in sparse graphs and when the number of communities is large, outperforming earlier spectral methods.

Other authors have investigated alternative matrix representations beyond the opposing Laplacian. One line of work uses the eigenvectors of the adjacency matrix~\citep{bonchiDiscoveringPolarizedCommunities2019} or of the Hashimoto (non-backtracking) matrix~\citep{zhongEfficientAlgorithmBased2020} to capture community structure directly from connectivity patterns. A related idea is to use the Gremban expansion, which embeds a signed graph into an unsigned one with twice as many nodes, so that standard spectral techniques apply without modification~\citep{fox2017numerical,foxInvestigationSpectralClustering2018}.

\subsubsection{Statistical inference}\label{subsubsec:statinf}
\begin{figure}
    \centering
    \includegraphics[width=\linewidth]{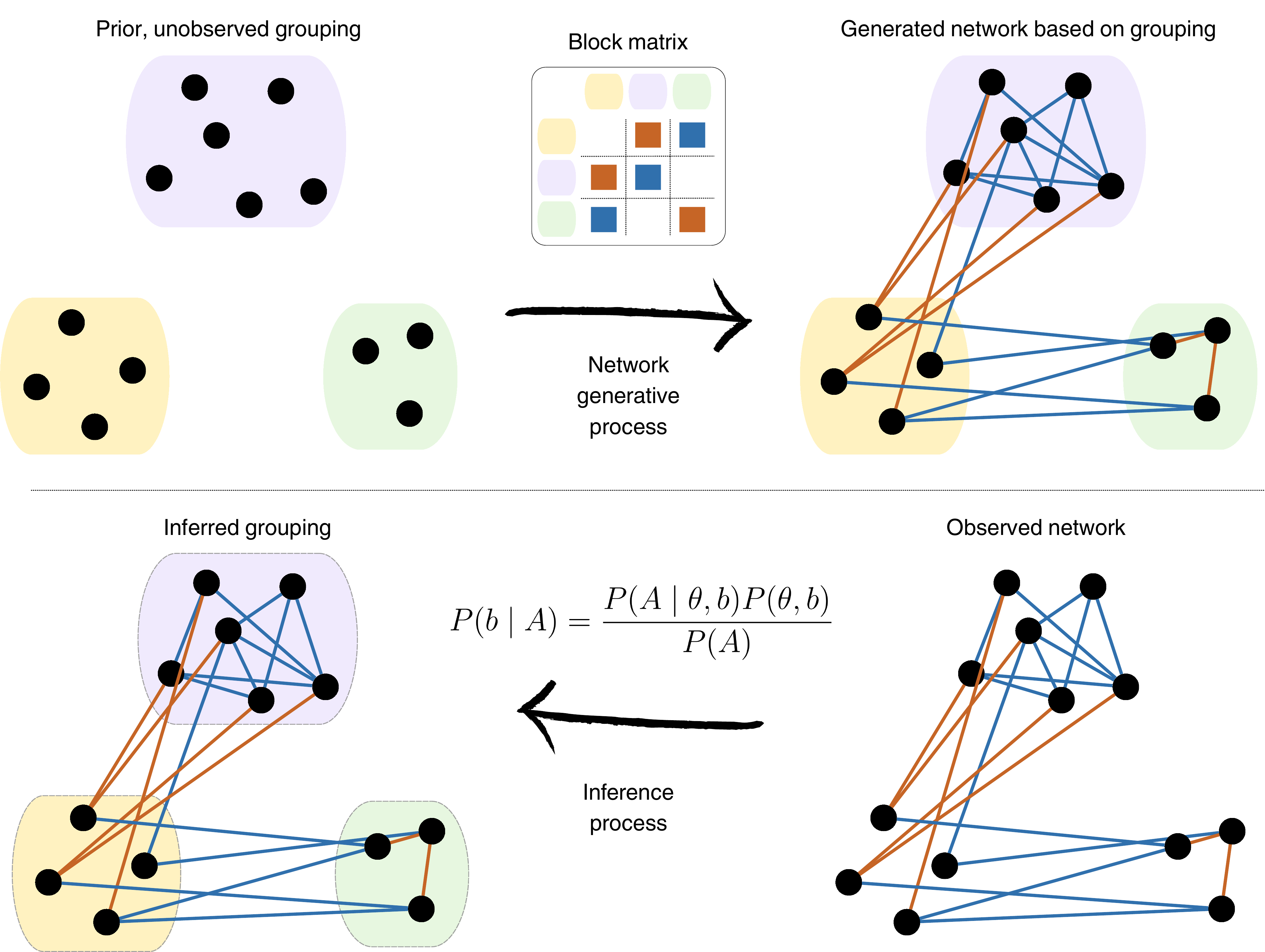}
    \caption{\textbf{Community detection via statistical inference.} Inferential methods assume the existence of an underlying, unobserved grouping of nodes (top-left panel) that influences the generative process of the network. In this view, the observed network structure (top-right panel) is partially shaped by this hidden block structure. In one of the simplest formulations, the generative process is governed by a block matrix, where each entry $rs$ specifies the expected number of edges between groups $r$ and $s$. Edges are then placed at random according to these probabilities, meaning that nodes within the same group have the same connection probabilities to the rest of the network. The goal is to recover this latent grouping from the observed network (bottom-right panel) by identifying the partition that maximises the posterior probability under the model’s assumptions (bottom-left panel).}
    \label{fig:comm-det-inference} 
\end{figure}
Most of the methods presented in previous sections had some common grounds: they relied on principles similar to structural balance and/or frustration, i.e., the aim was to find maximally balanced communities, thus with minimal frustration. A complementary approach to community detection in signed networks uses statistical inference~\citep{casellaStatisticalInference2024}. These methods assume that the network arises from a probabilistic model where edge signs depend on a latent group structure~\citep{newmanMixtureModelsExploratory2007, peixotoBayesianStochasticBlockmodeling2019}, not necessarily dictated by structural balance theory. The aim is to recover the partition that best explains the observed connections, often in terms of maximum likelihood or posterior probability. Unlike quality-function-based methods, statistical models allow uncertainty estimates and can incorporate prior knowledge, especially under Bayesian frameworks~\citep{peixotoDescriptiveVsInferential2023}. They also help prevent overfitting and avoid the resolution limit. An illustration of this inferential approach to community detection is depicted in Figure \ref{fig:comm-det-inference}.

The first attempt to apply statistical inference to signed networks was by Chen et al.~\citep{chenOverlappingCommunityDetection2014}. They detected communities by maximising a likelihood function under a stochastic block model, which assumed that positive links occur only within groups and negative ones only between groups. This formulation is too strict for most real networks, and as a consequence, this model is not able to detect communities in weakly balanced networks or when violations of structural balance are present. Jiang~\citep{jiangStochasticBlockModel2015} improved on Chen's model by relaxing its assumptions. Instead of forbidding positive links between groups and negative links within groups, their model allows both types of edges to occur anywhere in the network, but with different probabilities depending on the group membership. This makes the model more flexible and capable of handling unbalanced networks.

The likelihood function they optimise has the following form:
\begin{equation}
\mathcal{L} = \mathbb{P}(A|\omega^+, \omega^-, \theta, \phi) = \prod_{i<j} \left( \sum_{r=1}^k \sum_{s=1}^k\omega^+_{rs} \theta_{ri} \phi_{sj}  \right)^{A^+_{ij}}  \left( \sum_{r=1}^k \sum_{s=1}^k\omega^-_{rs} \theta_{ri} \phi_{sj}  \right)^{A^-_{ij}},
\label{eq:jiang-likelihood}
\end{equation}

where \(A^+\) and \(A^-\) are the adjacency matrices of the positive and negative subgraphs, respectively. The parameters \(\omega^+_{rs}\) and \(\omega^-_{rs}\) represent the probabilities of observing a positive or negative edge between latent groups \(r\) and \(s\). The terms \(\theta_{ri}\) and \(\phi_{sj}\) specify the probability that node \(i\) belongs to group \(r\) as a sender and that node \(j\) belongs to group \(s\) as a receiver, respectively~\footnote{The distinction between $\theta$ and $\phi$ is needed only when dealing with directed networks. }.

In general, the likelihood quantifies the probability of observing a given network under a particular set of model parameters. The assumption is that edges are conditionally independent given the latent variables; hence, the likelihood of the full network is written as the product of the likelihoods of individual edges.
Inside the product in Eq.~\ref{eq:jiang-likelihood}, the first term models the probability of observing a positive edge between nodes \(i\) and \(j\), while the second term models the probability of observing a negative edge. These probabilities depend on three components: the probability that node \(i\) belongs to group \(r\) as a sender (\(\theta_{ri}\)); the probability that node \(j\) belongs to group \(s\) as a receiver (\(\phi_{sj}\)); and the probability that nodes from groups \(r\) and \(s\) are connected by a positive or negative link (\(\omega^\pm_{rs}\)). 
Note that, unlike the hard constraint in~\citep{chenOverlappingCommunityDetection2014}, here the model allows different connection probabilities for positive and negative links between any pair of groups. This gives the model greater flexibility and makes it possible to account for structural noise or weak balance; for instance, by assigning a small but nonzero probability to positive edges between antagonistic groups or to negative edges within communities. Moreover, the model generalises beyond frustration-based community detection: it can capture overlapping community structure, directed interactions (via asymmetric \(\theta\) and \(\phi\)), and mesoscale patterns such as core–periphery or antibalanced structures.

To infer the parameter values \(\theta\), \(\phi\), and \(\omega^\pm\), the authors employ the maximum likelihood estimation method; i.e., they estimate the parameter values that make the observed network most probable under the model assumptions. The main challenge is that the group memberships of nodes are not observed, so the likelihood must be maximised over both the model parameters and the latent variables. To solve this, Jiang et al. use an expectation–maximisation (EM) algorithm: in the expectation step, they estimate the hidden group memberships of edges given the current parameter values; in the maximisation step, they update the parameters \(\theta\), \(\phi\), and \(\omega^\pm\) to maximize the expected log-likelihood. These two steps are repeated until convergence.

Another approach to handling negative edges within an inferential framework was proposed by Peixoto in~\citep{peixotoNonparametricWeightedStochastic2018}. Peixoto treats edge signs as covariates (i.e., edge attributes) decoupled from the underlying network model. The network structure is modelled with a standard, degree-corrected, SBM~\citep{peixotoNonparametricBayesianInference2017,peixotoBayesianStochasticBlockmodeling2019}, while the signs are accounted for independently. For real-valued signed weights, they are modelled with a Gaussian-like model; for binary signs ($\pm1$), a Bernoulli model; and for correlations in $[-1,1]$, the author suggests to transform the data to $(-\infty, \infty)$ using $y = 2\, \mathrm{arctanh}(x)$. In all cases, this allows for avoiding overfitting and enables hierarchical inference\footnote{This can be done by marginalising the weight distributions, or using instead microcanonical models that fix statistics like mean and variance}. 
Importantly, the model does not incorporate any assumption of structural balance. This is both a strength and a limitation. On one hand, it can reveal signed mesoscale structures beyond what balance theory predicts. Examples include core-periphery configurations and assortative structures where both positive and negative links are concentrated within the same groups. According to \cite{zhang2025meso}, these unbalanced patterns are prevalent in real-world networks, even those that appear balanced at the micro-scale level of triangles. On the other hand, the inferred communities may be difficult to interpret or relate to established theoretical notions, as explored in~\citep{gonzalez2025defines}.

Other extensions of SBM-based inference include maximum-entropy approaches ~\citep{galloAssessingFrustrationRealworld2024, di2025assessing}; models that generalise to edge weights from any exponential family~\citep{aicherLearningLatentBlock2015}; Bayesian approaches like~\citep{yangStochasticBlockmodelingVariational2017} and ~\citep{zhaoNetworkModellingVariational2018}; EM-based variants such as~\citep{zhaoStatisticalInferenceCommunity2017,liSSBMSignedStochastic2023}; and signed block $\beta$-models that account for degree heterogeneity while modelling both strong and weak balance~\citep{wangGegenNetSpectralConvolutional2025}.

In summary, statistical inference methods for signed networks are based on the assumption that the observed links arise from latent group structure within a probabilistic model. This perspective provides several advantages, such as uncertainty estimates, the ability to incorporate prior information, and protection against overfitting. At the same time, these models come with challenges: they can be computationally demanding in large networks, require careful specification of the model's assumptions, and may yield communities whose interpretation is less clear when balance theory does not apply. Despite these limitations, statistical inference offers a flexible and principled framework that complements balance-based and spectral approaches to community detection in signed networks.

\clearpage

\section{Dynamics}\label{chapter:dynamics}

So far, we have focused on the structural aspects of signed networks. We now turn to their dynamics, both \textit{on} and \textit{of} networks. Examining temporal evolution helps us understand how local interaction rules between nodes produce collective outcomes that a static description cannot capture, and how these outcomes, in turn, reshape the underlying structure. Links may strengthen, weaken, or switch sign, partially in response to node states, creating a feedback loop that couples structure and dynamics. 

Negative edges give rise to qualitatively distinct dynamics compared to positive edges. With respect to dynamic processes occurring on top of signed networked structures, negative links introduce antagonistic influences that affect the propagation of information in new ways. Their presence can disrupt alignment processes and drive the system toward states such as dissensus, oscillations, or fragmentation. A central question in this regard is how negative edges affect information spreading: do they block propagation, reverse the transmitted state, or induce inhibition? Regarding how network structures evolve, new characteristics emerge from the evolution of signed networks, beyond the standard ones in unsigned systems (homophily, reciprocity, differential popularity, etc.). Additional drivers, such as balance, status, and conflict avoidance, come into play, often in competition or synergy with the former. Disentangling these intertwined processes remains a key challenge.

This chapter is structured as follows. Section~\ref{subsec:dynamics-on-networks} focuses on dynamics \emph{on} networks, where the topology is fixed, and only node states evolve, with emphasis on opinion dynamics models and contagion models. Section~\ref{subsec:dynamics-of-networks} then examines the dynamics \emph{of} signed networks, in which the topology itself evolves over time and node-state dynamics are absent. We review both theoretical approaches, based on families of toy models guided by heuristic rules, and empirical studies that analyse real datasets (particularly in social contexts) to identify the mechanisms that shape network evolution. The chapter concludes with a discussion of co-evolutionary models that combine both perspectives (Section~\ref{subsec:coevolution}).

\subsection{Dynamics on networks}\label{subsec:dynamics-on-networks}

We first focus on dynamical processes that unfold \emph{on} signed networks. We explore theoretical models rooted in statistical mechanics, where some dynamical process occurs on top of a network topology. The dynamics are guided by specific mechanisms or rules that determine how the process unfolds and is affected by the specified topology. To illustrate how antagonistic interactions shape specific dynamics, we mostly focus on models of opinion dynamics and contagion, which are standard examples of dynamics on networks. Other classes of models are briefly discussed in Section~\ref{subsec:other_dynamics}. 
 
In opinion dynamics, nodes are agents, and their state is their opinion on a given topic. The network edges may encode friendships, acquaintances, or connections in a digital platform. Agents repeatedly adjust their opinion based on interactions with their neighbours. Over the years, the models have incorporated various mechanisms such as cognitive biases, advertising campaigns, or recommendation systems \citep{starniniOpinionDynamicsStatistical2025}. The central question when studying these models is whether all agents eventually agree (consensus), or whether they fragment into clusters with discordant opinions (polarisation). Early models of opinion dynamics were built on assimilative forces, as interactions between two agents brought their opinions closer to one another. It was not long until subsequent refinements incorporated repulsive effects~\citep{liVoterModelSigned2015}, by which interactions push agents away from each other’s views (a phenomenon often referred to as the \emph{backfire effect}~\citep{Nyhan2010}). Signed networks emerged as a natural framework to model this distinction: positive links promote convergence, while negative ones induce divergence or opposition. Within this framework, we distinguish two broad classes of models: \textit{diffusive dynamics} and \textit{voter models}.

Contagion models, in contrast, seek to describe how states or information spread between connected agents through discrete transmission events rather than gradual averaging. In signed networks, the contagion represents the spread of influence, information, or behaviour, and the sign of each link determines the transformation of what is being transmitted: positive links reinforce, while negative links invert or counter it. We consider three families of contagion models: \textit{compartmental}, \textit{threshold}, and \textit{cascade}.

\subsubsection{Diffusive dynamics}

In diffusive dynamics, opinions spread by local averaging, referring to the physical diffusion equation. These dynamics were initially introduced by French \citep{french} and Degroot \citep{degrootReachingConsensus1974}. Agents are endowed with opinions that they reassess at each step by taking the average of their neighbours' opinions. Eventually, a consensus is reached, provided there exists a node that can reach all the others\footnote{If not, a consensus is reached independently on each of the largest subnetworks to verify this property.}.

Altafini was the first to extend this framework to signed networks~\citep{altafiniDynamicsOpinionForming2012,altafiniConsensusProblemsNetworks2013}. In Altafini's model, the evolution of the opinion of an agent $i$, denoted by $x_i$, is described by the following equation:
\begin{align} \label{aux:Altafini}
    \frac{dx_i}{dt} &= \sum_j |A_{ij}|(x_i-\sgn(x_j)).
\end{align}
In this equation, agent \(i\) tends to align with the opinion of its positive neighbours \((A_{ij} = +1\) drives \(x_i \to x_j\)) and adopt the opposite of its negative neighbours' opinion (\(A_{ij} = -1\) drives \(x_i \to -x_j\)). Equation~\eqref{aux:Altafini} can be rewritten in terms of the opposing Laplacian (see Section \ref{subsec:signed-laplacian}) as $\frac{dx}{dt} = -L_ox$. As a consequence, the long-term behaviour of the dynamics depends exclusively on the structural balance of the network: if the graph is balanced, then each balanced faction reaches a consensus state, with opposite opinions between the factions\footnote{A special case of balanced graph is the unsigned graph, where there is only one non-empty balanced faction and all nodes converge to the same opinion.}. In contrast, if the graph is unbalanced, all opinions converge towards 0. This holds for all possible initial opinions. 

Scholars have also studied \emph{repelling} dynamics, where agents move away from the opinion of their foes:
\begin{align} \label{eq:repelling_opdyn}
    \frac{dx_i}{dt} &= \sum_j A_{ij}(x_j-x_i).   
\end{align}
This can be rewritten in terms of the repelling Laplacian (see Section \ref{subsec:signed-laplacian}): $\frac{dx}{dt}= -L_rx$. In this case, when the graph is balanced, the opinions within each faction move increasingly away from those of the opposing faction, eventually diverging. In unbalanced networks, the system may converge to consensus or diverge, depending on the network topology and the proportion of negative edges~\citep{cisneros-velardePolarizationFluctuationsSigned2021,shiDynamicsSignedNetworks2019,shiEvolutionBeliefsSigned2016}. 

These models have been extended in a number of ways, including time-varying topologies~\citep{proskurnikovOpinionDynamicsSocial2016}, non-linearities~\citep{fontanRoleFrustrationCollective2021}, and graphons~\citep{frascaOpinionDynamicsSigned2024}; they have also been applied to the study of the diversity of exposure to media outlets~\citep{quattrociocchiOpinionDynamicsInteracting2014}, as well as to the problem of opinion control~\citep{zhouFriedkinJohnsenModelOpinion2024}. We refer the interested reader to~\citep{proskurnikovTutorialModelingAnalysis2018,shiDynamicsSignedNetworks2019} for more exhaustive reviews of the literature on diffusive models.

\subsubsection{Voter model and random walks}

The voter model is one of the most widely studied approaches to opinion dynamics. In this model, agents randomly update their state by copying that of a randomly chosen neighbour. This model's popularity can be attributed to four key reasons. First, it offers a natural framework for herding behaviour and emergent phenomena: each agent updates its state by adopting a neighbour’s state, so large‐scale phenomena emerge from simple, local imitation. Second, the probability distribution of each agent's opinion is a linear combination of their neighbours' distributions, enabling analytical tractability. Third, the model has no control parameters, meaning that, apart from stochasticity, its behaviour is determined entirely by the network structure and the initial state. Finally, the model can reproduce macroscopical features of election outcomes, and retrieve individual opinions of online social media users (see \citep{vendevilleVoterModelCan2025} and references therein).

\begin{figure}
    \centering
    \includegraphics[width=.95\linewidth]{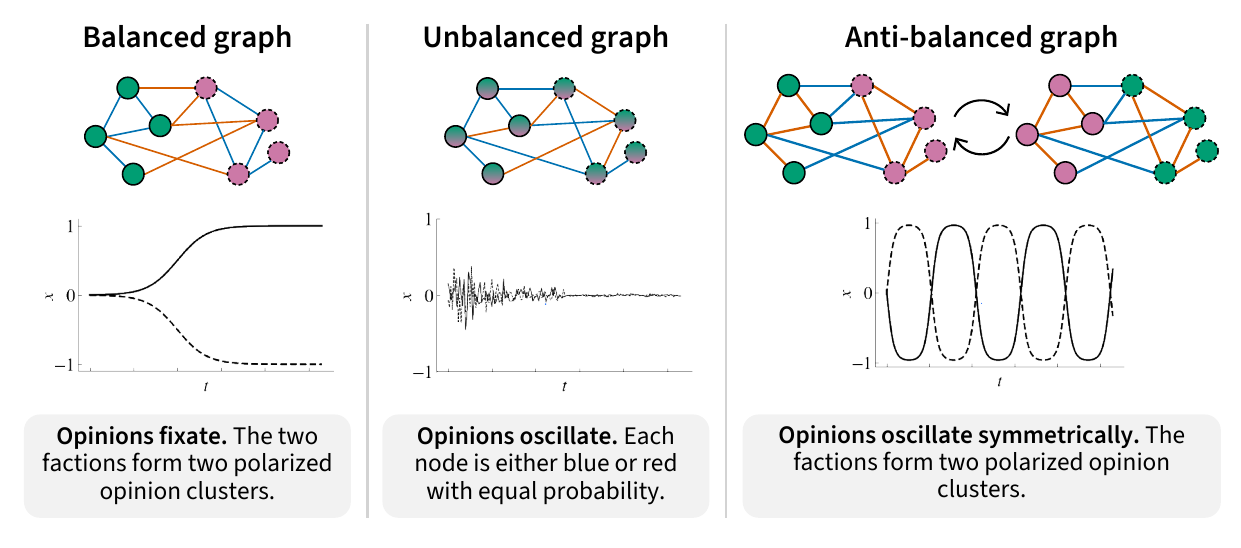}
    \caption{\textbf{Dynamics and asymptotic behaviour of the Voter Model with synchronous updates on signed networks}~\citep{liInfluenceDiffusionDynamics2013,liVoterModelSigned2015}. Node colours designate opinions in $\{-1,+1\}$. There are two factions, distinguished by full and dashed lines. We show the networks with equilibrium opinions as well as dynamics over time.}
    \label{fig:votermodel}
\end{figure}

Adaptations of the model to signed networks initially focused on the ``anti‑voter'' case, where all edges are negative~\citep{matloffErgodicityConditionsDissonant1977,matloffDissonantVotingModel1980,huberStationaryDistributionAntivoter2004}, and then extended to systems containing both positive and negative links~\citep{saadaModeleVotantMilieu1995,gantertVoterModelAntivoter2005}. In the signed voter model, the elementary update step is modified to account for both assimilative (positive) and repulsive (negative) interactions. Assuming two possible opinions $+1$ and $-1$, the dynamics unfold in successive steps as follows. A random node $i$ is selected, and a random neighbour $j$ of $i$ is selected. Then, $i$ adopts $j$'s opinion if the $(i,j)$ edge is positive; otherwise, it adopts the opposite opinion. Let $x(t)$ be the vector whose $i$‑th entry is the probability that agent~$i$ is in state $+1$ at time~$t$. The time evolution of $x(t)$ is governed by the equation:
\begin{align}  \label{eq:voter_model_master_eq}
x(t+1) = 
K^{-1}\!\Bigl(A^+\,x(t) \;+\; A^-\,( \mathbf{1}-x(t))\Bigr).
\end{align}
where $K$ is the underlying degree matrix. The first term in the equation captures the copying mechanism through positive edges, while the second term captures disagreement through negative edges. Because the dynamics are linear in \(x\), Eq.~\ref{eq:voter_model_master_eq} can be solved analytically and is formally equivalent to the evolution equation of a signed random  walk~\citep{liInfluenceDiffusionDynamics2013,liVoterModelSigned2015}\footnote{
Recent work~\citep{babul2025strong} distinguishes strong and weak random walks on signed networks. Weak walks can handle structures with more than two antagonistic factions.}. This equivalence allows the use of existing results on signed random walks~\citep{tianSpreadingStructuralBalance2024} to describe the opinion distribution. In particular, the long‑term behaviour depends exclusively on the network’s structural balance: for a balanced graph, the system converges to a static, perfectly polarised state; for an antibalanced graph~\citep{hararyStructuralDuality1957}, it also achieves perfect polarisation but with opinions flipping at each time step, so that the level of polarisation is preserved; and for a strictly unbalanced graph, the system settles into an equilibrium in which each node holds opinion 0 or 1 with probability $1/2$, irrespective of the initial conditions. Figure~\ref{fig:votermodel} illustrates these different asymptotic behaviours. Building on this, subsequent work has extended the voter model to incorporate more complex interaction patterns, like directed, weighted signed graphs~\citep{liInfluenceDiffusionDynamics2013,liVoterModelSigned2015}. Other extensions include non-linear interactions~\citep{louThresholdQvoterModel2021,krawieckiVoterModelIndependence2024}, mechanisms of free will~\citep{krawieckiVoterModelIndependence2024}, and settings with more than two possible opinions~\citep{linOpinionPropagationSigned2019}. The similar but deterministic majority-rule model has also received some attention~\citep{chujyo2025steering,diianni2023opinion}.

\subsubsection{Compartmental models}

\begin{figure}
    \centering
    \includegraphics[width=\linewidth]{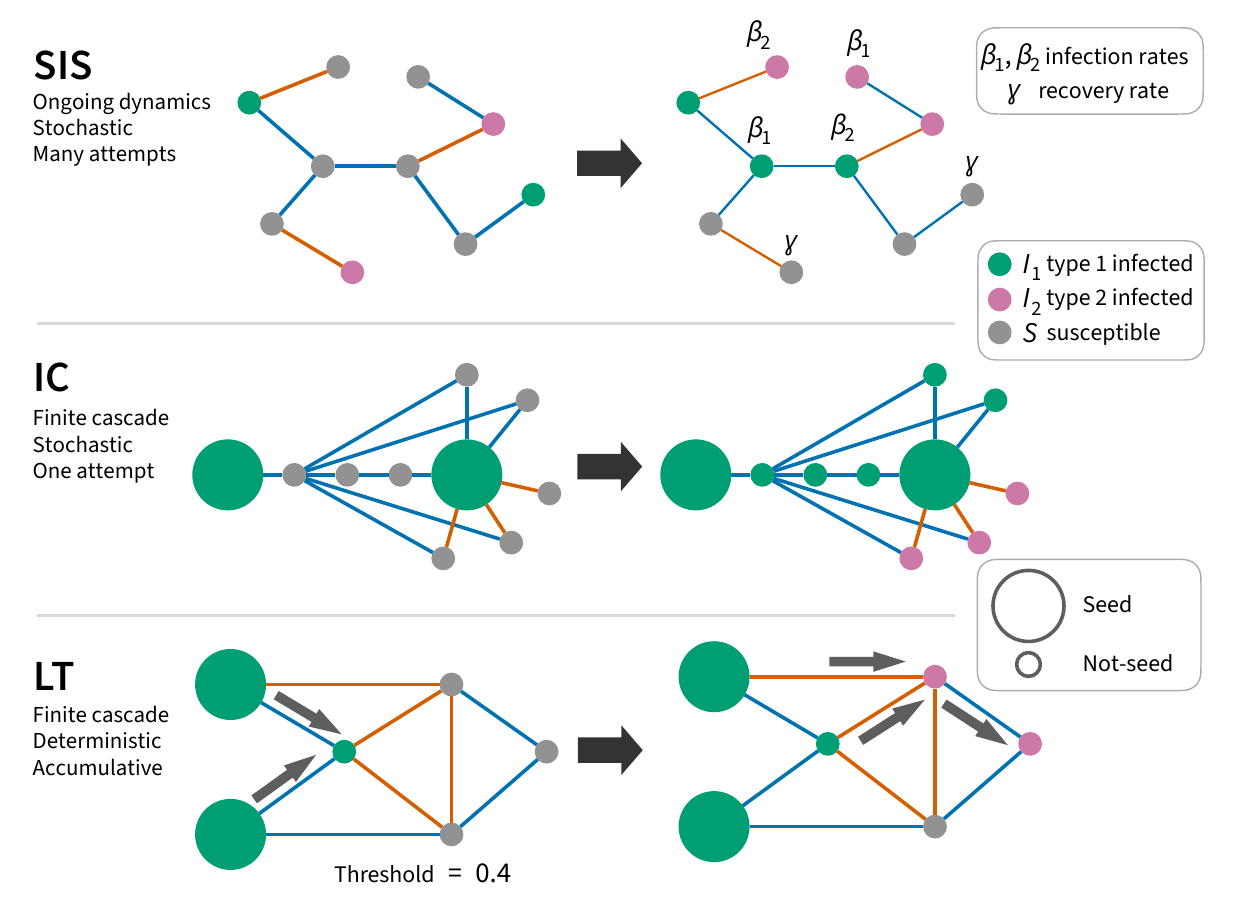}
    \caption{Evolution of the SIS (Susceptible-Infected-Susceptible), IC (Independent Cascade), and LT (Linear Threshold) models. Contagion-like models usually include two opposing infected states (represented here in green and pink) and one susceptible state (in grey). Transmission along a positive edge preserves the infected state of the source node, while transmission along a negative edge flips it to the opposite infected state. Black arrows indicate the time direction, with multiple time-steps between the left and right panels for all models. LT panel adapted from~\citep{srivastavaSocialInfluenceComputation2015}}
    \label{fig:contagion}
\end{figure}

In epidemic spreading models, populations are divided into health state compartments, typically Susceptible (not yet infected), Infected (currently spreading the disease), and Recovered (immune after infection). Compartmental models describe how agents move between discrete states, such as susceptible, infected, or recovered, through probabilistic transition rates that capture continuous-time spreading and recovery. One of the first applications of this framework to signed networks was proposed by~\citep{fanAnalysisOpinionSpreading2012}, who extended the Susceptible-Infected-Recovered model (SIR) model to homogeneous signed networks with a varying fraction of negative edges. Susceptible ($S$) individuals hold the neutral state and can be infected by either type 1 ($I_1$) or type 2 ($I_2$) infectious individuals. When the link $S - I$ is positive, the susceptible node will adopt the same state as the infected one with probability $\beta_1$; when it is negative, the node adopts the opposite state with probability $\beta_2$. Both types of infected state may transition to the recovered state at a rate $\gamma$. The authors find a threshold in the number of negative links beyond which contagion slows down, and the system polarises, while networks dominated by positive links converge rapidly to a consensus.

Another approach sees negative edges not as spreading an opposite state but as changing how easily a process, like adopting a new technique, spreads through the network. For example, in~\citep{liDynamicsEpidemicSpreading2021}, each edge in the signed network carries its own infection rate: a positive edge transmits with a rate proportional to the fraction of balanced triads in which it participates, while a negative edge transmits with a rate proportional to the fraction of unbalanced triads it belongs to.
Thus, contagion depends not only on the sign of a link but also on its local structural balance.

Signed compartmental models have also been extended to multiplex and higher-order settings. In multiplex networks, studies such as~\citep{fengHeterogeneousPropagationProcesses2024,SheOnANetworkedSIS2022} couple an unsigned physical-contact layer, where SIS contagion occurs, with a signed awareness layer, where information about the disease circulates. When all links in the awareness layer are positive, individuals reinforce each other's caution, which raises the epidemic threshold and reduces prevalence. When the awareness layer includes negative links, however, some individuals oppose their neighbours’ warnings, which weakens collective responses and traps the system in polarised states. These divisions reduce perceived risk and allow the epidemic to persist, even when awareness is widespread. In higher-order contagion models, contagion acts through both single links and group interactions. When all links are positive, infections within a group strengthen one another and favour collective activation. The presence of negative links changes this logic: negative links create feedback that can oppose, or under certain conditions, reinforce the spread.~\citep{kemmeterComplexContagionSocial2024} finds that all-negative triads slow contagion at low infection rates but accelerate it once the process becomes strong enough. Fully positive triads make bistability possible and keep the infection active, while configurations with two negative links block propagation.~\citep{maSocialContagionEmotional2024} reaches a similar conclusion: as antagonism becomes more common, collective adoption weakens, the epidemic threshold increases, and the transition to the disease-free state shifts from discontinuous to continuous.

\subsubsection{Cascade and threshold models}

Threshold models define adoption as a collective process in which a node changes state once the influence of its neighbours exceeds a fixed threshold.  In unsigned networks, this influence is additive, so contagion grows monotonically with connectivity. In contrast, in signed networks, negative links reverse or weaken the contribution of active neighbours~\citep{srivastavaSocialInfluenceComputation2015,heInformationDiffusionSigned2019}. The impact of these inhibitory effects depends on the network structure~\citep{muellerRoleNegativeLinks2023}: scale-free networks stay more resilient because of their hubs, while small-world networks are resilient only to a small fraction of negative links. Negative edges can also represent veto-power interactions~\citep{leeThresholdCascadeDynamics2023}, where a single antagonistic neighbour prevents activation.

On the other hand, in cascade models, each newly active node has one chance to activate its neighbours with a certain probability, after which it remains active permanently. In signed networks, activation follows the same cascade principle as in unsigned ones, except that negative links invert the influence of active neighbours, causing some nodes to adopt the opposite state of their influencers. A central question is how to choose the initial set of seed nodes to maximise the spread of activation across the network. Determining either the optimal seed set or the final number of activated nodes is a computationally intractable problem in both unsigned and signed networks. A common approach in unsigned networks is to use a greedy algorithm that iteratively selects the most influential node to approximate the optimal seed set. In signed networks, this strategy fails because influence can both increase and decrease through negative links, so the monotonicity required for the greedy approximation guarantee no longer holds. To address this limitation, Srivastava et al.~\citep{srivastavaSocialInfluenceComputation2015} propose a heuristic for two competing cascades. Their method selects an inactive node and assigns it a state ($+1$ or $-1$) based on which choice yields the largest expected increase in infections in the next time step. This strategy allows the counterintuitive strategy of seeding some nodes with the opposing opinion to ultimately amplify the target opinion. The authors also extended the algorithm to the linear threshold model. 

The Independent Cascade and Linear Threshold models have also been applied to describe information flow between antagonistic communities in social networks. For example, Vendeville et al.~\citep{vendevilleModelingEchoChamber2025} used them to show that echo chambers can arise not only in structurally balanced (polarised) networks but also in antibalanced ones. These results challenge the common assumption that structural polarisation and echo‐chamber formation are equivalent.

\subsubsection{Other dynamics} \label{subsec:other_dynamics}

So far, we have focused on social dynamics as a paradigmatic study case of dynamics on signed networks. However, the effects of negative links on other types of dynamics are relevant across many fields. Here, we briefly direct the reader to some of these studies, with a special focus on synchronisation.

Network synchronisation refers to the process by which the dynamical states of interacting nodes in a network become coordinated over time~\citep{arenasSynchronizationComplexNetworks2008}\footnote{Some examples of spontaneous synchronisation include neuron activity, chemical oscillators, and fireflies synchronising their flashes.}. Each node typically follows its own internal dynamics (e.g., an oscillator, neuron, or agent), but the coupling through the network causes their states to align. In full synchronisation, all nodes evolve identically; in partial synchronisation, only some subsets achieve coordinated behaviour.
One of the most paradigmatic models in synchronisation is the Kuramoto model, in which each node $v_j$ has an internal phase $\phi_j$, which evolves as:

\[
\phi_j \;=\; \omega_j \;+\; K \sum_{k=1}^{N} A_{ij} \sin\!\bigl(\phi_k - \phi_j\bigr),
\]

where $\omega_j$ is the natural frequency and $K$ the coupling strength.

As with other types of dynamics, the presence of negative edges fundamentally alters synchronisation. At the local level, two oscillators connected by a negative link tend to oscillate out of phase. However, if the network is structurally unbalanced, this tendency creates frustration in the collective dynamics. This frustration can lead to slow, glassy relaxation dynamics when the coupling is sufficiently strong. More importantly, the existence of a fully phase-synchronised state (all phases equal) is no longer guaranteed, in contrast with unsigned networks, and one can observe frequency synchronisation without full phase alignment~\citep{restrepoSynchronizationLargeDirectedNetworks2006}. Interestingly, under certain conditions, negative links can actually facilitate synchronisation, similar to what happens in consensus dynamics~\citep{nishikawaNetworkSynchronizationLandscape2010}. A particularly studied situation involves oscillators known as \textit{contrarians}, which interact only through negative couplings with the rest of the network~\citep{hongKuramotoModelCoupledOscillators2011, hongConformistsContrariansKuramoto2011}. More recent literature focuses on synchronisation in signed bipartite networks, establishing that structural balance is necessary for bipartite synchronisation~\citep{zhaiPinningBipartiteSynchronizationAntagonisticSwitching2016, yaghmaieBipartiteCooperativeOutputSynchronizations2017, liuBipartiteSynchronizationLureNetwork2018, liuBipartiteSynchronizationCoupledDelayedNeural2018, xuBipartiteSynchronizationSignedNetworksAperiodicallyIntermittent2021}.

Other dynamics explored in signed networks include spiking integrate and fire with excitatory and inhibitory synapses~\citep{brunelDynamicsSparselyConnectedExcitatoryInhibitorySpikingNeurons2000, litwinKumarSlowDynamicsHighVariability2012}, Hopfield-like dynamics~\citep{sompolinskyChaosRandomNeuralNetworks1988, mezardMeanFieldMessagePassingHopfield2016}, and Boolean regulatory networks~\citep{thomasBooleanFormalizationGeneticControlCircuits1973, remyGraphicRequirementsMultistabilityAttractive2008, richardPositiveNegativeCyclesBoolean2019}.

\subsection{Dynamics of networks: evolving networks}\label{subsec:dynamics-of-networks}

Most systems represented as signed networks are inherently dynamic, evolving. Static analyses (based on single snapshots or an aggregation over time of the network) therefore provide only a coarse approximation of their behaviour, and they carry implicit assumptions about how the system evolves. For instance, when evaluating the level of structural balance in a network, we are often assuming implicitly that some tendency toward balance has shaped the network’s current configuration. However, it is notoriously a difficult task to incorporate explicit temporal evolution into a model. In general, a deep level of domain-specific knowledge is needed beyond the application of the signed networks framework, as the mechanisms governing the evolution of the underlying complex system are rarely universal. Historically, most research within the signed networks framework has focused on social systems and the concept of balance. However, evidence is mixed regarding whether signed networks become more balanced over time, and the dynamics of such systems are likely shaped by additional mechanisms and constraints. In fact, a major challenge is to disentangle the effects of balance from other potentially correlated mechanisms that also shape the evolution of signed networks. In this section, we seek to shed further light on these problems.

Before proceeding, we must clarify the term \textit{mechanism}, which plays a central role in the study of network dynamics. In this context, a mechanism is a rule or process that explains how the structural properties of a network change over time, offering a causal interpretation. Mechanisms translate abstract assumptions or theories into specific terms of network evolution. For example, a \textit{balancing mechanism} translates Heider's hypothesis that people feel tension in unbalanced social configurations into specific rules that stipulate how links evolve in response to imbalance. Any analysis of the dynamics of a network in practical terms always involves operationalising any underlying hypothesis or theory in the form of concrete mechanisms. For this reason, we will use the word mechanism to refer both to concrete rules or processes and to theories in the abstract.

In the remainder of this section, we will explore evolving networks from two complementary perspectives. First, we cover toy models of network dynamics, whose main advantages are the analytical and numerical tractability and their connections with fields like statistical mechanics. Then, we discuss empirical analysis of evolving networks, focusing on the main mechanisms identified by each study.

\subsubsection{Toy models of signed networks dynamics} \label{subsubseq:toymodels}

Toy models, usually developed within the context of statistical mechanics, offer a simplified but effective framework to study how local interaction rules can give rise to global structure. In the context of signed networks, they are most often used to investigate how global balance may emerge from specific microscopic mechanisms. Most models in this setting follow a common strategy: starting from an initial signed network, they define a rule or model that governs how links update to reduce imbalance. The resulting dynamics are then studied, either analytically or through simulations, to understand whether the system converges to a balanced configuration and what factors determine the outcome.

\paragraph{Discrete-time models}

The paradigmatic work in this line of research, i.e., toy models with simple local rules, studied from a theoretical perspective in terms of absorbing states, phase transitions, and long-term behaviour, is the paper by Antal et al.~\citep{antalDynamicsSocialBalance2005}. In this model, the authors propose two variants of a triad-based balancing mechanism: Local Triad Dynamics (LTD) and Constrained Triad Dynamics (CTD). At each time step, an unbalanced triangle may flip the sign of one of its links to reach a balanced state. In both variants, if the triangle is of the $(---)$ type, it updates to a $(+--)$ triangle with probability one. However, if the triangle is of the $(++-)$ type, it updates to the $(+++)$ type with probability $p$ and to $(+--)$ with probability $1-p$. In LTD, the updates are always accepted, while in CTD, the updates are only accepted if they do not create new unbalanced triangles elsewhere in the network. Under LTD and in the limit of infinite sizes, the system exhibits a dynamic phase transition: as $p$ varies, the network shifts from a dynamic steady state to a frozen one where all links become positive. In finite systems, the network always reaches a balanced state, whose nature depends on $p$. A similar behaviour is found with CTD, although the dynamics may get trapped in jammed states: configurations that cannot evolve anymore under the update rules, but are not structurally balanced. While earlier models also explored balance dynamics (e.g.,~\citep{wangSentimentSocialMitosis2003}), Antal et al.'s work remains the most influential example, introducing key ideas for subsequent research like jammed states~\citep{marvelEnergyLandscapeSocial2009,chatterjeeSocialBalanceNetworks2022} and balance-unbalance transitions ~\citep{antalSocialBalanceNetworks2006,arabzadehLifetimeLinksInfluences2021,rabbaniMeanfieldSolutionStructural2019}. Subsequent extensions have incorporated continuous edge weights~\citep{dengStudySignAdjustment2012} and alternative mechanisms, including homophily~\citep{gorskiHomophilyBasedFew2020} and status dynamics~\citep{gorskiCoexistenceBalanceHierarchies2025}. Finally, some models explore a more sophisticated implementation of a balancing mechanism. For instance, the model proposed in~\citep{hummonDynamicsSocialBalance2003} couples Heider's notion of individual-level balance with Harary's group-level balance, and explores the interaction between both. Under this model, the system can become frozen in states where local configurations are balanced, but the global structure remains unbalanced. 

It is important to bear in mind that there is no empirical evidence that LTD, CTD, or other mechanisms proposed in these works represent realistic processes driving networks toward balance. What they do provide, however, are important insights into the kinds of behaviours that can emerge depending on the nature and characteristics of the specific mechanisms employed. 

\paragraph{Continuous-time models}

Parallel to the development of Antal et al.'s discrete model, continuous-time approaches also emerged. These models describe the evolution of link weights through differential equations rather than local discrete rules, encoding the drive toward balance in the equation's structure. A foundational example is~\citep{kulakowskiHEIDERBALANCECONTINUOUS2005} (see also~\citep{gawronskiHeiderBalanceSocial2005}), which proposes 
\(\frac{dA}{dt} = A^2\)
 as the evolution equation for a fully connected signed network, where \( A \) is the adjacency matrix. Simulations show that the system always converges to a balanced state. A more detailed exploration of this model appears in the work of Marvel et al.~\citep{marvelContinuoustimeModelStructural2011}, who show that, for generic initial conditions, the system always converges either to a fully positive network or to a polarised state with two antagonistic factions. Building on this framework,~\citep{summersActiveInfluenceDynamical2013} introduces the idea of targeted influence: the authors show that a single agent can steer the system toward any desired outcome by locally adjusting its friendliness levels, and they provide a method to calculate how much influence is needed to do so.

\paragraph{Game-theoretical models}

Given the nature of relationships in a signed network, in which interactions can be hostile or friendly, it is natural to introduce concepts from game theory, such as cooperation and defection, into dynamic models of signed network evolution. The ways to do this are multiple: one can either analyse whether certain games favour the formation of certain structures, or whether certain structures favour the adaptation of players with different strategies. 
~\citep{traagDynamicalModelsExplaining2013} explored the interplay between balance dynamics and game theory by taking a continuous-time model in which the network evolves driven by different balancing mechanisms. They studied which type of network structures favour a better payoff when the agents play a Prisoner's Dilemma, assuming that a positive link from $i$ to $j$ represents that $i$ considers $j$ to have a good reputation, and therefore would choose to cooperate with $j$, and a negative link would lead the agent to defect. Another example is~\citep{righiEMOTIONALSTRATEGIESCATALYSTS2014}, in which agents play a Prisoner's Dilemma in a signed network, and the network co-evolves with the agents' strategies (see Section \ref{subsec:coevolution} for other co-evolving models). The authors show that when the strategy depends on the emotional content embedded in the signed network, favourable conditions are created for the emergence of unconditional cooperation (see also~\citep{tanEvolutionaryGameApproach2016} and~\citep{hillerFriendsEnemiesModel2017}). Finally,~\citep{heEvolutionCooperationSigned2018} develops a model in which the coupled dynamics between cooperation and balance are studied. The main difference in this model is that the authors explore the possibility that the signs of the relationships influence the payoffs received in the games considered and study the coupled evolution of strategies and relationships.

\paragraph{Hamiltonian models}
\begin{figure}[ht]
\centering
\includegraphics[width=1.0\linewidth]{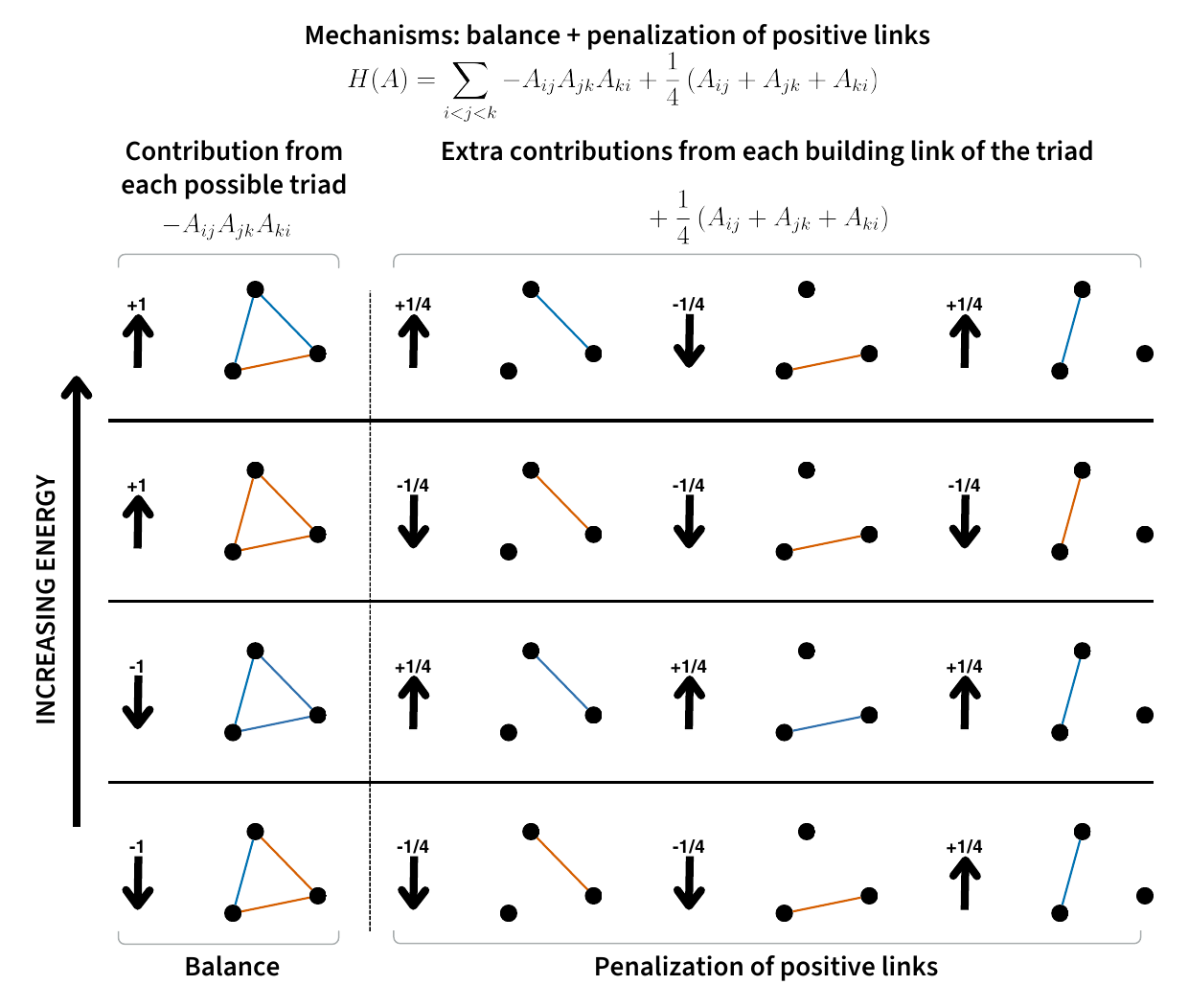}
\caption{\textbf{Exemplary Hamiltonian model.} Illustration of a specific Hamiltonian model that incorporates two distinct dynamical mechanisms, each represented by a separate term in the Hamiltonian function. The first term favours balanced triads by assigning them lower energy, while the second term favours the formation of negative links, also by lowering their energy. By combining both terms, we obtain a complete energy-based ordering of the four possible triad types in a fully connected signed network. For example, the triad with two negative and one positive link has the lowest energy, $-\tfrac{5}{4}$ (computed as $-1 - \tfrac{1}{4} - \tfrac{1}{4} + \tfrac{1}{4}$). The same logic applies to the other three triads, establishing a ranking based on their total energy. At equilibrium, we expect the relative abundance of each triad type to reflect this ordering: the lower the energy, the more frequently the triad should appear. Conversely, if the coefficients in the Hamiltonian (e.g., $-1$ and $+\tfrac{1}{4}$) are treated instead as free parameters, they can be fitted to empirical data. In this case, the observed abundance of triads can be interpreted as evidence for the underlying mechanisms assumed in the model, useful to evaluate their relative importance in the network dynamics. }  
\label{fig:hamiltonian}
\end{figure}

Finally, we turn to models that describe networks using a Hamiltonian function. While these models are not strictly dynamic models of network evolution, they incorporate implicit dynamical concepts, as we will see. These models originate from statistical physics and assume that each possible local network configuration (e.g., dyads or triads) is assigned an \emph{energy} or \emph{cost} that reflects how likely that local configuration is. This energy is specified by a Hamiltonian function \( H(s) \), where \( s \) represents the network's state (e.g., its pattern of connections)~\citep{parkStatisticalMechanicsNetworks2004,squartiniAnalyticalMaximumlikelihoodMethod2011}. The model then assumes that configurations with lower energy are more likely, following a probability distribution of the form \( P(s) \propto e^{-\beta H(s)} \). The parameter \( \beta \) controls how strongly the system favours low-energy states: high \( \beta \) enforces strong preference for optimal configurations, while low \( \beta \) allows greater randomness\footnote{Although here we focus on Hamiltonian approaches to model the dynamics of signed networks, these methods have also played an important role in the signed community detection (see Section \ref{subsec:comm-det}) and in defining null models for signed networks (see Section \ref{subsec:null-models})}. It is worth mentioning that, beyond the context of signed networks specifically, Hamiltonian approaches are closely connected to Exponential Random Graph Models~\citep{lusher2013exponential} (see Section \ref{subsec:null-models}).

This approach was pioneered by Axelrod~\citep{axelrodLandscapeTheoryAggregation1993} and Marvel et al.~\citep{marvelEnergyLandscapeSocial2009}, but one of the most comprehensive overviews can be attributed to Belaza et al.~\citep{belazaStatisticalPhysicsBalance2017}. The Hamiltonian they propose is
\[
H(A) = \sum_{i<j<k} 
-\alpha\, A_{ij} A_{jk} A_{ki}
- \gamma \left( A_{ij} A_{jk} + A_{ji} A_{ki} + A_{jk} A_{ki} \right)
+ \omega \left( A_{ij} + A_{jk} + A_{ki} \right),
\]

Where $A$ represents the adjacency matrix. Each element accounts for the cost associated with different dyadic and triadic configurations, weighted by the parameters $\alpha$, $\gamma$, and $\omega$. These parameters determine the influence of each configuration, with $\alpha$ controlling triadic interactions, $\gamma$ the dyadic interactions, and $\omega$ the edge-level contributions. As we can see, the main advantage of this approach is that it provides flexibility, as each term in $H$ can model a different mechanism controlling network dynamics. Another advantage is that the parameters $\alpha, \gamma, \omega$ can be computed from empirical data and used to evaluate the relative importance of the corresponding mechanism. An example is depicted in Figure~\ref{fig:hamiltonian}. Other works using Hamiltonian models include~\citep{shojaeiPhaseTransitionNetwork2019},~\citep{belazaSocialStabilityExtended2019},~\citep{rabbaniMeanfieldSolutionStructural2019} or~\citep{manshourDynamicsSocialBalance2021}, each introducing different Hamiltonian functions associated with specific phenomena under investigation. 

As we mentioned, an important but subtle detail is that these models do not necessarily describe how networks evolve dynamically. Instead, they describe the macroscopic characteristics of the network when it reaches a dynamical equilibrium, an assumption that has been shown to hold in some real networked systems~\citep{gonzalez-casadoEvidenceEquilibriumDynamics2025} but cannot be taken for granted. However, the terms in the Hamiltonian include assumptions about the mechanisms that make the networks evolve in time, favouring some configurations over others. Therefore, it can be applied to real systems to extract insights about the implicit dynamics. In fact, once a Hamiltonian is specified, it is possible to define a dynamics that makes the system evolve toward such an equilibrium~\citep{landauGuideMonteCarlo2021}. The usual approach is to let the system evolve by randomly proposing small changes (like flipping a link sign) and deciding whether to accept them based on the change in the system's energy. For example, the Metropolis rule~\citep{metropolisEquationStateCalculations1953} always accepts changes that lower the energy of the system, and accepts energy-increasing transitions with a probability $e^{-\beta \Delta E}$ (where $\Delta E$ is the increase in energy).

\subsubsection{Data-driven approaches}

Toy models offer explainability and analytical tractability, but they rarely account for the complexity of real systems. For that, we need to turn to data-driven approaches that provide a more realistic view of the dynamics underlying real network evolution. However, they also face significant challenges. Arguably, the most notable obstacle is the difficulty of collecting longitudinal data on the temporal evolution of networks. In the following, we summarise the main signed network empirical analyses of real-world data, focusing on the mechanisms each study could characterise. We postpone the details concerning the methods used to gather data to Chapter \ref{chapter:collection-inference-data}. We will restrict our discussion to social systems, as these were the first domain in which signed networks were applied, and where the mechanisms governing edge dynamics are better understood. We do not cover mechanisms from ecological, neural, or other types of networks.

There is a wide range of mechanisms beyond structural balance that can operate in parallel, and those governing unsigned dynamics may interact in nontrivial ways with those governing the dynamics of signed links. Before proceeding, we should clarify a subtle point: a stable level of a mechanism does not imply that the structures it generates are becoming more common over time. Rather, it means that these structures remain more frequent than would be expected by chance. In practice, other mechanisms may simultaneously compete to dismantle these structures, stabilising certain configurations.

\paragraph{Offline social networks}

One recurring example in the study of signed network dynamics involves interpersonal relations within relatively small groups in closed environments, such as schools, universities, or monasteries. These settings are especially appealing because they make it relatively easy to track relationships over time and collect longitudinal data. The most influential datasets in this area are the Sampson and Nordlie-Newcomb studies. The Sampson dataset~\citep{sampsonNovitiatePeriodChange1968} captures the evolution of relations among a group of monks, while the Nordlie-Newcomb dataset~\citep{nordlieLONGITUDINALSTUDYINTERPERSONAL1958,newcombAcquaintanceProcess1961} records the evolving relationships among college students from their first encounter over a period of several weeks. These early works laid the empirical foundation for much of the subsequent research on signed network dynamics.

Analyses of the Nordlie-Newcomb dataset (such as those by~\citep{nakaoLongitudinalApproachSubgroup1993},~\citep{doreianBriefHistoryBalance1997}, and~\citep{doreianPretransitiveBalanceMechanisms2001}) identify three key mechanisms that shape network evolution: reciprocity (tendency for mutual links), transitivity (the tendency to form closed triads), and social balance. These studies show that reciprocity is present early on, while transitivity and balance become increasingly prominent over time. Later analyses also reveal the role of differential popularity, where certain individuals consistently attract more positive or negative links, possibly due to personality traits or group-level perceptions. Differential popularity also shows how mechanisms can interact in misleading ways: it can create an excess of balanced or unbalanced triads, potentially leading researchers to mistakenly infer the presence of a structural balance mechanism if these effects are not properly controlled for. Similar insights emerge from the analysis of the Sampson dataset in~\citep{doreianEvolutionHumanSigned2004}, which observes a gradual shift toward more balanced, transitive, and reciprocal structures over time.

Apart from these two famous datasets, most of the work on offline social networks has focused on schools, where the structure of students’ personal relationships is analysed. Early studies considered only networks of directed positive links.~\citep{katzConceptConfigurationInterpersonal1959} can be considered as a foundational work in this direction, and they analyse for the first time the temporal evolution of such a social network as a Markov chain. This quickly became the state-of-the-art tool to analyse network dynamics~\citep{snijdersStatisticalModelsSocial2011}. This work was followed by others which explored the action of different mechanisms at the level of positive links, like transitivity and reciprocity, and their consequences in the network structure~\citep{davisStructurePositiveInterpersonal1967,hollandMethodDetectingStructure1970,hollandTransitivityStructuralModels1971,hallinanCommentHollandLeinhardt1972,hallinanStructuralModelSentiment1974,hallinanFriendshipPatternsOpen1976,hallinanDevelopmentChildrensFriendship1977}. The work by Hallinan et al.~\citep{hallinanStructuralModelSentiment1974} stands out
because it reintroduces the role of the cognitive perspective in the analysis, as Heider suggested (see Section \ref{subsec:Heiderian_balance}). Specifically, it develops the idea that, in a network, it is the nodes that experience tension. This idea, which seems obvious, is overlooked, for example, when the analysis is focused on triads, telling us whether they are balanced or unbalanced, and seeing how unbalanced ones can evolve into more balanced states. Hallinan suggests that to understand this evolution, it is necessary to put oneself in the perspective of the nodes, not the triangles, and see how the triad tension affects them. In fact, they use this idea to explain the abundance of certain types of triads in the data that other authors had not been able to explain. 

Most studies focusing on schools collect longitudinal data exclusively on positive links. However, a few works have gathered longitudinal network data in schools as closed environments, incorporating also negative links~\citep{huitsingVictimsBulliesTheir2014,stadtfeld2020emergence,escribanoEvolutionSocialRelationships2021,escribano2023stability,gonzalez-casadoEvidenceEquilibriumDynamics2025}. Among them,~\citep{gonzalez-casadoEvidenceEquilibriumDynamics2025} is particularly relevant, as it shows that the network of social relationships remains in dynamical equilibrium in the statistical–mechanical sense. In other words, while the microscopic dynamics are highly active, the macroscopic properties of the network remain stable. In the context of signed networks, it implies that, even with balancing mechanisms operating, the overall level of balance in the network remains constant despite strong volatility in negative links. Hence, balancing mechanisms coexist and compete with other processes that act to stabilise the level of balance in the system.

\paragraph{International relations}

Another domain in which signed networks evolution has received attention is international relations. These data are relatively easy to obtain from historical event databases like Correlates of War~\citep{palmerMID5Dataset201120142022,giblerInternationalMilitaryAlliances2008} (see Section \ref{subsec:common-data}). From them, one can reconstruct temporal signed networks where nodes represent states, positive links indicate alliances, and negative links capture tensions, rivalries, or wars.

The first to apply this type of analysis was Harary, who used balance theory to offer a new perspective on conflicts in the Middle East~\citep{hararyStructuralAnalysisSituation1961}. A similar analysis was later conducted by Moore~\citep{mooreInternationalApplicationHeiders1978,mooreStructuralBalanceInternational1979}. While pioneering, these works have been criticised (e.g.~\citep{doreianStructuralBalanceSigned2015}) for relying on arbitrary country selections, short time windows, and limited attention to broader context. They also argued that focusing on particularly acute events may create an illusion of balance that does not reflect broader or long-term dynamics. As an alternative,~\citep{doreianStructuralBalanceSigned2015} analysed the international system from 1946 to 1999 using the Correlates of War dataset and found that balance levels fluctuate rather than steadily increase. Later studies, such as~\citep{diaz-diazMathematicalModelingLocal2024}, examined these fluctuations at the level of individual countries, finding correlations with major historical events occurring in the corresponding countries and years. However, as previously noted, a stable level of balance does not necessarily imply the absence of structural balance as an active mechanism in the system. Its presence is better reflected in the persistent over-representation of certain structures, compared to what would be expected by chance. In this regard, Kirkley et al.~\citep{kirkleyBalanceSignedNetworks2019} provided compelling evidence that international relations networks exhibit significant levels of balance.

It should be noted that the analysis of signed networks of international relations is much less developed than in the context of offline social networks. Most existing studies adopt an interpretative perspective, with relatively few efforts aimed at modelling or predicting network evolution. A notable exception is the work by Askarisichani et al.~\citep{askarisichani19952018GlobalEvolution2020}, who propose a Markov model of triad transitions with demonstrable predictive power. Mechanisms beyond structural balance have also received little attention. In this regard, the international relations literature offers a valuable example of how to critically assess structural balance in relation to alternative theories~\citep{healyBalancePowerInternational1973a,maozWhatEnemyMy2007}.
There is also a growing need to bridge network-based approaches with traditional geopolitical analysis, for example by combining structural balance theory with factors like geographical proximity or economic ties, and some challenges are still open, such as the inclusion of memory effects in the system (what effect past events of the network have on the present and to what extent they can be modelled as Markovian systems), if the length of cycles plays a role on the balance mechanism, whether link directionality plays a significant role.

\paragraph{Online social media}

A third approach to gathering empirical signed networks involves extracting data from online platforms. This strategy has gained popularity thanks to the ease of data collection and the ability to access systems involving thousands of agents. One of the earliest studies in this direction is~\citep{kunegisSlashdotZooMining2009}, which analyses the Slashdot Zoo network but does not focus on temporal evolution.  A more complete dynamic analysis appears in~\citep{szellMeasuringSocialDynamics2010, szellMultirelationalOrganizationLargescale2010, szellSOCIALDYNAMICSLARGESCALE2012}, which examines the multilayer network of Pardus, a massive multiplayer online game with over 300,000 nodes. The structural analysis, though static, already points to key mechanisms in the network. They determine three different ways in which positive and negative links differ: first, positive links are highly reciprocal, whereas negative links are not; second, negative links follow a power-law degree distribution, unlike positive ones; and finally, positive layers exhibit higher clustering. In addition, the hubs in the positive and negative subgraphs are uncorrelated. The network also appears more balanced than expected by chance. In their analysis of network dynamics, the authors examine the network evolution over three years. Once again, they find qualitatively different behaviour in the link dynamics depending on their sign: negative links follow preferential attachment, whereas positive ones do not. In addition, they distinguish between two types of negative links: private enemies, which are typically local and uncorrelated with balance, and public enemies, who act as negative hubs and are often jointly targeted by friends, thus playing a central role in the formation of balanced triads.
A follow-up study~\citep{sadilekAsocialBalanceHow2018} develops a generative model to simulate Pardus’ dynamics. They show that, while the dynamics of positive links are relatively autonomous, the dynamics of negative links are tightly coupled to the positive layer. Other examples of such analyses are~\citep{liModelingOnlineSocial2018,askarisichaniStructuralBalanceEmerges2019,liuEvolutionStructuralBalance2020}.

\subsubsection{Mechanisms governing signed network dynamics}\label{subsubsec:mechanisms-governing}

\begin{figure}[ht]
\centering
\includegraphics[width=1.0\linewidth]{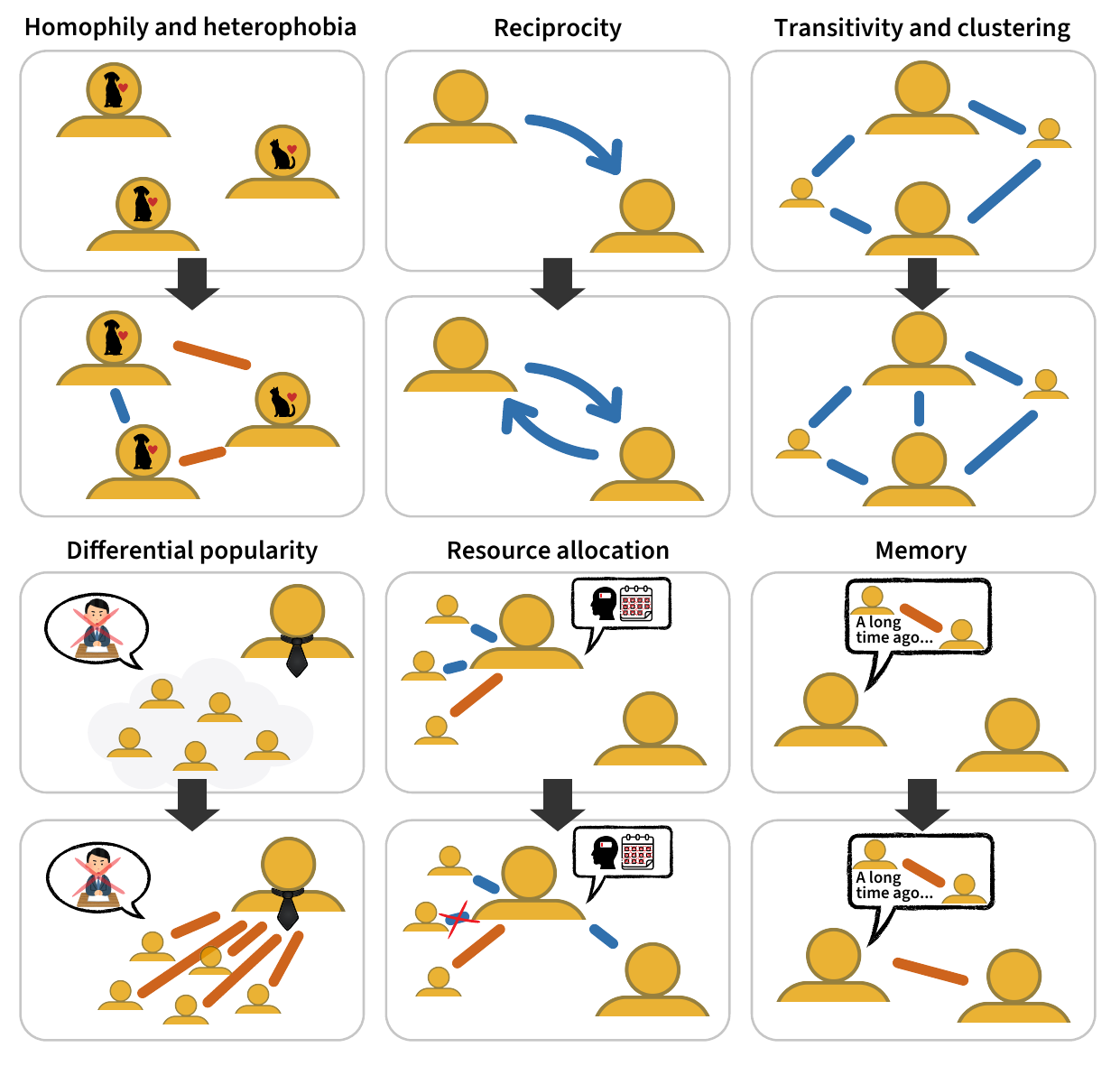}
\caption{\textbf{Some mechanisms governing social network dynamics.} Depiction of six distinct mechanisms that have been identified as influential in the dynamics of social signed networks. Each mechanism is illustrated using two panels: the top panel shows the network state before the mechanism takes effect, and the bottom panel shows the resulting state after the mechanism operates. \textit{Homophily and heterophobia:} Three individuals with specific pet preferences form positive ties with others who share their preferences, and negative ties with those who differ. \textit{Reciprocity:} An agent reciprocates a positive directed link after receiving one. \textit{Transitivity and clustering:} Two individuals who share two common friends form a new connection, closing two open triads. \textit{Differential popularity:} Five individuals, each with a negative view of authority, form negative ties toward their boss. \textit{Resource allocation:} An individual with limited cognitive resources or time dissolves an existing connection to establish a new one. \textit{Memory:} A person re-establishes a negative relationship due to the resurfacing of a past conflict.}  
\label{fig:mechanisms}
\end{figure}

As we have seen, many empirical analyses agree that the network evolution is shaped by mechanisms beyond structural balance. Identifying these mechanisms has proven difficult, partly because they are often specific to the system under study and partly because they interact in complex, nontrivial ways. In addition, common simplifications such as removing link directionality can obscure important aspects of the dynamics. More fundamentally, the mechanisms we have covered so far, which we will name \textit{endogenous mechanisms}, are just one part of the picture. In general, we can group the factors driving network change into three types~\citep{riveraDynamicsDyadsSocial2010}. \textit{(i) Endogenous mechanisms:} some links appear or disappear by the simple presence of other links in the network, so that the current structure of the network constrains and modifies its future structure. These mechanisms include balance, reciprocity, or triadic closure. \textit{(ii) Node-based attributes} that are not encoded in the network structure but affect the way the network evolves over time. Examples include personality, social status, or shared interests. (iii) \textit{Contextual and external factors}, coming from the environment in which the network is embedded. These include institutional settings, cultural norms, or platform algorithms. 

The combination of these three factors makes the network evolve over time, and they operate locally in the network through what we call mechanisms. In what follows, we review some well-established mechanisms that shape the evolution of signed social networks. An illustration of some of these mechanisms is also provided in Figure \ref{fig:mechanisms}.

\begin{enumerate}
    \item \textbf{Homophily}: the tendency to form positive links with others who share similar characteristics, such as interests, beliefs, or social status~\citep{riveraDynamicsDyadsSocial2010}. This mechanism is closely related to, and often hard to separate from, \textit{influence}, which is the tendency for connected individuals to become more similar over time. Homophily can have three variations: homophobia/heterophily/heterophobia, which would be the tendency to form negative/positive/negative links with those nodes with similar/different/different characteristics, respectively. These variations can be generalised with the concept of assortativity.

    \item \textbf{Reciprocity}: the tendency for directed links to become bidirectional. This results in an overrepresentation of mutual links compared to a null model. Moreover, when asymmetrical links do not become reciprocal, they tend to be unstable and eventually disappear~\citep{riveraDynamicsDyadsSocial2010}.

    \item \textbf{Transitivity and clustering}: the tendency to complete the connection  \( p \rightarrow q \) given the existing connections \( p \to o \) and \( o \to q \). In undirected networks, this mechanism is known as clustering or triadic closure, the tendency to close open triangles. This effect seems to be cumulative and grows with the number of common neighbours shared by two nodes.~\citep{riveraDynamicsDyadsSocial2010}.

    \item \textbf{Differential popularity}: the tendency of networks to display heterogeneous degree distributions for both positive and negative links~\citep{riveraDynamicsDyadsSocial2010}. This mechanism includes two distinct dimensions: \textit{activity}, which refers to variation in the number of outgoing links, and \textit{popularity}, which refers to variation in the number of incoming links. At a dynamic level, differential popularity can appear by means of preferential attachment or detachment, a mechanism through which the rates of gain or loss of links depend on the degree of the node~\citep{barabasiEmergenceScalingRandom1999a}. It should be noted that the different ideas encompassed within the generic term ``differential popularity'' may be related to each other, giving rise to theories such as visibility theory, popularity theory, activity theory, or karma theory. All of these possibilities are collected in~\citep{yapWhyDoesEverybody2015}.

     \item \textbf{Balance}: the tendency to form consistent triads, either in the Heiderian sense or in the structural sense (see Chapter~\ref{chapter:balance}). 

    \item \textbf{Resource allocation}: forming and maintaining links requires time, attention, and cognitive effort, which are all limited. This constraint shapes how many links a person can have and how these social bonds are structured~\citep{powellOrbitalPrefrontalCortex2012,tamaritCognitiveResourceAllocation2018,tamaritDunbarCirclesContinuous2022}. These resource constraints shape the degree distributions for both positive and negative links, and can explain why negative links are often unstable (conflict is costly). It also helps explain why seemingly stable structures can disappear: when links dissolve, the freed resources can be reinvested in new connections, and conversely, existing links may need to be dropped to make room for others.

    \item \textbf{Memory}: while many models assume Markovian dynamics (i.e., memoryless evolution), empirical evidence suggests that past interactions can influence future ones. For instance, nodes may be more likely to re-establish past positive links or avoid reconnecting after a negative link has disappeared. A similar reasoning has been suggested to influence the process of link removal~\citep{kleinbaumReorganizationTieDecay2018}.
\end{enumerate}

Combinations of mechanisms can yield patterns worth examining on their own, like hierarchical structures. Hierarchies can appear at the individual or group level, and there is no clear consensus on why and how these hierarchies are formed. At the most basic level, asymmetric links may be tentative or aspirational links in a relationship in the process of formation and have not yet had time to be consolidated. However, if they are not reciprocated or dissolved, they can lead to the formation of hierarchies. Some authors suggest hierarchies could appear as a combination of reciprocity, transitivity, and differential popularity~\citep{davisStructurePositiveInterpersonal1967,hollandMethodDetectingStructure1970,hollandTransitivityStructuralModels1971,hallinanCommentHollandLeinhardt1972,hallinanStructuralModelSentiment1974}. This pseudo-mechanism provides the theoretical basis behind status theory~\citep{leskovecSignedNetworksSocial2010}~\citep{leskovecPredictingPositiveNegative2010}~\citep{leskovecPredictingPositiveNegative2010}. However, there are other possible explanations behind the formation of asymmetric links. Thresholds in the definition of positive or negative versus neutral relationships across nodes may cause this asymmetry. It may also be a combination of differential popularity and a resource constraint that prevents a reciprocation of links, even if there is a willingness to do so. Finally, some findings suggest it may be a combined effect of the tendency to reciprocate and a long self-perceived centrality on the part of the nodes~\citep{anOriginsAsymmetricTies2015}. This discussion highlights, among other things, the importance of developing a comprehensive domain-specific knowledge of the system under study, which can be integrated into the mathematical framework to study the dynamics rigorously.

\paragraph{Interaction between mechanisms}

As noted by~\citep{riveraDynamicsDyadsSocial2010}, mechanisms cannot be understood in isolation, even though understanding how different mechanisms interact is as important as understanding the mechanisms themselves. They may produce overlapping or even opposing effects, making them hard to disentangle. To what extent does homophily exist in a network because actors bring themselves to contexts in which certain characteristics are shared? Is homophily still valid when I connect with my friends’ friends? How does balance operate if nodes are averse to conflict? Some interactions between mechanisms have been studied in unsigned networks. For instance, in~\citep{blockReciprocityTransitivityMysterious2015}, the authors explore how transitivity and reciprocity operate together, finding a somewhat counter-intuitive result: asymmetric links are more stable when they are structurally embedded.~\citep{schaeferFundamentalPrinciplesNetwork2010} studies the joint effect of reciprocity, differential popularity, and transitivity, giving rise to the concept of structural cascading. 

For signed networks, some studies model interactions between mechanisms in purely theoretical terms. The most notable research in this direction concerns the interaction between balance and homophily~\citep{dengInfluenceStructuralBalance2016, gorskiHomophilyBasedFew2020, phamEffectSocialBalance2020, phamBalanceFragmentationSocieties2021, phamEmpiricalSocialTriad2022}. These papers extend existing models of balance by introducing homophilic preferences in link formation.

Many other papers study the competition between mechanisms at the empirical level. An early example of this is~\citep{leskovecSignedNetworksSocial2010}, which compares the predictive power of balance theory and status theory. It finds that both theories make some contradicting predictions, and balance better explains triads with reciprocal links, while status is more relevant for asymmetric, hierarchical ones. This result suggests that both mechanisms operate simultaneously but affect different aspects of the structure. Other early examples are ~\citep{facchettiComputingGlobalStructural2011} and~\citep{doreianTestingTwoTheories2014}, which study the influence of differential popularity on balance. The authors' study examines the effect that the presence of positive and negative hubs has on balance. They find that when there is a negative link hub, there is a good chance that this node is involved in many balanced triangles. Two nodes connected by a positive link will have a better chance of both being connected to the hub, regardless of how balance is operating in the network. The opposite applies if the hub is positive. Finally, one of the most systematic empirical comparisons of mechanism interplay is given in~\citep{yapWhyDoesEverybody2015}. Specifically, that paper analyses a large corpus of networks and quantifies the percentage of the data's variance explained by balance theory, by homophily, and by status theory. This paper shows how it is necessary to include the combined effect of all these mechanisms to explain the observed structure. Another recent interesting contribution based on the usage of ERGMs is~\citep{fritz2025exponential}.

While most of the above studies take a heuristic or model-driven approach, more systematic methodologies are emerging. A promising direction is the use of Stochastic Actor-Oriented Models (SAOMs)~\citep{snijdersStatisticalModelsSocial2011}, which enable statistical inference about multiple interacting mechanisms from longitudinal data. Applications of SAOMs to signed networks, such as~\citep{lernerStructuralBalanceSigned2016,huitsingVictimsBulliesTheir2014, rambaranDevelopmentAdolescentsFriendships2015,bergerCompetitionEnvySnobbism2013,toroslu2022avoidance,kros2021avoidance}, offer a rigorous framework for testing how mechanisms operate jointly.

\subsection{Co-evolving dynamics} \label{subsec:coevolution}
A recent line of research has taken an interest in co-evolving dynamics, which combine the evolution \textit{on} and \textit{of} signed networks. Compared to opinion dynamics on static signed networks, agents still strive to agree with their friends and disagree with their foes, but they may also adjust the sign of their relationships based on opinion similarity. These co-evolving dynamics seek to explain how opinion polarisation and structural balance interact over time.

The first co-evolving model of opinion dynamics with signed links was introduced in~\citep{saeedianAbsorbingPhaseTransition2019}. In this work, each node holds a binary opinion, and each link can be positive or negative. A connection is satisfying when a positive link joins nodes with the same opinion or a negative link joins nodes with opposite opinions; otherwise, either one node flips its opinion or the link changes sign to restore consistency. Through this mutual adjustment, the system evolves toward either global consensus (with only positive links) or polarisation into two antagonistic groups connected by negative edges.
In a follow-up paper~\citep{saeedianAbsorbingstateTransitionCoevolution2020}, the authors allowed unsatisfying links to rewire instead of merely changing sign, so that disagreeing nodes could remove hostile links and form new friendly ones. When this rewiring mechanism dominates, the network fragments into two cohesive factions; otherwise, the network is driven toward full consensus. 

A significant fraction of co-evolving models is based on Hamiltonian-based approaches. In particular, a series of works~\citep{phamEffectSocialBalance2020,phamBalanceFragmentationSocieties2021,phamEmpiricalSocialTriad2022,galesicExperimentalEvidenceConfirms2025,castro2025heterophobic} studies Hamiltonians of the form: 
\begin{equation} \label{eq:hamiltonian}
    H = -\sum_{i\neq j} A_{ij}s_is_j -\sum_{i\neq j\neq k} A_{ij} A_{jk} A_{ik},
\end{equation}
where $A$ is the signed adjacency matrix, and $s_i\in\{-1,+1\}^d$ are vectors containing the agents' binary opinions on multiple topics\footnote{Hamiltonian formulations have also proved useful for the study of the dynamics of signed networks, without node-level features (see Section~\ref{subsubseq:toymodels}), and for opinion dynamics on static signed networks~\citep{liBinaryOpinionDynamics2019,Baldassarri2023}.}. This equation encodes two mechanisms driving the evolution of stress in the system: the correspondence between connections and opinion similarity and structural balance. Agents seek to reduce their social stress by maximising these two quantities, i.e.\ by minimising $H$. In the model of~\citep{phamEffectSocialBalance2020}, the edges evolve alongside opinions to reduce social stress, while in~\citep{phamBalanceFragmentationSocieties2021,phamEmpiricalSocialTriad2022,castro2025heterophobic}, they are determined directly by the similarity of agents' opinions. In both cases, the system exhibits a first-order transition between polarised and unpolarised states depending on the model parameters and the network topology. Perhaps the most important result is that, even when removing the explicit need for structural balance, i.e., the second term in Eq.~\eqref{eq:hamiltonian}, dyadic interactions modelled by the first term of $H$ are sufficient to entail structural balance~\citep{phamEmpiricalSocialTriad2022}. These theoretical results are validated on a real-life network extracted from an online multiplayer game, and through a participant experiment~\citep{galesicExperimentalEvidenceConfirms2025}. In particular, the dyadic interaction model matches better with empirical observations.

Beyond Ising-like models, another central contribution is by Schweighofer et al.~\citep{schweighoferWeightedBalanceModel2020,schweighoferRaisingSpectrumPolarization2024}, who proposed an empirically validated model to explain opinion polarisation and issue alignment in multi-dimensional opinion spaces. The system consists of agents who evaluate their affect toward others on the basis of the similarity of their multi-dimensional opinions. In other words, the nature of the signed link between any two agents is a function of their individual (signed) links with a third party, namely the political issue of interest (e.g.\ support or opposition to environmental policies). The authors validate this micro-level hypothesis with data from the 2016 American National Election Study (ANES). The dynamics of the model proceed as the agents repeatedly re-evaluate their relationships, then nudge their opinions closer to those of their friends and further away from those of their foes. After a transient state of consensus, the model generates polarised structures with aligned issues. These dynamics can reproduce trends of polarisation among ANES respondents over 40 years of data~\citep{schweighoferRaisingSpectrumPolarization2024}.

Coevolution has also been studied in the context of contagion, where the system updates both node states and link signs to reduce social tension at the same time that the contagion advances. One way to encode this coupling is by penalising both unbalanced triads (social tension) and friendly exposure between infected and non–infected nodes (spreading pressure). 
This coevolution typically ends either in a balanced network or in jammed local minima~\citep{saeedianEpidemicSpreadingEvolving2017}. In both cases, unfriendly links act as an immunisation that prevents infection even if some friendly cross-links remain. A natural extension adds an “alert” compartment to capture behavioural response~\citep{kalimzhanovCoEvolution_of_Viral_Processes_and_Structural_Stability_in_Signed_Social_Networks2023}. Under this assumption, the network self-organises into two clusters (one alerted and one infected) interconnected mostly by unfriendly links, shrinking the magnitude of the epidemic.

\clearpage

\section{Collection, inference and availability of signed data}\label{chapter:collection-inference-data}

As discussed in the previous chapters, signed networks offer a powerful framework to study a broad range of systems, since they capture phenomena that cannot arise when all connections are positive. However, all these insights are only possible if we have access to signed data, which is often scarce in real-world systems. Ideally, whatever the system under study (whether social relationships in a high school or in a subreddit community), we would like to collect explicit signed relations between individuals. In practice, though, we can often only access other types of data and need to infer the sign or bias of the links. 

Section~\ref{subsec:signet-mining} reviews existing methods for inferring signed networks from different empirical settings, including bipartite projections, aggregated unsigned interactions, and text annotations. We discuss the methodological assumptions underlying each approach and emphasise that inference should be treated as a central analytical step rather than a preprocessing task. Section~\ref{subsec:common-data} then surveys existing available datasets that provide either explicit signed information or allow for the inference of signed links under minimal assumptions. These datasets, while useful, are limited in number and often outdated. This scarcity highlights the importance of developing new, standardised, high-quality, and openly available signed network data. 

\subsection{Mining signed networks from raw data}\label{subsec:signet-mining}

In many cases, signed relations are not directly observed but must be inferred from other types of interaction data. Depending on the context, we may have access to (i) unsigned interaction data, which can be aggregated as a weighted unsigned directed or undirected network, (ii) co-occurrence data, which can be represented through a bipartite network, (iii) ordered preferential data (i.e. ranking data) or (iv) signed interaction data aggregated as a weighted signed network, which might be too noisy depending on the intended application. 

For example, in social networks, we rarely observe direct information on interpersonal relations. Instead, we rely on measurements of individual interactions that we expect to align with the underlying relationship between pairs of users (which have assumptions that are very dependent on context, such as online or in-person communities). Alternatively, we could have access to event attendance or vote agreements, being able to infer a consistent ``encounter/agreement'' or ``avoidance/disagreement'' that can hint towards positive or negative relations. 

\begin{figure}[ht]
    \centering
    \includegraphics[width=0.85\linewidth]{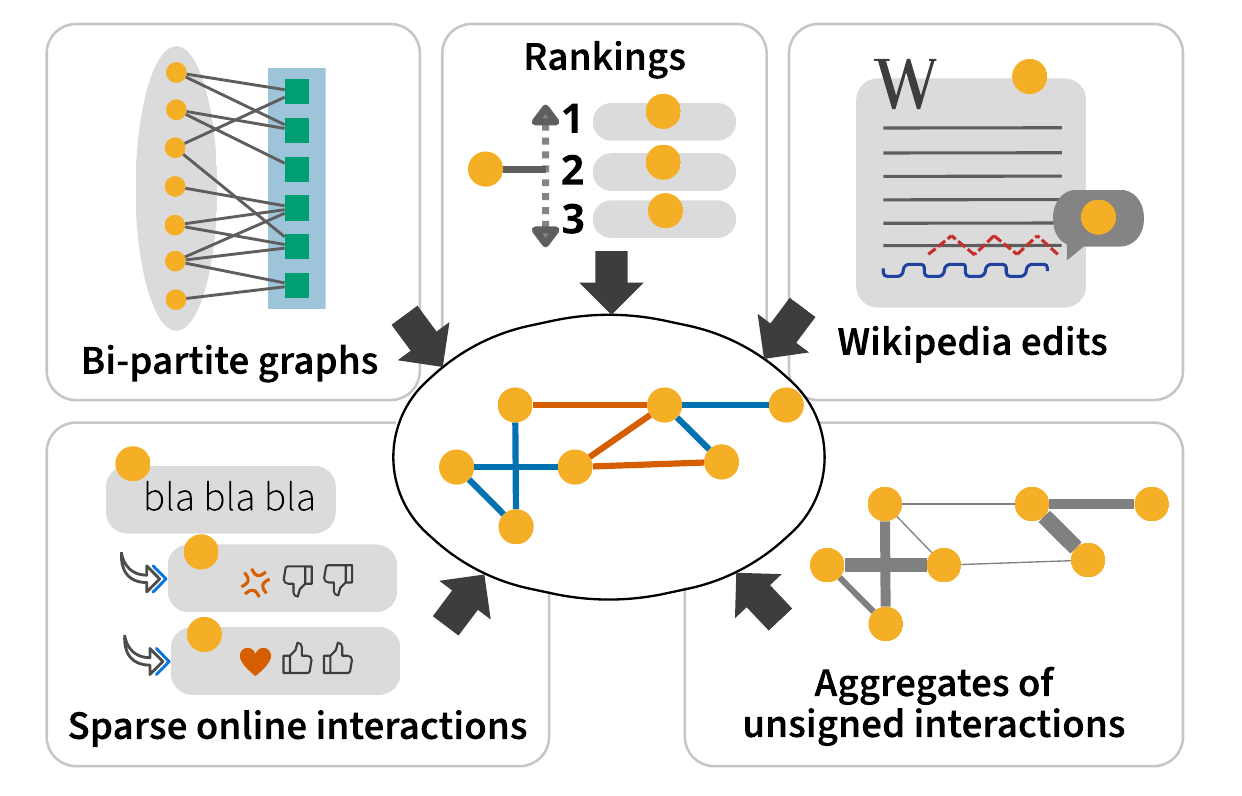}
    \caption{\textbf{Data sources for signed network inference.} The figure illustrates different types of data that can be used to infer signed networks, including bipartite projections, aggregated unsigned interactions, online textual or voting interaction data, and co-edit patterns in Wikipedia. }
    \label{fig:data-sources}
\end{figure}

Below, we describe several methodological approaches used to infer signed relationships from different types of interaction data, as shown in Figure~\ref{fig:data-sources}. All of these methods have their limitations. Some struggle to handle ambivalent links when simplifying weighted or directed signed networks. Others assume that positive relations increase interaction and negative relations reduce it, which may not hold. In many cases, there is also no clear conceptual justification for why sharing artefacts (for example, event attendance) should indicate friendship or enmity.

Therefore, the preparation and processing of the data to obtain signed networks, while being useful in expanding our set of use cases, must be treated as part of the methodology and examined with care, as it can have a strong effect on the results of the posterior analysis.

\subsubsection{Ranking data}

A popular approach to infer signed relations relies on ego network rankings, where each participant is asked to rank others in terms of their preference to interact with them. One of the first well-known examples is the Newcomb dataset~\citep{newcombAcquaintanceProcess1961}, in which a group of students was asked to rank the remaining group members across several time points. From such rankings, signed relations can be inferred by selecting thresholds that determine which positions in the ranking reflect positive, negative, or neutral links~\citep{doreianBriefHistoryBalance1997}. For instance, the top-ranked individuals might be interpreted as positive links, the bottom-ranked as negative, and the rest as neutral. The main limitation of this method is that details on the mapping from ordinal ranks to signed links (e.g., the choice of thresholds) can affect the resulting network structure.

However, as Hallinan~\citep{hallinanAnalysisIntransitivitySociometric1975} points out, fixing the number of allowed alters (e.g., limiting participants to only select a fixed number of positive and negative links or forcing the selection a posteriori from ranking data) can lead to artefacts in structural analysis. In particular, constraints on the number of declared links reduce the number of possible closed triangles, potentially underestimating the level of balance in the network. One proposed solution is to allow participants to list as many positive or negative links as they wish, which results in a more accurate reflection of the social structure and increases the chance of observing structurally balanced triads.

\subsubsection{Unsigned bipartite networks}\label{subsub:bipartite-to-signed}

Many real-world datasets are naturally bipartite\footnote{The terms unipartite and one-mode (respectively, bipartite and two-mode) are often used interchangeably in the literature.}, meaning they contain two types of entities (called agents and artefacts) where connections exist only across the two sets. Examples include authors writing papers, legislators sponsoring bills, or customers reviewing products they bought. From these data, researchers frequently infer a co-occurrence, unipartite network, where two agents are connected if they share artefacts (e.g., two senators co-sponsoring the same bill). This process, known as bipartite projection, typically results in a positively weighted network, with edge weights reflecting the number of shared artefacts. Mathematically, given a bipartite network represented as the rectangular matrix \(B \in \{0,1\}^{n_1 \times n_2}\), where each row represents an agent, and each column represents an artefact, the one-mode projection onto the set of agents is defined as
\(W = B B^{\top},\)
where \(W_{ij}\) counts the number of shared artifacts between agents \(i\) and \(j\).

This raw co-occurrence count method has several problems, as the projected network is often very dense, and it usually contains spurious data that does not really reflect the relation between agents. For example, if two senators co-sponsor a bill that everyone supports, that connection does not provide meaningful information about their relationship. Moreover, if the original bipartite network is unsigned, its projection will only include positive edges, while a signed representation could sometimes better capture the underlying relationships. Therefore, it is important to implement a statistical framework to test which connections are stronger or weaker than expected by chance. This process is called \textit{backbone extraction}. Here, we outline the most popular approaches; for more details, please refer to \citep{nealBackboneBipartiteProjections2014}.

The simplest way to extract a backbone is to apply a global threshold, removing all connections with weights below the defined value. This method assumes that raw co-occurrence counts are directly comparable across all pairs of agents. Its main advantage is its easy implementation, no need for statistical modelling, and its computational efficiency. However, it suffers from important limitations ~\citep{nealBackboneBipartiteProjections2014}, such as the arbitrary choice of the threshold, as different thresholds can yield very different network structures. A low threshold produces a dense, almost complete graph, while a high threshold may leave very few edges. Moreover, the resulting network is often biased. For example, thresholded graphs typically have high clustering, even when the original bipartite network is entirely random \citep{zalesky2012use}, making properties such as heavy-tailed distributions likely to be statistical artefacts. 

More principled approaches compare the observed co-occurrences to what would be expected under a random null model (see Section~\ref{subsec:null-models}). The null models used for backbone extraction often preserve the degree of each agent and each artefact. The constraints can be enforced either exactly, such as the Fixed Degree Sequence Model (FDSM) \citep{zweigSystematicApproachOnemode2011}, or drawn from a degree distribution, such as the Stochastic Degree Sequence Model (SDSM) \citep{nealBackboneBipartiteProjections2014}. The former model has been used, for instance, to infer signed social networks from the bipartite projection of play interactions among preschoolers \citep{nealInferringSignedNetworks2022}, while the latter has been applied to extract signed networks of political relationships among members of the U.S. Congress \citep{nealSignTimesWeak2020}.

In what follows, we outline the main steps for extracting a signed network from bipartite data using the Stochastic Degree Sequence Model (SDSM). 
\begin{enumerate}
    \item Take an empirical bipartite network, represented by a rectangular matrix \( B \), where each row corresponds to an agent and each column to an artefact. 
    \item Compute the degrees of agents ($k_i^{\text{ag}}$) and artefacts ($k_i^{\text{art}}$), given by the row and column sums of \( B \), respectively.
    \item Obtain the one-mode projection of the bipartite network applying $W=BB^{\top}$.
    \item Generate random bipartite networks whose agent and artefact degrees match, \textit{on average}, those of the original network. This is the only step where the procedure differs from the Fixed Degree Sequence Model (FDSM), where the degrees must match exactly. Then, fit a binary outcome model (e.g., logit or scobit) to predict the entries \( B_{ij} \) from the degrees of the corresponding agent and artefact, as well as their product: \[
    B_{ij} = f\big(\beta_1 k_i^{\text{ag}} + \beta_2 k_j^{\text{art}} + \beta_3 k_i^{\text{agent}} k_j^{\text{art}} \big) + \epsilon_{ij} := p_{ij} + \epsilon_{ij},
    \]
    where \( f \) is the chosen link function, $\epsilon_{ij}$ are the residuals of the model, and the parameters \( \beta_i \) are estimated from the data. \( p_{ij} \) is the probability of observing an edge between agent \( i \) and artefact \( j \).
    \item Compute the probabilities \( p_{ij} \) using the fitted model. 
    Then, perform a Bernoulli trial with probability \( p_{ij} \) for each pair \((i,j)\) to decide whether an edge exists. This produces a random bipartite network, encoded in a matrix $B^*$, whose expected agent and artefact degrees match the empirical degrees, but which differs from the original network in any single realisation. 
    \item Repeat this process $M$ times to generate an ensemble of random networks. Project the \( M \) generated null-model networks onto the set of agents: \(W^* = B^* B^*\ ^{\top}\). 
    \item Assess the statistical significance of each edge in the projection. Fix a significance level \( \alpha \) and for each pair of agents \((i,j)\), compare the observed co-occurrence \( W_{ij} \) with the distribution of \( W_{ij}^* \) values from the \( M \) random projections. If \( W_{ij} > W_{ij}^* \) in more than \( \alpha M / 2 \) realizations, assign a {positive edge} between agents \( i \) and \( j \). If \( W_{ij} < W_{ij}^* \) in more than \( \alpha M / 2 \) realizations, assign a {negative edge}. Otherwise, no edge is assigned between \( i \) and \( j \). This completes the construction of the signed backbone. 
\end{enumerate}

In summary, both the FDSM and the SDSM test whether an observed co-occurrence between two agents is larger or smaller than expected under a degree-preserving (or degree-distribution-preserving) null model. 
The SDSM approach is typically the preferred one, since it is faster, scales better, and avoids over-conditioning the null model (see ~\citep{nealComparingAlternativesFixed2021} for a more in-depth comparison of the different models). A ready-to-use implementation of this method with different null model choices can be found in the R \textit{backbone} package~\citep{nealBackbonePackageExtract2022}. A modified version of SDSM that incorporates edge constraints is presented in~\citep{nealStochasticDegreeSequence2024}, and a recent extension by Gallo et al.~\citep{gallo2025statistically} adapts the method to bipartite signed networks.

Another relevant line of work has explored the inference of signed links from a bipartite structure of behavioural data. Yang et al.~\citep{yangFriendFrenemyPredicting2012} propose the \textit{Behaviour–Relation Interplay} (BRI) model, which predicts the sign of social links, in this case trust or distrust, by jointly modelling user behaviour (agents) with artefacts or items (e.g., news content) and social interactions. The framework combines a series of latent factor models for user–item decisions and social relations to determine edge signs automatically, even in unsupervised settings. The fundamental assumption is that behavioural agreement is an indicator of positive links and disagreement is a signal of negative ones. Based on this assumption, the model captures the mutual influence between behaviour and relationships. They test their method on Epinion and Yahoo! and show that it provides stronger predictive power for behavioural modelling than standard homophily-based approaches.

\subsubsection{Aggregates of unsigned interaction data}

A particularly challenging case of signed network reconstruction arises when the observed data consist of contact frequencies rather than direct measures of social relationships. In this case, inference methods take as input a collection of unsigned and repeated interactions between individuals instead of a bipartite structure. They usually rely on the assumption that, in social systems, a statistical over-representation of unsigned interactions signals a positive relation, while an under-representation signals a negative relation. In such settings, the absence of interaction between two individuals is inherently ambiguous: it may arise from limited opportunity for contact rather than deliberate avoidance, and failing to account for this distinction risks misclassifying absent ties as negative ones. \citep{ferenczi2026inferring} address this issue with a Bayesian generative model that encodes interaction opportunities through latent interaction groups, thereby separating chance non-contact from socially meaningful avoidance. Their MCMC-based inference framework reconstructs signed networks while quantifying uncertainty in inferred edge signs, and performs especially well in detecting negative edges relative to baseline methods on synthetic benchmarks. Applied to face-to-face contact data from a French high school, the approach recovers signed structures consistent with questionnaire-derived friendship patterns, with posterior predictive checks supporting the adequacy of the model.

An alternative approach that infers weighted signed relations is the one from \citep{andresReconstructingSignedRelations2023}. Their method requires data on interactions and partial data on ground truth relations. Similar to the SDSM~\citep{nealBackboneBipartiteProjections2014}, this method works by comparing the observed interaction counts with a network null model, in this case, the hypergeometric ensemble of random graphs (HypE)~\citep{casiraghiConfigurationModelsUrn2021}. HypE incorporates the heterogeneity in the activities and attractiveness of individuals, and has the advantage of being analytically tractable. 

Based on this null model, a weighted signed edge between two individuals is assigned according to the marginal probability that the observed number of interactions is higher or lower than its expected value. To determine the signed relation, these probabilities are weighted with community-dependent variables $a$ and $b$, learned through the declared ground truth relations in the datasets. 
This method is applied to five datasets with different features (size, directed/undirected, type of interactions, form of surveyed relations, etc.): a windsurfer community~\citep{freemanHumanSocialIntelligence1988}, the known Zachary's Karate Club dataset~\citep{zacharyInformationFlowModel1977}, face-to-face interactions in a French high school~\citep{mastrandreaContactPatternsHigh2015}, the Nethealth project~\citep{liuNetworkAnalysisNetHealth2018}, and Epinions~\citep{hamedaniTrustRecEffectiveApproach2021}.

Similarly, in~\citep{gursoyExtractingSignedBackbone2021}, they provide a method to extract the signed backbone of intrinsically dense weighted networks, which assumes that all nodes are aware of other nodes\footnote{Because of this assumption, this method might be inappropriate for tasks like extracting signed networks from large-scale social media data.}. They show applications to datasets related to migration, voting, temporal interaction, and species similarity networks. The method extracts positive and negative links whose weights significantly and substantially deviate from the random expectation under a null model that preserves node strength.
They introduce two filters to identify backbone links: the \textit{significance filter}, which removes links not significantly different from expectation (using separate thresholds for positive and negative cases); and the \textit{vigour filter}, which applies stricter thresholds based only on link intensity. This method accepts both directed and undirected networks. Their software is available in Python and R as \texttt{signed.backbones}.

\subsubsection{Sparse online interactions}\label{subsubsec:online}

Most of the methods explained above are centred on extracting signed networks of personal relations in systems where it is assumed that all individuals are aware of the existence of other individuals, and then choose to interact more or less with each of them depending on the nature of their relation. In these settings, the absence of an edge between two nodes usually implies a neutral relationship between the two. This assumption correctly applies to elite political systems, international relations, or co-authorship networks of small fields. However, it becomes problematic when applied to relations formed on social media. This is because in large-scale digital spaces, it is difficult to distinguish if a non-relation is the result of a conscious decision to avoid the other user or simply of a lack of exposure due to the algorithm or the user activity. This generates two types of neutral relations: those in which there has been interaction, but it does not lean towards positive or negative, and the cases in which there has been no opportunity for the interaction to happen. Yet both of these situations are categorised as non-edge. 
Moreover, in these platforms, the decision of what content or users one is exposed to is tightly defined by recommendation algorithms, which may have a strong effect on the final interactions and how these are distributed along the network. 
Finally, these interactions are also platform-specific: in some cases, we can find explicitly signed data, text reply data, or endorsement only, such as likes, retweets, or follows. The choice of how to translate these data structures to signed networks is then particular to each study, and we can find a large variation in collection and inference methods.

When collecting signed interactions, one typically aggregates all events between two users into a single signed weighted edge. However, this aggregation approach can generate ambiguities. For instance, a pair with only a few interactions seems identical to a pair with many interactions that cancel out (i.e., equal numbers of positive and negative signs). To address this, alternative methods have been developed to aggregate multiple signed interactions into more robust representations of user relationships.

One such method uses a Bayesian model that treats each signed interaction as a Bernoulli random variable~\citep{fraxanetUnpackingPolarizationAntagonism2024}. This model incorporates a beta prior to account for platform-level asymmetries between positive and negative interactions. Similar to the significance and vigour filter in~\citep{gursoyExtractingSignedBackbone2021}, given the posterior probability, only pairs with a clear bias towards one type of interaction and low variance will form a signed relationship, which will be determined by the direction of the bias.

\paragraph{Inferring Signed Interactions from Textual Content} 
A common approach to generate signed data from social platform interactions is to extract the agreement and disagreement from text, such as interactions through comments, with (Large) Language Models. In \citep{pougue-biyongDEBAGREEMENTCommentreplyDataset}, they introduce an annotated comment-reply dataset from Reddit, with labels of \textit{agree}, \textit{disagree}, and \textit{neutral}, finding that state-of-the-art pre-trained models struggle most at capturing neutral interactions. Interestingly, they also propose as future work the combination of unsigned network data and text annotations to perform edge sign predictions with Graph Neural Networks (see  Section~\ref{subsubsec:ML-link-pred}). 

Another work~\citep{keucheniusWhyItImportant2021} uses the \texttt{fastText} algorithm for text classification to classify Twitter mentions into positive and negative, and uses a majority vote system to infer relations between users. They show that, compared to analysing the Twitter debate with just unsigned interactions, the addition of antagonistic interactions enriches the understanding of the online debate by allowing for the separation of conflict and indifference.

Also using Twitter data, \citep{tacchiSignedEgoNetwork2022} infer signed interactions from sentiment analysis using the VADER framework~\citep{huttoVADERParsimoniousRuleBased2014}. Their approach infers relations relying on the psychological theory of the \textit{golden interaction threshold}, in which one relationship is considered positive as long as there is one or fewer negative interactions per five positive interactions (i.e., roughly 17\%).

While text-based inference allows signed modelling in settings where explicit signs are absent, it can introduce substantial noise and ambiguity due to sarcasm, context dependence, or classifier bias.

\paragraph{The Case-Study of Wikipedia}

Wikipedia has been the source of numerous studies that extract signed networks from co-editing behaviour and user interactions~\citep{brandesNetworkAnalysisCollaboration2009,sepehriradLeveragingEditorCollaboration2012, sepehri-radIdentifyingControversialWikipedia2015,nakamuraClusteringEditorsWikipedia2013,maniusilviuBuildingSignedNetwork,bogdanovCommunityDiscoverySigned2010,lernerFreeEncyclopediaThat2020}.
Some of these approaches focus on building signed networks based on text interactions and try to obtain a clustering of the contributors to analyse their relationships~\citep{brandesNetworkAnalysisCollaboration2009, nakamuraClusteringEditorsWikipedia2013}. Others incorporate both textual and behavioural signals, including interactions across articles and adminship elections, to construct signed and directed networks used for tasks such as link prediction or article quality estimation~\citep{maniusilviuBuildingSignedNetwork}.

Furthermore, Sepehri-Rad et al. \citep{sepehriradLeveragingEditorCollaboration2012,sepehri-radIdentifyingControversialWikipedia2015} used co-editing patterns to predict user attitudes and detect controversial material within articles. They define individual, directional, and mutual features based on collaborators who co-edit the same articles, and use the adminship votes as ground truth labels to train a model that detects which features are more predictive of the ``real'' relationship amongst contributors. Other, less conventional approaches assess the sign of the interaction based on the direction of the change in distance within a latent space of a topic model (i.e., if user $A$ moved the paragraph in the same direction as user $B$, they agree)~\citep{bogdanovCommunityDiscoverySigned2010}. 

Lastly, \citep{lernerFreeEncyclopediaThat2020} introduced signed networks based on \textit{undo} and \textit{re-do} actions between users and explored the effects of balance and reputation theories on editor interactions. The authors also considered potential actions to account for implicit interactions (e.g, a text from A could have been deleted by B, but it has not) with a decay weight. The relation between two users was then explained by latent probabilities specified by logistic regression models based on past interactions.

\subsection{Common data sources}\label{subsec:common-data}

In the following, we describe publicly available datasets of signed networks. We focus on four categories: in-person social communities, online platforms, political systems, and country systems. Other relevant research fields leveraging signed networks, e.g.\ ecology, have been discarded due to the unavailability of public databases in signed network format. Summarised information can be found in \autoref{tab:data} of the Appendix, alongside links to online data repositories.

\subsubsection{In-person social communities}

The description of human relationships, categorised as friendly or hostile, is a natural use case of signed networks. Datasets have been collected in the context of identified social communities, often through surveys, where the friendly or hostile nature of the links is self-reported. Often, these datasets are collected in waves at different points in time, allowing us to study the temporal evolution of the networks.

\begin{itemize}
    \item \textit{Sampson's monastery data}~\citep{sampsonNovitiatePeriodChange1968} records the evolution of relations between 18 monks during a political crisis within a US cloister, across three temporal waves. Each wave, the monks were asked to select the three others they esteemed the most and the one they esteemed the least. Similar ratings were collected for liking, influence, and praise instead of esteem.
    \item  \textit{Nordlie-Newcomb's fraternity dataset}~\citep{nordlieLONGITUDINALSTUDYINTERPERSONAL1958,newcombAcquaintanceProcess1961} is the result of an experiment of social psychology conducted at the University of Michigan. Seventeen male college students were housed together, and the authors periodically collected data via surveys to assess the evolution of friendships and antagonisms between the students. A similar study collected data about college female students in the US \citep{Lemann1952}.
    \item \textit{Spanish high school students}~\citep{ruiz-garciaTriadicInfluenceProxy2023}. Recently, Ruiz-Garcia et al.~\citep{ruiz-garciaTriadicInfluenceProxy2023} surveyed more than three thousand students (11-15 years old) in 13 Spanish schools, asking them about which other students they liked or disliked. The result is a directed, weighted network with over 60,000 edges, with additional information about individual features and personality traits. 
    \item \textit{Behavioural observations on preschoolers}~\citep{nealInferringSignedNetworks2022}. Neal et al.~\citep{nealInferringSignedNetworks2022} inferred positive and negative links between children in two classrooms of preschoolers ($N=35$), based on behavioural patterns during playtime. To do so, they leverage the backbone procedure described in \citep{nealBackboneBipartiteProjections2014} (see Section \ref{subsub:bipartite-to-signed}).
\end{itemize}

Other relevant data sources include signed networks of relationships between antique \textit{Philosophers}~\citep{Lemann1952}, alliances and antagonisms between $N=16$ \textit{tribes of the Highlands region in New Guinea}  \citep{readCulturesCentralHighlands1954}, as well as relationships between inhabitants of $N=32$ \textit{Honduran villages} collected during a clinical trial. We finally mention another widely used source of data, the \textit{Urban Communes Dataset}, which records the nature of relationships between groups of inhabitants in several major US cities, alongside socio-demographic indicators and individual beliefs, across multiple waves. Formerly freely available, it is now accessible on request.

\subsubsection{Online platforms}\label{subsubsec:online-platforms}
The availability of large-scale, high-quality signed datasets for studying online social interactions remains limited. A few datasets have become standard benchmarks, in particular those hosted in the SNAP repository~\citep{snapnets}, and have supported a wide range of work on sign prediction, community detection, and structural balance:

\begin{itemize}
    \item \textit{Epinions trust network}~\citep{leskovecSignedNetworksSocial2010}. This dataset originates from the Epinions product review platform, where users could explicitly express trust or distrust toward other members. The resulting signed network captures these declared trust relationships among users.
    \item \textit{Slashdot Zoo}~\footnote{\url{www.slashdot.org}}~\citep{gomez2008,leskovecSignedNetworksSocial2010}. Slashdot is a technology-oriented news platform featuring user-submitted and editor-evaluated stories related to science and technology. In 2002, the site introduced the \textit{Zoo} feature, allowing users to tag others as friends (those whose comments they appreciated) or foes (those whose comments they found uninteresting). The dataset includes three temporal snapshots of this signed social network.
\end{itemize}

Both Epinions~\citep{liuPredictingTrustsUsers2008, leskovecPredictingPositiveNegative2010, hamedaniTrustRecEffectiveApproach2021, ahmadalinezhadSignPredictionSigned2018, papaoikonomouPredictingEdgeSigns2014} and Slashdot~\citep{kunegisSlashdotZooMining2009, ahmadalinezhadSignPredictionSigned2018, papaoikonomouPredictingEdgeSigns2014} have been widely used in the literature for sign prediction and structural balance studies.

\begin{itemize}
    \item \textit{Bitcoin Alpha} and \textit{Bitcoin OTC}~\citep{kumarCommunityInteractionConflict2018}. These datasets represent who-trusts-whom networks extracted from two Bitcoin trading platforms. Users rate each other following transactions, creating directed weighted links that indicate trust or distrust. Since participants are anonymous, the reputation systems underlying these platforms aim to mitigate fraud and risk.
    \item \textit{Wikipedia Requests for Adminship (RfA)}~\citep{leskovecPredictingPositiveNegative2010,leskovecSignedNetworksSocial2010,west2014exploiting}. This dataset captures the voting process through which Wikipedia editors decide whether to promote a candidate to administrator status. Each vote can express support, opposition, or neutrality. The initial version, based on a 2008 dump, includes nearly $2{,}800$ elections. A later extended version by West et al.~\citep{west2014exploiting} spans the entire period from 2003 to 2013 and also contains the textual content of votes (i.e., the accompanying comments). Besides RfA, several studies have inferred signed relations from co-editing behaviour and interaction patterns among editors (see Section~\ref{subsubsec:online}).
    \item \textit{Reddit Hyperlink Network}~\citep{kumarCommunityInteractionConflict2018}. This dataset represents interactions between subreddits over a 2.5-year period (2014--2017). It is constructed from hyperlinks in user posts that reference content in other subreddits. Each edge originates from a source subreddit and points to a target subreddit, with sentiment annotations derived from the text of the linking post, resulting in a signed inter-community network.
\end{itemize}

Besides the datasets collected in the SNAP repository, recent works have made available several other data sources.
\begin{itemize}
    \item \textit{DEBAGREEMENT}~\citep{pougue-biyongDEBAGREEMENTCommentreplyDataset}. This dataset consists of 42,894 annotated comment–reply pairs collected from five politically and socially active subreddits: r/BlackLivesMatter, r/Brexit, r/climate, r/democrats, and r/Republican. Each pair is labelled as agree, neutral, or disagree, creating a signed interaction network among users based on the tone of their replies. 
    \item \textit{Birdwatch} (now Community Notes)~\citep{yasseri2023can,allenBirdsFeatherDont2022}. 
    This platform was launched by Twitter in January 2021 as a crowd-sourced fact-checking platform where selected users rated the trustworthiness of Tweets through evaluative notes. 
    Signed networks have been extracted from user-note ratings from Birdwatch data to study political polarisation and antagonism~\citep{yasseri2023can,allenBirdsFeatherDont2022}.
    As of this writing, anonymised Birdwatch (now Community Notes) data remains publicly available. 
    \item \textit{DerStandard Forum Interactions}~\citep{fraxanet2025decade,derstandard_data}. \textit{DerStandard}~\footnote{\url{https://www.derstandard.at/}} is an Austrian online newspaper that features a popular comment section integrated with its articles. The dataset captures user interactions on the platform over nearly ten years (2013--2022), including explicit agreement and disagreement votes on comments. These interactions can be used to build temporal signed networks across diverse topics such as football, national elections, or COVID-19 policy~\citep{fraxanetUnpackingPolarizationAntagonism2024}.
    \item \textit{Meneame.net}. Meneame~\footnote{\url{https://www.meneame.net/}}, a Spanish news aggregator with a similar structure to Reddit, was studied by~\citep{candellone2025community,candellone2025negative} to investigate the role of negative interactions in highlighting polarised content. The platform allows users to submit news stories and interact through upvotes, downvotes, and comments. 
    Moreover, unlike many platforms, Menéame applies no personalised recommendations, so any observed phenomenon arises purely from user behaviour.
\end{itemize}

Some additional platforms have supported signed or sign-inferable interactions, though the corresponding research does not provide public access to the datasets. Examples are from \textit{Yahoo Answers}~\citep{yangFriendFrenemyPredicting2012}, where the authors infer trust and distrust through question-and-answer behaviour and voting patterns; \textit{Essembly.com}~\citep{brzozowskiFriendsFoesIdeological2008}, where users can label others as ``friends'' or ``foes''; \textit{Pardus}~\citep{szellMultirelationalOrganizationLargescale2010}, a multi-relational online game, included both cooperative and antagonistic links.

\subsubsection{Political systems} \label{subsubsec:political-data}
The formalism of signed networks has raised interest as an effective tool to describe and analyse cooperation and competition between political actors. For example, one can represent a Parliament as a signed network of members of parliament (MPs) linked by positive or negative edges, where the sign represents the similarity of their voting behaviours. This can be used to study the dynamics of polarisation among political elites. Such data has been studied in the context of the EU, US, British, and Brazilian Parliaments, as well as the United Nations General Assembly \citep{arinikMultiplePartitioningMultiplex2020}; publicly available datasets include signed networks of the EU Parliament \citep{arinikMultiplePartitioningMultiplex2020} and the UK House of Commons \citep{intalDissentRebellionHouse2021}. An approach based on bill co-sponsorship in the US Congress has also been proposed \citep{nealSignTimesWeak2020}. The authors released a collection of signed networks, one for each US congressional session from 1973 (Senate) and 1979 (House of Representatives) up to 2016. The nodes represent MPs, and the sign of the edges is inferred using the SDSM method~\citep{nealBackboneBipartiteProjections2014} described in Section~\ref{subsub:bipartite-to-signed}. Another interesting data source includes a weighted signed network describing the self-reported similarity between Slovene political parties in 1994 \citep{Kropivnik1996}. 

\subsubsection{Country systems}\label{subsubsec:country-data}

International relations studies seek to understand patterns of cooperation, conflict, and interdependence between states. Signed networks provide a natural framework for this, where nodes represent states and signed edges represent cooperative or antagonistic relationships, such as formal alliances, defence agreements, or military disputes~\citep{askarisichani19952018GlobalEvolution2020,maozWhatEnemyMy2007,diaz-diazMathematicalModelingLocal2024}. Networks of international relations have also been used as benchmarks to evaluate the performance of balance indices and community detection algorithms~\citep{doreianStructuralBalanceSigned2015, estradaWalkbasedMeasureBalance2014, galloTestingStructuralBalance2024,jiangStochasticBlockModel2015, kirkleyBalanceSignedNetworks2019}.

One of the most prominent resources for quantitative international relations research is the \textit{Correlates of War} (COW) Project\footnote{\url{https://correlatesofwar.org/} (accessed October 14, 2025.)}, a collection of cross-national data on war and cooperation, including several datasets, such as:
\begin{itemize}
    \item The \textit{Militarised Interstate Disputes} dataset (v5.0) records instances in which one state threatened, displayed, or used force against another, covering the period 1816--2014 \citep{palmerMID5Dataset201120142022}.
    \item The \textit{War Data} series categorise inter-, intra-, extra-, and non-state wars between 1816--2007 \citep{sarkeesResortWar181620072025}.
    \item The \textit{Defense Cooperation Agreement Dataset} (DCAD)~\citep{kinne2020defense} covers bilateral defence treaties between 1980--2010, documenting defence-related collaboration such as policy coordination, joint military exercises, peacekeeping operations, or joint research and procurement.
    \item The \textit{Formal Alliances} dataset (v4.1)~\citep{giblerInternationalMilitaryAlliances2008} identifies all formal alliances between at least two states between 1816 and 2012, classifying them into defence pacts, neutrality or non-aggression treaties, and entente agreements. This dataset can be used to build positive edges between countries.
\end{itemize}

Another approach to mapping international relations relies on automated event data collection. The \textit{Integrated Crisis Early Warning System} (ICEWS) \citep{icews} compiled millions of coded socio-political interactions between international actors, extracted from news reports. ICEWS was discontinued in 2023 and succeeded by the \textit{POLitical Event Classification, Attributes, and Types} (POLECAT) dataset \citep{polecat}, which employs a machine-learning-based coder. POLECAT provides continuously updated, multilingual, and geolocated records of political events from 2018 onward.

Finally, another line of work infers signed relations between countries from cultural, rather than political, interactions \citep{garciaMeasuringCulturalDynamics2013}.  One example uses Eurovision voting data to measure affinity or antagonism between countries. 
The authors define a \textit{Friend-or-Foe} (FoF) coefficient from voting patterns and compare it with cultural distances computed using Hofstede's method~\citep{hofstede1984culture}, finding a significant negative correlation: country pairs with stronger affinities in Eurovision tend to be culturally closer.
\clearpage

\section{Myths and pitfalls}\label{chapter:pitfalls}

Signed network science is a highly interdisciplinary field, with ideas shared across sociology, physics, computer science, neuroscience, and many other domains. This interdisciplinarity is one of the field's greatest strengths, but it also creates challenges: terminology can vary across disciplines, concepts sometimes get extended in ways that obscure their original meaning, and the same words may refer to different ideas in different contexts. In this chapter, we discuss claims and assumptions (myths and pitfalls) that we believe benefit from clarification. These are loosely grouped into four themes. Items 1–4 concern the function and interpretation of negative edges. Items 5–8 address the relationship between signed networks and other network types, such as directed, weighted, or multilayer graphs. Items 9–13 deal with claims related to cognitive and structural balance. Finally, items 14–15 focus on other structural aspects. For each, we state the claim, offer our assessment, and discuss why the claim may be problematic, where ambiguity arises, and in which cases, if any, it may hold.

\begin{enumerate}
    \item \textbf{The underlying unsigned structure matters more than the sign.}

    \textit{Usually false}. In many complex systems, positive and negative links display different structural behaviours \citep{szellMeasuringSocialDynamics2010}, have distinct functional properties \citep{fox_PNAS_2005}, and are driven by different dynamical mechanisms (see Chapter \ref{chapter:dynamics}). The sign is therefore indispensable. Ignoring it removes essential information about inhibition, competition, or conflict, leading to misleading interpretations of the system's organisation and function. As shown in Section \ref{subsec:definition_signed_networks}, the coupling between positive and negative subnetworks cannot be treated independently, and as detailed in Section \ref{subsec:basic_metrics}, many network metrics (clustering, centralities, distances) fundamentally change their interpretation when signs are introduced. While in some contexts focusing on topology alone may be justified, in most applications the sign is a defining element of network structure, behaviour, and function.
    
\item \textbf{Positive and negative edges are equivalent and interchangeable.}

    \textit{False}. Positive and negative edges must be treated differently because they differ structurally, conceptually, and in how they emerge and change over time. Mathematically, positive and negative edges are not interchangeable: swapping all signs in a network fundamentally changes its structural and spectral properties (Section \ref{subsec:signed-laplacian}), modifies levels of balance (Section \ref{sec:balance_indices}), and alters centralities (Section \ref{subsubsect:centralities}) and mesoscale organisation (Section~\ref{subsec:comm-det}) in nontrivial ways. Conceptually, they encode different types of interaction. For example, alliances and wars are not mirror images of each other, and excitatory and inhibitory interactions in biology have complementary, not opposing, functions—they serve distinct roles in maintaining system stability and function rather than simply negating each other. Finally, empirical studies have shown that the evolution of positive and negative links is driven by different mechanisms (Section \ref{subsubsec:mechanisms-governing}). In social systems, they carry different cognitive weights: humans tend to process and remember hostility more vividly but also to resolve or forget it more quickly, making negative edges typically more volatile, short-lived, and less embedded in the network than positive ones (see Section \ref{subsec:dynamics-of-networks}).

\item \textbf{To turn a signed network into an unsigned one, one can add a constant value to all entries of the adjacency matrix to make it non-negative\footnote{One example of this operation would be to add one to all edges, which turns negative edges into non-edges, non-edges into weight-one edges, and positive edges into weight-two edges.}.} 

\textit{False}. The resulting matrix no longer represents the same network. The transformation creates spurious edges between unconnected nodes and treats negative edges as the absence of edges. As a result, the network's properties become distorted: degree distributions change (Section~\ref{subsec:basic_metrics}), mesoscale organisation is altered (Section~\ref{subsec:comm-det}), and dynamical processes defined on the network (Chapter~\ref{chapter:dynamics}) behave fundamentally differently. If a non-negative representation is required (for example, to apply unsigned spectral methods as in Section~\ref{subsubsec:spectral}), one should instead use mathematically consistent transformations. One example is the Gremban expansion~\citep{gremban1996combinatorial, diazdiazGremban}, which transforms a signed graph into an unsigned two-layer graph. This method duplicates the node set into two identical layers containing the positive edges, and represents each negative edge as a cross-layer link between the corresponding node copies. This transformation preserves structural properties like balance, mesoscale structure, and walk statistics, while making possible the use of techniques designed for unsigned networks.

\item \textbf{When constructing a network from correlation data, one may threshold all values below a chosen value 
$\theta$.}

\textit{Usually false}. Thresholding a signed correlation matrix can be done in two ways, and both approaches introduce systematic distortions. One option is to remove all values below a given threshold without considering their sign, which effectively discards all negative correlations. From the perspective of signed networks, this is particularly problematic: it eliminates all antagonistic or inhibitory relationships that are essential to the system’s structure and dynamics. For instance, in the brain, functional connectivity dynamically partitions into bipartite community structures that coalesce and dissolve over time~\citep{sporns2021dynamic}; thresholding away negative correlations would eliminate the anticorrelations that define these bipartitions. The second option is to threshold correlations by absolute value, discarding links of small magnitude but keeping negative values if they are sufficiently strong. While better than erasing all negative links, this approach also presents several issues: it can remove weak but important interactions and systematically bias the inferred network toward denser and more clustered configurations~\citep{zalesky2012use}. As a result, thresholding-based network reconstruction is discouraged, since the resulting topology rarely reflects the true organisation of the system~\citep{peixoto_PRX_2025}.

\item \textbf{Signed and directed networks encode the same information.}

\textit{False.} Signs and directions capture independent but complementary structural dimensions. Signs encode valences (positive or negative), while directions encode asymmetries (who acts on whom). A signed edge does not imply orientation, and a directed edge does not imply positivity or negativity. A possible source of confusion is the different interpretation of negative entries in different matrix representations (Section~\ref{subsec:definition_signed_networks}). When representing a network through an adjacency matrix, negative entries represent negative edges and asymmetric entries ($A_{ij}\neq A_{ji}$) represent directional links. However, when representing the network through an \textit{incidence matrix}~\citep{biggsAlgebraicGraphTheory1993}, positive and negative entries are used to represent the orientation of the edges ($+1$ for head, $-1$ for tail), not their sign. Although most developments in signed network theory have focused on undirected graphs, directionality is present in many real systems. In geopolitical contexts, for instance, both an armed conflict and an invasion may be represented as negative links, yet the direction of interaction carries very different meanings. Directionality can be essential for modelling the dynamical evolution of a network or for interpreting its structural properties. It is not the same for a node to \emph{send} many negative links as to \emph{receive} them: in the former case, the node may be conflictive or aggressive, whereas in the latter, it may simply be unpopular. Confounding senders and receivers may thus lead to incorrect conclusions about the network’s functional and dynamical behaviour.

\item \textbf{Signed networks are a subclass of weighted networks.}

\textit{Misleading}. Signed networks can be viewed as weighted networks whose edges take both positive and negative values. However, most of the literature on weighted networks assumes that weights are strictly positive, representing intensity, frequency, or capacity. This difference has deep methodological consequences. Many tools designed for weighted networks break down when negative edges appear, since the underlying mathematics rely on properties that do not hold in the presence of negative links. For example, distance-based centralities require non-negative path lengths (Section~\ref{subsec:basic_metrics}), and diffusion models require non-negative weights to ensure convergence to a unique and nontrivial stationary state (Section~\ref{subsec:dynamics-on-networks}). As a result, even though signed networks are technically a subclass of weighted networks, the presence of negative links requires dedicated methods that respect the fundamentally different nature of negative links.

\item \textbf{A signed network can be viewed as a multiplex network, with one layer for positive edges and another for negative ones.} 

\textit{Misleading}. A multiplex network consists of the same set of nodes connected through two or more layers, each representing a distinct type of interaction. In a signed network, one can, in principle, separate positive and negative edges as two layers of a multiplex, and this can be useful in some contexts, like analysing structural differences between distinct types of cooperation and conflict (e.g., friendship, communication, trade, enmity, aggression, punishment). 
However, this representation ignores the interplay between positive and negative links, and as we noted in Section~\ref{subsec:definition_signed_networks}, the coupling between positive and negative subnetworks should not be treated independently. Additionally, the interaction between positive and negative links is a key driver of emergent behaviour in signed networks (see Chapters~\ref{chapter:balance} and ~\ref{chapter:dynamics} on balance theory and dynamics), thus disregarding it can lead to misleading conclusions.

\item \textbf{Correlation matrices can be directly treated as signed networks.}

\textit{Misleading}. While correlation matrices encode pairwise signed relationships, they do not automatically constitute networks with meaningful structure. That said, network-based methods can be used to extract modules from a correlation matrix~\citep{macmahonCommunityDetectionCorrelation2015}, interpreting correlations as positive/negative network links. This approach is, however, only methodological: it employs a network representation to facilitate the analysis of the data, but does not imply that the underlying system is itself a network. Note, however, that it is possible to \textit{infer} a signed network from a correlation matrix, but only after applying appropriate statistical methods such as Bayesian inference~\citep{peixoto_PRX_2025}.

\item \textbf{Harary's definition of balance is a generalisation of Heider's notion of balance.}

\textit{False}. Heider’s formulation of balance is rooted in psychological consistency and aims to describe how individuals strive for cognitive consistency locally, among their positive and negative attitudes (Section~\ref{subsec:Heiderian_balance}). Although Heider’s theory is often illustrated through triads, it also applies more broadly to small systems of relations from a specific local perspective. Harary later translated some elements of Heider’s qualitative framework into mathematical form, redefining balance as a global property of a network in a mathematical sense, rather than a local property in a cognitive sense (Section~\ref{subsec:Hararian_balance}). In doing so, he omitted key aspects of Heider’s original theory. As discussed in Section~\ref{subsec:hei-har}, some predictions of Heider's theory are not captured by Harary's formulation---for instance, conflict aversion may favour dissolving negative links even when they participate in structurally balanced cycles. One should therefore avoid claiming balance in the Heiderian sense when applying approaches based on Harary's framework.

\item \textbf{Balance implies harmony, peace, or social cohesion.} 

\textit{False}. Structural balance describes a mathematical property of network organisation, not the quality of social relations or the absence of conflict. A perfectly balanced network naturally organises into two (or more, under weak balance) mutually antagonistic factions with exclusively positive edges within groups and negative edges between them (Section~\ref{subsec:math-propr-balanced}). Empirical studies of real social networks show that balanced states often correspond to the most polarised or fragmented configurations~\citep{talagaPolarizationMultiscaleStructural2023,diaz-diazMathematicalModelingLocal2024,vendevilleModelingEchoChamber2025}. Conversely, networks with low or moderate balance may exhibit more blended dynamics where positive and negative relationships coexist in less extreme patterns, potentially indicating more complex but less polarised social structures~\citep{talagaPolarizationMultiscaleStructural2023}. The apparent paradox—that "balance" produces polarisation—arises from conflating the colloquial meaning of balance (harmony, moderation) with its precise structural definition (even numbers of negative edges in cycles). As discussed in Section~\ref{subsec:hei-har}, this confusion partly comes from the different perspectives of Heiderian cognitive balance (which does emphasise psychological consistency and reduced tension at the individual level) versus Hararian structural balance (which characterises global network organisation that may produce group-level antagonism). Interpreting structural balance as evidence of peaceful coexistence or social harmony is misleading, and we should instead recognise it as a signature of structural division.

\item \textbf{Looking only at triangles is enough to assess structural balance.}

\textit{False}. According to Harary’s definition of structural balance (Section~\ref{subsec:Hararian_balance}), a network is balanced if and only if the product of signs along \textit{every} cycle is positive. This includes both triangles and longer cycles. Therefore, having only balanced triangles is a necessary but by no means sufficient condition for a graph to be balanced. As an example, consider a network consisting of a $(+--)$ triangle and a $(+---)$ square that share one of the negative edges. Clearly, every triangle in the graph is balanced, yet the 4-cycle is unbalanced. Restricting the analysis to triangles can thus overlook imbalances that arise in larger loops. In some cases, the domain-specific knowledge of the system under study may lead researchers to disregard longer cycles. If that is the case, assessing only triads might be justified, but it is essential to be precise with terminology. One should avoid referring to such measures as structural balance, which is a mathematically defined global property, and instead acknowledge the scale of the analysis.

\item \textbf{Structural balance in signed networks can be evaluated directly from empirical data without using null models.}

\textit{Only partially true}. Structural balance measures how easily a network can be divided into antagonistic groups with positive internal and negative external links, and a network is close to balance when only a few edges or cycles violate this pattern. This descriptive view does not require comparison with a null model and expresses how far from perfect balance a signed network is. However, null models become necessary when the goal is to assess \textit{significance} rather than describing the network. They indicate whether the observed balance goes beyond what a random sign assignment would produce, and thus whether the pattern reflects genuine structure or mere noise. The design of null models must also match the system: for example, if links can be created and destroyed, both topology and signs should be randomised, whereas if connections are fixed, only signs should be (see Section~\ref{subsec:null-models}). In short, both views are complementary: descriptive measures quantify how balanced a network is, while null models test whether this level of balance is meaningful. This distinction mirrors the descriptive versus inferential approaches to community detection~\citep{peixotoDescriptiveVsInferential2023}.

\item \textbf{Balance theory is a universal mechanism that dominates the evolution of all signed networks.}

\textit{False}. This misconception conflates three distinct claims about balance theory's scope, dominance, and dynamics.
Balance as an evolutionary mechanism does not appear universally in all signed networks. It relies on cognitive psychology assumptions (Chapter~\ref{chapter:balance}), which hold to varying extents in social systems but become meaningless in gene regulation, neural connectivity, or ecological networks. Nevertheless, many non-social systems display balance-like patterns arising not from cognitive consistency but from stability requirements or optimisation principles.
In what concerns mechanisms, even in social systems where balance operates, it is not necessarily dominant. Other forces, such as homophily, differential popularity, transitivity, reciprocity, and resource constraints, also shape evolution and often compete with balance (Section~\ref{subsec:dynamics-of-networks}). Some mechanisms may produce the same motif patterns as balance, leading to misattribution. 
In temporal dynamics, balancing mechanisms also do not guarantee that balance increases over time. Many signed networks do not move steadily toward higher balance, as balancing forces are counteracted by other factors like volatility of negative edges, external shocks, or other processes that reintroduce imbalance.

\item \textbf{Centrality measures in signed networks quantify influence or importance in the same way as in unsigned networks.}

\textit{False}. In signed networks, different centrality measures usually have different interpretations. For example, eigenvector centrality no longer measures overall importance but rather reflects polarisation: nodes with high positive values belong to one faction, and those with high negative values to the opposite one. Katz centrality is better seen as a measure of power, since it aggregates positive and negative walks that start at a given node. Nodes with many effective enemies (those connected through numerous negative walks to others) have lower centrality than nodes whose relations are mostly positive. Another clear example is subgraph centrality, which in signed networks measures local balance, as it compares the number of positive and negative closed walks that pass through a given node. Centralities in signed networks, therefore, can describe different characteristics of a given node (polarisation, power, or balance) instead of a single notion of importance.

\item \textbf{Communities in signed networks are found using the same definitions and algorithms as in unsigned networks.}

\textit{Usually false}. This misconception operates at two levels: conceptual (what communities are) and methodological (how to find them). \textit{Conceptually}, most signed network literature defines communities as groups with predominantly positive edges within and negative edges between them, following principles of structural balance (Section \ref{subsec:comm-det}). This differs fundamentally from unsigned networks, where communities are simply densely connected subgraphs without considering edge signs. However, some works apply unsigned definitions to signed data, leading to confusion when methods correctly identify signed communities (factions) but are dismissed as incorrect~\citep{knyazevSpectralPartitioningSigned2018,foxInvestigationSpectralClustering2018}. A terminological proposal~\citep{diazdiazGremban} helps clarify: use "communities" for densely connected groups regardless of sign, and "factions" for groups with positive internal and negative external edges. This separation makes explicit that dense connectivity and sign-based cohesion are distinct structural properties. \textit{Methodologically}, no single algorithm is universally correct for detecting mesoscale structure in signed networks, just as in unsigned networks. The appropriate method depends on the goal of the analysis and the assumptions about the data. For example, spectral methods are well-suited when the aim is to find factions that follow structural balance and are approximately the same size. In contrast, stochastic block models are more flexible and can detect a wider range of patterns, including signed core–periphery structures and other configurations that depart from strict balance.

\end{enumerate}

\clearpage

\section{Conclusions}\label{chapter:conclusions}

Throughout this book, we have built a comprehensive landscape of signed networks: their mathematical foundations, the organising principles of balance and frustration, methods for structural inference and dynamics, approaches to data collection, and the pitfalls that can undermine analysis. This foundation positions us to address a crucial question: where should the field go from here? 

The rapid development of signed network research—spanning sociology, psychology, computer science, physics, mathematics, ecology, and neuroscience—reflects both its maturity and its growing relevance. Yet, this expansion has created challenges. The absence of integration efforts means that synthesising findings, resolving conflicting results, and identifying productive research directions become increasingly difficult.

While individual domains have developed sophisticated methods and accumulated valuable findings, the field lacks systematic reviews and meta-analyses that can synthesise existing findings, validate or discard conflicting results, and guide future work. Different fields have developed their own terminology, methodological preferences, and evaluation standards, which can obscure common principles and lead to duplicated efforts. As a result, it becomes difficult to assess when findings conflict due to genuine differences in the systems studied versus differences in definitions, null models, or analytical choices.

This book has aimed to provide such integration, consolidating foundational theory, clarifying methodological choices, and identifying common misconceptions across domains. Yet, integration is an ongoing process. For the remainder of this last chapter, we identify challenges that remain open and opportunities to move forward.

\subsection*{Methodological and computational challenges}

\textbf{Scalability.} With the increasing availability of online relational data, signed networks can easily reach millions of nodes and edges. Yet many methods for analysing structural balance or detecting community structure do not scale well to these settings. Algorithms like simulated annealing or exact frustration minimisation are computationally intractable for large networks, and even spectral methods can become prohibitively expensive. Developing approximation algorithms, parallel implementations, or sampling-based approaches that maintain theoretical guarantees while scaling to massive networks remains an important direction.

\paragraph{Neutral and absent edges.} Most real-world signed networks are sparse, including a large number of links that are either unreported, non-existent, or explicitly neutral. However, many models treat all absent links as equivalent or irrelevant, which can bias their analysis. The absence of a link between two individuals who have never met is not the same as a neutral link between individuals who interact regularly but express neither affinity nor antagonism. Developing models that differentiate these cases is an important direction for both inference and benchmarking. 

\paragraph{Dynamics of negative edges.} In most dynamic models, negative edges are operationalised as sign-flipping mechanisms: in consensus dynamics, negative neighbours push opinions in opposite directions; in contagion models, negative edges transmit the opposite state. However, real-world antagonistic relations can produce more complex effects: avoidance (reducing contact altogether), suppression (blocking transmission), conflict escalation (amplifying rather than reversing signals), or noise injection. Moving beyond simple sign-flipping assumptions to capture these richer dynamics would improve our ability to model influence, diffusion, synchronisation, and learning in signed networks.

\paragraph{Signed motifs.} While balance at the triadic level has received significant attention, larger motifs involving combinations of positive and negative edges remain poorly understood. Three main challenges limit progress. First, the combinatorial explosion: the coexistence of positive and negative edges greatly increases the number of distinct motifs. Second, the inferential challenge: assessing statistical relevance requires appropriate null models that reflect system-specific constraints (Section~\ref{subsec:null-models}). Third, the interpretive challenge: providing causal explanations for motif abundance. At present, Heider’s theory of balance offers the dominant framework for explaining balanced motifs, but it applies primarily to social systems, and other mechanisms (Section~\ref{subsec:dynamics-of-networks}) may generate similar patterns, complicating identification of underlying processes.

\subsection*{Theoretical and conceptual challenges}

\paragraph{Extending core network metrics.} Some fundamental quantities still lack accepted signed versions (e.g., assortativity). Others admit multiple possible generalisations: the Laplacian can be either opposing or repelling, node degree can be positive or negative depending on whether one considers signed, net, or total degree, each of which suits different contexts. In many cases, familiar properties break down when signs are introduced: shortest-path distances no longer satisfy metric axioms, and random-walk transition rates can become negative~\citep{tianSpreadingStructuralBalance2024}. Even when a measure can be generalised, its interpretation often changes fundamentally. For instance, Katz centrality describes power relations and may become negative, and subgraph centrality encodes local balance rather than importance~\citep{diaz-diazMathematicalModelingLocal2024}. Developing principled frameworks for when to use which generalisation, and clarifying their distinct interpretations, would help standardise practice across the field.

\paragraph{Defining and measuring polarisation.} The term ``polarisation'' is used ambiguously in signed network research, encompassing distinct phenomena. Elite polarisation (in legislative or parliamentary contexts) differs from mass polarisation (in online platforms). Structural polarisation (the emergence of differentiated groups in relationship patterns) differs from affective polarisation (emotional or attitudinal hostility between groups)~\citep{iyengar2012affect}. Moreover, signed networks enable distinguishing between mere separation, cohesive but disconnected groups, and explicit antagonism through negative ties. Understanding how these definitions relate and how empirical results differ when only endorsing actions are considered versus when negative interactions are also included is crucial~\citep{candellone2025negative}, and what each reveals about underlying social or political processes remains an important open question.

\paragraph{Systematic evaluation across datasets and domains.} Currently, studies use different measures of balance, different null models, and different datasets, making comparisons difficult and generalisation unreliable. A coordinated effort to evaluate balance levels and other signed network properties under standardised assumptions would help clarify when and where balance emerges, how it relates to the type of relationships represented by edges (friendship, trust, alliances, animosity), and what theoretical insights different metrics actually capture. A similar systematic evaluation would benefit community detection methods, which currently lack comprehensive benchmarking.

\subsection*{Empirical and data challenges}

\paragraph{Moving beyond balance as the dominant lens.} Structural balance has received enormous attention because of its historical origins, elegant mathematical properties, and intuitive appeal. However, real systems are often driven by mechanisms that balance theory alone cannot capture. Homophily, differential popularity, reciprocity, status-seeking, resource constraints, and memory effects all shape signed network evolution (Section 5.2.3), and balance effects may be confounded with or dominated by these alternative mechanisms. Some of these can produce motif patterns indistinguishable from those generated by balance-seeking behaviour, leading to misattribution. Moreover, in non-social systems (ecological networks, gene regulatory circuits, financial correlations), patterns consistent with balance often arise from stability requirements or optimisation principles rather than cognitive consistency. A more complete theory of signed networks must consider how multiple mechanisms interact, reinforce, or compete with one another, and must develop methods to disentangle their effects empirically.

\paragraph{Context-aware application of methods.} Signed network research demands domain-specific understanding alongside general methodological knowledge. For instance, when modelling friendship networks, triadic mechanisms like transitivity and reciprocity may dominate, making longer cycles less meaningful for understanding dynamics. In contrast, geopolitical alliance networks may involve more strategic, long-range considerations where longer cycles matter. Similarly, assumptions about temporal dynamics (Markovian vs. non-Markovian), the role of directionality, and the interpretation of balance must be guided by knowledge of the specific system. Researchers need frameworks for determining which assumptions and methods suit which contexts: a form of "meta-knowledge" about when different tools apply.

\paragraph{When to use signed networks. } A foundational question remains often overlooked: what kinds of systems are best modelled as signed networks? While the framework is widely used, its application sometimes lacks clear justification. Not all oppositional or divergent relationships are best captured via edge signs. For example, there are serious concerns about treating correlation matrices as signed networks without prior processing~\citep{peixoto_NatComm_2022}. Distinguishing between situations where signed networks provide genuine insight versus where other representations (multilayer networks, attributed graphs, hypergraphs, time-varying networks~\cite{aleta2025multilayer}) might be more appropriate would help guide methodological choices and avoid misapplication.

\subsection*{Opportunities for progress and paths forward}

These challenges reflect both gaps in current knowledge and opportunities for the field to mature. The growing availability of temporal signed data from diverse sources enables testing theories of network evolution. Advances in machine learning, particularly graph neural networks and representation learning, offer powerful tools for sign prediction, community detection, and pattern recognition, though their application must be guided by theoretical understanding to ensure interpretability. The field is positioned to address important questions about online discourse, international relations, neural dynamics, and ecological stability: questions that require tools described in this book alongside careful attention to context and mechanisms. Yet tools and data alone are insufficient. Progress will require deeper integration across disciplines, clearer conceptual frameworks that go beyond balance to encompass multiple mechanisms, systematic evaluation that uses comparable methods and assumptions, and sustained attention to when and why different approaches apply. It requires recognising that signed networks are not a single method but a family of related approaches, each suited to different questions and contexts. It requires balancing the desire for general theory with the need for domain-specific knowledge about what structures mean and how they arise.
 
This book has sought to consolidate what we know, clarify what remains uncertain, and identify paths forward. Signed networks offer a powerful framework for understanding how positive and negative relationships jointly shape complex systems: from individual minds to global politics, from human behaviour to ecological webs. The opportunity now is to use that framework more effectively: with greater rigour, clearer integration, and deeper awareness of both its strengths and its limitations. The challenges remain substantial, but as this book demonstrates, so is the potential for meaningful advances in signed network science.

\clearpage

\section*{Author contributions}

A.S.T. and F.D.-D. coordinated the preparation of the book and ensured consistency across chapters; \\
Chapter 1 was written by A.S.T. and revised by F.D.-D.; \\
Chapter 2 was written by F.D.-D. and revised by A.V. and A.S.T.; \\
Chapter 3 was written by F.D.-D. and M.A.G.-C. and revised by F.D.-D., M.A.G.-C., E.F., and A.S.T.; \\
Chapter 4 was written by F.D.-D., E.C., and E.F., and revised by F.D.-D., E.C., and A.S.T.; \\
Chapter 5 was written by M.A.G.-C., I.F., and A.V., and revised by F.D.-D. and A.S.T.; \\
Chapter 6 was written by E.F. and A.V., and revised by F.D.-D., E.C., and A.S.T.; \\
Chapters 7 and 8 were written and revised jointly by F.D.-D., M.A.G.-C and A.S.T.; \\
Figures were prepared by E.F., E.C., M.A.G.-C., and A.V.; \\
All authors read and approved the final version of the book.


\subsection*{About the book and the FRIENDS project}
The \href{https://signet-friends.github.io/}{\textit{FRIENDS}} (Frameworks, Research and applIcations in complEx Networks with signeD edgeS) project is an ongoing initiative to promote the study of signed networks. This book is part of this effort. Additionally, on our website, we provide educational resources for learning and teaching signed-network concepts. We also organize satellites and workshops at complex systems and network science conferences. So far, we have been present at CCS 2024, NetSci 2025, and NetSci 2026.

\section*{Acknowledgements}
We thank the Winter Workshop on Complex Systems, and in particular the 2023 and 2024 editions, for creating the synergies that made this collaboration possible. We sincerely thank João Brázia, Violeta Calleja-Solanas, Gonzalo Contreras-Aso, Anna Gallo, Javier Garcia-Bernardo, Peter Gerbrands, Renaud Lambiotte, Francesco Paolo Nerini, Anxo Sánchez, and Mahdi Shafiee Kamalabad for their valuable feedback, with special thanks to Renaud Lambiotte for his extensive and insightful comments. We also gratefully acknowledge the discussions and panels at the FRIENDS Satellites of CCS 2024 and NetSci 2025, which inspired several parts of this review. We thank Nicole Samay for her careful work in refining the figures and ensuring a consistent visual style throughout the manuscript.
F.D.-D. thanks funding from Program for Units of Excellences Maria de Maeztu (CEX2021-001164-M/10.13039/501100011033)
and project APASOS (PID2021-122256NB-C22). M.A.G.-C. acknowledges support from the Comunidad de Madrid through the grants for the hiring of pre-doctoral research personnel in training (reference PIPF-2023/COM-29487). Both F.D.-D. and M.A.G.-C. thank grant PID2022-141802NB-I00 (BASIC) funded by MCIN/AEI/10.13039/501100011033 and by ‘ERDFA way of making Europe’. E.F. acknowledges funding from the Maria de Maeztu Units of Excellence Programme CEX2021-001195-M, funded by MICIU/AEI /10.13039/501100011033. I.F acknowledges support by Grant No. PRE2019-090279 (MCIN/AEI/10.13039/501100011033), Spanish Grant No. PID2021-128005NB-C22, funded by
MCIN/AEI/ 10.13039/501100011033. A.S.T. acknowledges support by FCT – Fundação para a Ciência e Tecnologia – through the LASIGE Research Unit, ref. UID/408/2025.
E.C., M.A.G.-C., and A.S.T acknowledge the support of the AccelNet-MultiNet program, a project of the National Science Foundation (Award \#1927425 and \#1927418).

\subsubsection*{Declaration of generative AI and AI-assisted technologies in the writing process}
During the preparation of this work, the author(s) used genAI for copy-editing and to improve readability. After using this tool/service, the authors reviewed and edited the content as needed and take(s) full responsibility for the content of the publication.

\printbibliography
\clearpage

\appendix
\setcounter{table}{0}
\renewcommand{\thetable}{A\arabic{table}}

\setcounter{figure}{0}
\renewcommand{\thefigure}{A\arabic{figure}}

\begin{landscape}
\section{Appendix}
\begin{longtblr}[
  caption = {Node embeddings methods on signed networks.},
  label = {tab:node-embeddings},
]{
  colspec = {l l X l},
  width = \linewidth,
  rowhead = 1,
}
\hline
\textbf{Author(s), Year} & \textbf{Method} & \textbf{Short description} & \textbf{Code} \\
\hline
Kermarrec \& Thraves (2014)~\citep{kermarrecSignedGraphEmbedding2014} & Graph Drawing & Euclidean embeddings aligning geometry with balance theory by placing positively connected nodes close together and negatively connected ones farther apart. & \\
Yuan et al. (2017)~\citep{yuanSNESignedNetwork2017} & SNE & Extension of the skip-gram objective to signed networks, enforcing proximity for friends and separation for enemies. & \\
Wang et al. (2017)~\citep{wangSignedNetworkEmbedding2017} & SiNE & Deep learning model for signed network embedding, enforcing structural balance with pairwise constraints. & \href{https://github.com/InzamamRahaman/SiNE}{InzamamRahaman/SiNE} \\
Islam et al. (2018)~\citep{islamDistributedRepresentationsSigned2018} & sign2vec & Node2vec-style random walks adapted with edge signs and balance-aware context sampling. & \href{https://github.com/raihan2108/signet}{raihan2108/signet} \\
Javari et al. (2018)~\citep{javariStatisticalLinkLabel2018} & STLGM & Statistical latent graph model for sign prediction and embeddings in signed networks. & \\
Derr et al. (2018)~\citep{derrSignedGraphConvolutional2018} & SGCN & Graph convolutional network adapted for signed graphs, incorporating balance theory in aggregation. & \href{https://github.com/benedekrozemberczki/SGCN}{benedekrozemberczki/SGCN} \\
Schaub et al. (2019)~\citep{schaubMultiscaleDynamicalEmbeddings2019} & Multiscale Dyn. & Embeddings derived from diffusion dynamics across temporal and structural scales, capturing community and polarisation patterns. & \href{https://github.com/michaelschaub/DynamicalEmbeddings}{michaelschaub/DynamicalEmbeddings} \\
Huang et al. (2019)~\citep{huangSignedGraphAttention2019} & SiGAT & Signed graph attention networks learning embeddings by weighting neighbours with balance-aware attention. & \href{https://github.com/huangjunjie-cs/SiGAT}{huangjunjie-cs/SiGAT} \\
Mara et al. (2020)~\citep{maraCSNEConditionalSigned2020} & CSNE & Conditional signed network embeddings using a maximum-entropy principle constrained by balance statistics. & \href{https://github.com/aida-ugent/CSNE}{aida-ugent/CSNE} \\
Shen \& Chung (2020)~\citep{shenDeepNetworkEmbedding2020} & DNE-SBP & Deep network embedding model for signed bipartite projections. & \href{https://github.com/shenxiaocam/Deep-network-embedding-for-graph-representation-learning-in-signed-networks}{shenxiaocam/Deep-network-embedding} \\
Song et al. (2021)~\citep{songHyperbolicNodeEmbedding2021} & HSNE & Hyperbolic embeddings on the Poincaré ball, well-suited for hierarchical and polarised signed structures. & \\
Yan et al. (2021)~\citep{yanMUSEMultifacetedAttention2021} & MUSE & Multifaceted attention signed GNN capturing different interaction types and their influence on embeddings. & \\
Shu et al. (2021)~\citep{shuSGCLContrastiveRepresentation2021} & SGCL & Contrastive representation learning for signed directed graphs using sign-aware graph augmentations and self-supervised objectives to learn robust node embeddings. & \href{https://github.com/xi0927/SGCL}{xi0927/SGCL}\\
Huang et al. (2021)~\citep{huangSignedBipartiteGraph2021} & SBGNN & Signed bipartite graph neural network extending GNNs to heterogeneous signed settings. & \href{https://github.com/huangjunjie-cs/SBGNN}{huangjunjie-cs/SBGNN} \\
Huang et al. (2022)~\citep{huangPOLEPolarizedEmbedding2022} & POLE & Random-walk–based embeddings preserving signed autocovariance, explicitly modelling multi-step polarisation. & \href{https://github.com/zexihuang/POLE}{zexihuang/POLE} \\
He et al. (2022)~\citep{heSSSNETSemiSupervisedSigned2022} & SSSNET & Semi-supervised signed GNN for clustering and embeddings with balance constraints. & \href{https://github.com/SherylHYX/SSSNET_Signed_Clustering}{SherylHYX/SSSNET\_Signed\_Clustering} \\
Pougué-Biyong et al. (2022)~\citep{pougue-biyongLearningStanceEmbeddings2022} & SEM & Stance embeddings combining signed node2vec walks with topical signals to capture ideological polarisation. & \href{https://github.com/lejohnyjohn/learning-stance-embeddings-from-signed-social-graphs}{lejohnyjohn/stance-embeddings} \\
Nakis et al. (2023)~\citep{nakisCharacterizingPolarizationSocial2023} & SLDM-SLIM & Latent distance embeddings mapping nodes to a simplex, representing polarisation through archetypal positions. & \href{https://github.com/Nicknakis/SLIM_RAA}{Nicknakis/SLIM\_RAA} \\
Nakis et al. (2023)~\citep{nakisHybridMembershipLatent2023} & sHM-LDM & Hybrid membership latent distance model allowing mixed affiliations and variable polarisation strength. & \href{https://github.com/Nicknakis/HM-LDM}{Nicknakis/HM-LDM} \\
Chen et al. (2023)~\citep{chenBalancingAugmentationEdgeUtility2023} & SiGAug & Data augmentation strategy for robust signed GNN embeddings. & \\
Huang et al. (2023)~\citep{huangSDGNNLearningNode2023} & SDGNN & Signed dual GNN learning embeddings through dual message passing. & \href{https://github.com/huangjunjie-cs/SiGAT}{huangjunjie-cs/SiGAT} \\
Kim et al. (2023)~\citep{kimTrustworthinessDrivenGraphConvolutional2023} & TrustSGCN & Trustworthiness-driven signed GCN enhancing robustness and interpretability. & \href{https://github.com/kmj0792/TrustSGCN}{kmj0792/TrustSGCN} \\
Sharma et al. (2023)~\citep{sharmaRepresentationLearningContinuousTime2023} & SEMBA & Continuous-time signed network embeddings via temporal point processes. & \href{https://github.com/claws-lab/semba}{claws-lab/semba} \\
Ko et al. (2023)~\citep{koUniversalGraphContrastive2023} & UGCL & Universal graph contrastive learning using magnetic Laplacian perturbations and spectral graph convolutions to learn self-supervised node embeddings across unsigned, directed, and signed graphs.  & \href{https://github.com/twko05/UGCL}{twko05/UGCL}\\
Li et al. (2023)~\citep{liSignedLaplacianGraph2023} & SLGNN & Signed Laplacian graph neural network combining low-pass and high-pass spectral filters with adaptive gating to jointly model positive and negative relations. & \\
He et al. (2023)~\citep{hePyTorchGeometricSigned2023} & pyTorch Geom. Sign. Dir. & Open-source PyTorch Geometric extension implementing signed and directed graph neural network models, datasets and evaluation tools within a unified framework. & \href{https://pytorch-geometric-signed-directed.readthedocs.io}{pytorch-geometric-signed-directed.readthedocs.io}\\
Babul \& Lambiotte (2024)~\citep{babulSHEEPSignedHamiltonian2024} & SHEEP & Spectral Hamiltonian embeddings based on a repelling Laplacian, emphasising separation of antagonistic groups. & \href{https://github.com/saynbabul/SHEEP}{saynbabul/SHEEP} \\
Diaz-Diaz \& Estrada (2025)~\citep{diazdiaz2025} & communicability & Communicability-based embeddings capturing signed geometric relationships. & \href{https://github.com/fernandodiazdiaz/signed_graphs_via_communicability_geometry}{signed\_graphs\_via\_communicability\_geometry} \\
Wang et al. (2025)~\citep{wangGegenNetSpectralConvolutional2025} & GegenNet  & Spectral graph neural network using Gegenbauer polynomial filters to learn expressive node embeddings on signed graphs while capturing both homophilic and heterophilic interactions. & \href{https://github.com/wanghewen/GegenNet}{wanghewen/GegenNet}\\
Zhou et al. (2026)~\citep{zhouAdversarialRobustnessLink2026} & BA-SGCL & Balance-augmented signed graph contrastive learning that improves the adversarial robustness of node embeddings by preserving balance-related information during contrastive training.  & \href{https://github.com/JialongZhou666/BA-SGCL}{JialongZhou666/BA-SGCL}\\
\hline
\end{longtblr}
\end{landscape}

\begin{landscape}
\begin{longtblr}[
  caption = {Sign prediction methods on signed networks.},
  label = {tab:sign-prediction},
]{
  colspec = {l l X l},
  width = \linewidth,
  rowhead = 1,
}
\hline
\textbf{Author(s), Year} & \textbf{Method} & \textbf{Short description} & \textbf{Code} \\
\hline
Guha et al. (2004)~\citep{guhaPropagationTrustDistrust2004} & Structural balance propagation & Sign inference from balance theory using products of signs along walks; negative influence propagates at most once. & \\
Kunegis et al. (2009, 2010)~\citep{kunegisSlashdotZooMining2009,kunegisSpectralAnalysisSigned2010} & Exponential and diffusion kernels & Predicts signs using matrix powers of $A$ with penalization ($c_k = \beta^k$ or $1/k!$); variants include effective resistance and diffusion kernels. & \\
Chiang et al. (2011)~\citep{chiangExploitingLongerCycles2011} & Extended structural balance & Incorporates higher-order cycles (up to $k$) into sign prediction, capturing longer balance patterns beyond triads. & \\
Singh \& Adhikari (2017)~\citep{singhMeasuringBalanceSigned2017} & Walk-based balance & Uses weighted sums of positive/negative walks to compute balance scores for edge signs. & \\
Kirkley \& Newman (2019)~\citep{kirkleyBalanceSignedNetworks2019} & Balance-based inference & Extends matrix-based formulations with factorial decays to estimate consistent sign assignments. & \\
Leskovec et al. (2010)~\citep{leskovecPredictingPositiveNegative2010} & Logistic regression & Treats sign prediction as supervised classification; uses degree-based and triad-based structural features. & \\
Chiang et al. (2011)~\citep{chiangExploitingLongerCycles2011} & Logistic regression & Improves Leskovec’s approach by incorporating cycle-based features (e.g., 4-cycles) for low-embeddedness edges. & \\
Papaoikonomou et al. (2014)~\citep{papaoikonomouPredictingEdgeSigns2014} & SVM with ego-graph motifs & Uses frequent subgraph patterns (motifs) in local neighbourhoods as features for sign classification. & \\
Derr et al. (2019)~\citep{derrBalanceSignedBipartite2019} & Bipartite motif classifier & Adapts motif-based features to bipartite signed networks (no triangles), using “caterpillars” and “butterflies.” & \href{https://github.com/DSE-MSU/signed-bipartite-networks}{DSE-MSU/signed-bipartite-networks} \\
Forsati et al. (2015)~\citep{forsati2015pushtrust} & Matrix factorization (MF) & Learns latent vectors $U, V$ minimizing a sign-consistent loss; predicts sign by $\operatorname{sign}(u_i^\top v_j)$. & \\
Hsieh et al. (2012)~\citep{hsiehLowRankModeling2012} & Low-rank matrix completion & Recovers missing signs by completing the adjacency matrix under a low-rank constraint (using trace norm). & \\
Ye et al. (2013)~\citep{ye_predicting_2013} & Transfer matrix factorisation & Jointly embeds source and target signed networks to transfer structural knowledge for improved sign prediction. & \\
Ruiz-García et al. (2023)~\citep{ruiz-garciaTriadicInfluenceProxy2023} & Neural imbalance predictor & Neural network predicting edge signs from triadic influence; finds structural information more predictive than node attributes. & \href{https://github.com/miguel-rg/triadic-influence}{miguel-rg/triadic-influence}\\
Luo et al. (2021)~\citep{luoInterpretableSignedLink2021} & SIHG & Hyperbolic GNN with attention separating positive/negative neighbours, aligning with balance and status theories. & \href{https://github.com/Luoyadan/SIHG}{Luoyadan/SIHG} \\
Fang et al. (2023)~\citep{fangSignedSubgraphEncoding2023} & SELO & Deep encoder learning edge embeddings from subgraphs, capturing higher-order signed structures. & \href{https://github.com/fang98/SELO}{fang98/SELO} \\
Chen et al. (2023)~\citep{chenBalancingAugmentationEdgeUtility2023} & SiGAug & Graph augmentation framework filtering unreliable negatives to enhance robustness in GNN-based sign prediction. & \\
\hline
\end{longtblr}
\end{landscape}

\begin{landscape}
\begin{table}[htbp]
\centering
\caption{Quality-function optimisation methods for community detection in signed networks.}
\label{tab:community-detection-quality}
\begin{tabularx}{\linewidth}{l l X l}
\hline
\textbf{Author(s), Year} & \textbf{Method} & \textbf{Short description} & \textbf{Code} \\
\hline
Doreian \& Mrvar (1996)~\citep{doreianPartitioningApproachStructural1996} & Blockmodel & Minimizes in-group disagreement and out-group agreement (Eq.~\ref{eq:doreian-mrvar-commdet}). & \href{https://github.com/schochastics/signnet}{schochastics/signnet} \\
Mrvar \& Doreian (2009)~\citep{mrvarPartitioningSignedTwoMode2009} & Two-mode blockmodel & Extension of blockmodels to bipartite signed networks. & \href{https://github.com/schochastics/signnet}{schochastics/signnet} \\
Doreian \& Mrvar (2009)~\citep{doreianPartitioningSignedSocial2009} & Relaxed blockmodel & Generalization allowing more than two factions. & \href{https://github.com/schochastics/signnet}{schochastics/signnet} \\
Gómez et al. (2009)~\citep{gomezAnalysisCommunityStructure2009} & Signed modularity & Extension of modularity with separate null models for positive and negative links (Eq.~\ref{eq:signed-modularity}). & \\
Traag \& Bruggeman (2009)~\citep{traagCommunityDetectionNetworks2009} & Spinglass & Potts model Hamiltonian unifying frustration and modularity (eq.~\ref{eq:traag_bruggerman}). & \href{https://github.com/igraph}{igraph} \\
Anchuri \& Magdon-Ismail (2012)~\citep{anchuriCommunitiesBalanceSigned2012} & Double quality function & Shows frustration minimisation and modularity maximisation are equivalent. & \\
Amelio \& Pizzuti (2013)~\citep{amelioCommunityMiningSigned2013} & SN-MOGA & Multi-objective evolutionary algorithm optimising modularity and frustration. & \\
Pougué-Biyong et al. (2024)~\citep{pougue-biyongSignedLouvainLouvainSigned2024} & SignedLouvain & Modification of Louvain allowing exploration via negative links. & \href{https://github.com/lejohnyjohn/signed-louvain}{lejohnyjohn/signed-louvain}\\
Mendonça et al. (2015)~\citep{mendoncaRelevanceNegativeLinks2015} & Parallel ILS & Iterative local search minimising frustration. & \href{https://github.com/CompNet/NetVotes}{CompNet/NetVotes} \\
Aref et al. (2019, 2020)~\citep{arefBalanceFrustrationSigned2019,arefModelingComputationalStudy2020,arefDetectingCoalitionsOptimally2020} & Integer programming & Exact optimization of frustration and coalition detection. & \\
\hline
\end{tabularx}
\end{table}
\end{landscape}

\begin{landscape}
\begin{table}[htbp]
\centering
\caption{Spectral methods for community detection in signed networks.}
\label{tab:community-detection-spectral}
\begin{tabularx}{\linewidth}{l l X l}
\hline
\textbf{Author(s), Year} & \textbf{Method} & \textbf{Short description} & \textbf{Code} \\
\hline
Kunegis et al. (2010, 2014)~\citep{kunegisSpectralAnalysisSigned2010,kunegisApplicationsStructuralBalance2014} & Signed Laplacian & Eigen-decomposition minimizing signed ratio cut (Eq.~\ref{eq:signed-ratio-cut}). & \\
Chiang et al. (2012)~\citep{chiangScalableClusteringSigned2012} & Balance Ratio Cut & Spectral relaxation handling multiple groups. & \\
Cucuringu et al. (2019)~\citep{cucuringuSPONGEGeneralizedEigenproblem2019} & SPONGE & Combines Laplacians of positive/negative subgraphs with parameters. & \href{https://github.com/alan-turing-institute/signet}{alan-turing-institute/signet} \\
Bonchi et al. (2019)~\citep{bonchiDiscoveringPolarizedCommunities2019} & EIGENSIGN & Adjacency-based spectral method for polarised groups. & \href{https://github.com/egalimberti/polarized_communities}{egalimberti/polarized\_communities} \\
Zhong et al. (2020)~\citep{zhongEfficientAlgorithmBased2020} & SNBT & Non-backtracking matrix eigen-decomposition for signed networks. & \\
Fox et al. (2017, 2018)~\citep{fox2017numerical,foxInvestigationSpectralClustering2018} & Gremban expansion & Embeds signed graphs into unsigned graphs for standard spectral clustering. & \\
\hline
\end{tabularx}
\end{table}
\end{landscape}

\begin{landscape}
\begin{table}[htbp]
\centering
\caption{Statistical inference methods for community detection in signed networks.}
\label{tab:community-detection-stat}
\begin{tabularx}{\linewidth}{l l X l}
\hline
\textbf{Author(s), Year} & \textbf{Method} & \textbf{Short description} & \textbf{Code} \\
\hline
Chen et al. (2014)~\citep{chenOverlappingCommunityDetection2014} & SPM & Signed probabilistic mixture model, restrictive assumption of perfect balance. & \\
Jiang (2015)~\citep{jiangStochasticBlockModel2015} & SBM variant & Relaxes Chen’s model, allows positive/negative edges with different probabilities (Eq.~\ref{eq:jiang-likelihood}). & \\
Peixoto (2018)~\citep{peixotoNonparametricWeightedStochastic2018} & Weighted SBM & Edge signs treated as covariates in hierarchical SBM. & \href{https://graph-tool.skewed.de/}{graph-tool.skewed.de} \\
Aicher et al. (2015)~\citep{aicherLearningLatentBlock2015} & Generalised SBM & Handles signed weights via exponential family distributions. & \\
Yang et al. (2017)~\citep{yangStochasticBlockmodelingVariational2017} & Bayesian SBM & Variational Bayesian inference for signed networks. & \\
Zhao et al. (2017, 2018)~\citep{zhaoStatisticalInferenceCommunity2017,zhaoNetworkModellingVariational2018} & EM/Bayesian SBM & EM-based and Bayesian approaches for signed networks. & \\
Li et al. (2023)~\citep{liSSBMSignedStochastic2023} & SSBM & Signed SBM with explicit sign probabilities. & \\
Nakis et al. (2023)~\citep{nakisHybridMembershipLatent2023} & sHM-LDM & Latent distance model embedding nodes in a simplex. & \href{https://github.com/Nicknakis/HM-LDM}{Nicknakis/HM-LDM} \\
Gallo et al. (2024)~\citep{galloAssessingFrustrationRealworld2024} & SSBM-BIC & Maximum entropy SBM with BIC-based parameter selection. & \\
Wang et al. (2025)~\citep{wangGegenNetSpectralConvolutional2025} & Signed block $\beta$-model & Extends the signed stochastic block model to heterogeneous networks by modelling community structure and degree heterogeneity. & \\
\hline
\end{tabularx}
\end{table}
\end{landscape}

\begin{landscape}
\begin{table}[htbp]
\centering
\caption{Other methods for community detection in signed networks.}
\label{tab:community-detection-other}
\begin{tabularx}{\linewidth}{l l X l}
\hline
\textbf{Author(s), Year} & \textbf{Method} & \textbf{Short description} & \textbf{Code} \\
\hline
Hsieh et al. (2012)~\citep{hsiehLowRankModeling2012} & Matrix completion & Low-rank modelling for signed adjacency matrices. & \\
Esmailian et al. (2015)~\citep{esmailianCommunityDetectionSigned2015} & Infomap & Extension of Infomap to signed networks. & \href{https://github.com/pouyaesm/signed-community-detection}{pouyaesm/signed-community-detection} \\
Chu et al. (2016)~\citep{chuFindingGangsWar2016} & k-OCG & Find $k$ cohesive subgraphs with high intra-graph cohesion and large inter-graph opposition. & \\
Chang et al. (2017)~\citep{changCommunityDetectionSigned2017} &  Hamming distance & Breadth-first search, bit-flipping algorithm, and a belief propagation algorithm. & \\
Su et al. (2017)~\citep{suAlgorithmBasedPositive2017} & SRWA & Random walk strategies for signed networks. & \\
Che et al. (2020)~\citep{cheMemeticAlgorithmCommunity2020} & MACD-SN & Based on evolutionary theories, a memetic algorithm & \\
\hline
\end{tabularx}
\end{table}
\end{landscape}

\begin{landscape}
\begin{table}[ht]
\caption{Overview of available signed network datasets. This list only includes already processed data in a network format. The number of nodes and edges is given as averages across datasets that contain multiple networks. The ``Multiple'' column marks datasets that contain multiple networks. RfA: Request for Adminship.} 
\label{tab:data}
\begin{tabular}{llccccllr}
\toprule
\multicolumn{1}{l}{\textbf{Category}} & \textbf{Name} & \textbf{Directed} & \textbf{Weighted} & \textbf{Temporal} & \textbf{Multiple} & \textbf{\#Nodes} & \textbf{\#Edges} & \textbf{URL} \\
\midrule
\multirow{4}{*}{Political Systems} & A sign of the times &  &  & \checkmark & \checkmark & 550 & 25,000 & \href{https://figshare.com/articles/dataset/A_Sign_of_the_Times/8096429/3?file=20758701}{Figshare} \\
 & NetVotes '17 &  & \checkmark & \checkmark & \checkmark & 75 & 1,350 & \href{https://figshare.com/articles/dataset/NetVotes_2017_-_iKnow_17/5785833}{Figshare} \\
 & House of Commons &  & \checkmark &  & \checkmark & 639 & 203,841 & \href{http://doi.org/10.7910/DVN/OR05MA}{Dataverse} \\
 & Slovene parties &  & \checkmark &  &  & 10 & 45 & \href{https://datarepository.wolframcloud.com/resources/Slovene-Parliamentary-Parties-Network/}{Wolfram Data} \\
 \midrule
\multirow{13}{*}{Online Systems} & Epinions & \checkmark &  &  &  & 131,828 & 841,372 & \href{https://snap.stanford.edu/data/soc-sign-epinions.html}{SNAP} \\
 & Slashdot 081106 & \checkmark &  &  &  & 77,357 & 516,575 & \href{https://snap.stanford.edu/data/soc-sign-Slashdot081106.html}{SNAP} \\
 & Slashdot 090216 & \checkmark &  &  &  & 81,871 & 545,671 & \href{https://snap.stanford.edu/data/soc-sign-Slashdot090216.html}{SNAP} \\
 & Slashdot 090221 & \checkmark &  &  &  & 82,144 & 549,202 & \href{https://snap.stanford.edu/data/soc-sign-Slashdot090221.html}{SNAP} \\
 & Wikipedia RfA 2008 & \checkmark &  &  &  & 7,000 & 100,000 & \href{https://snap.stanford.edu/data/wiki-Elec.html}{SNAP} \\
 & Wikipedia RfA 2003 - 2013 & \checkmark &  & \checkmark &  & 10,835 & 159,388 & \href{https://snap.stanford.edu/data/wiki-RfA.html}{SNAP} \\
 & Wikipedia Edit Wars 2009 &  &  & \checkmark &  & 118,100 & 2,917,785 & \href{https://icon.colorado.edu/networks}{ICON} \\
  & Wikipedia User Interaction 2011 &  &  &  &  & 138,592 & 740,397 & \href{https://icon.colorado.edu/networks}{ICON} \\
 & Bitcoin OTC & \checkmark & \checkmark & \checkmark &  & 5,881 & 35,592 & \href{https://snap.stanford.edu/data/soc-sign-bitcoin-otc.html}{SNAP} \\
 & Bitcoin Alpha & \checkmark & \checkmark & \checkmark &  & 3,783 & 24,186 & \href{https://snap.stanford.edu/data/soc-sign-bitcoin-alpha.html}{SNAP} \\
 & Reddit Hyperlinks & \checkmark &  & \checkmark &  & 55,863 & 858,49 & \href{https://snap.stanford.edu/data/soc-RedditHyperlinks.html}{SNAP} \\
 & Reddit Debagreement & \checkmark & \checkmark & \checkmark &  & 23,575 & 42,894 & \href{https://drive.google.com/drive/folders/14Qo7OIOIodcp73ynG5u8WdPp90mh1un6}{Google drive} \\
 & Meneame & \checkmark & \checkmark & \checkmark &  & 8,604 & 3,113,86 & \href{https://zenodo.org/records/15682068}{Zenodo} \\
 & DerStandard & \checkmark & \checkmark & \checkmark &  & 247,863 & 412,511,165 & \href{https://doi.org/10.82201/RVQGUG}{Dataverse} \\
 \midrule
\multicolumn{1}{l}{\multirow{7}{*}{In-person social communities}} & Preschoolers & \checkmark &  &  & \checkmark & 17 & 76 & \href{https://osf.io/q7nh6/overview}{OSF} \\
\multicolumn{1}{l}{} & Honduras & \checkmark &  &  & \checkmark & 5,773 & 350 & \href{https://clinicaltrials.gov/study/NCT01672580}{Clinical Trials} \\
\multicolumn{1}{l}{} & Spanish schools & \checkmark & \checkmark &  &  & 3,395 & 60,566 & \href{https://zenodo.org/records/7647000#.Y-5eDtLMJH4}{Zenodo} \\
\multicolumn{1}{l}{} & Sampson's monastery & \checkmark &  & \checkmark & \checkmark & 18 & 104 & \href{https://doi.org/10.6084/m9.figshare.12152628}{Figshare} \\
\multicolumn{1}{l}{} & New Guinea &  &  &  &  & 16 & 116 & \href{https://networks.skewed.de/net/new_guinea_tribes}{Skewed} \\
\multicolumn{1}{l}{} & College preferences & \checkmark & \checkmark &  & \checkmark & 19 & 86 & \href{https://doi.org/10.6084/m9.figshare.12152628}{Figshare} \\
\multicolumn{1}{l}{} & Newcomb's fraternity & \checkmark &  & \checkmark & \checkmark & 17 & 102 & \href{https://doi.org/10.6084/m9.figshare.12152628}{Figshare} 
\\
\bottomrule
\end{tabular}
\end{table}\end{landscape}

\clearpage

\end{document}